 \documentclass[onecolumn, nofootinbib, aps]{revtex4}
\usepackage{CJK}
\usepackage{graphicx}% Include figure fixles
\usepackage{dcolumn}% Align table columns on decimal point
\usepackage{bm}% bold math
\usepackage{amsmath}
\usepackage{amsthm}
\usepackage{amssymb}
\usepackage{physics}
\usepackage{diagbox}
\usepackage{dsfont}
\usepackage[normalem]{ulem}
\usepackage{multirow}
\usepackage[shortlabels]{enumitem}
\usepackage[colorlinks]{hyperref}
\usepackage[toc,page]{appendix}
\usepackage{hyperref}
\usepackage{adjustbox}
\usepackage{xcolor}
\usepackage{wrapfig}
\usepackage{setspace}
\usepackage{tcolorbox}

\usepackage[noend]{algpseudocode}
\usepackage{algorithm}

\usepackage{bibunits}

\usepackage{caption}
\usepackage{subcaption}

\algnewcommand{\Initialize}[1]{%
  \State \textbf{Initialize:}
  \Statex \hspace*{\algorithmicindent}\parbox[t]{.8\linewidth}{\raggedright #1}
}

\hypersetup{colorlinks, citecolor=blue, filecolor=blue, linkcolor=blue, urlcolor=magenta}

\def\setelem#1{\expandafter\def\csname myarray(#1)\endcsname}
\def\dispsymbol#1{\csname myarray(#1)\endcsname}

\newcolumntype{?}{!{\color{black}\vrule width 1pt}}

\newcolumntype{C}[1]{>{\centering\let\newline\\\arraybackslash\hspace{0pt}}m{#1}}
\newcolumntype{N}{@{}m{0pt}@{}}

\def\squareforqed{\hbox{\rlap{$\sqcap$}$\sqcup$}}
\def\qed{\ifmmode\squareforqed\else{\unskip\nobreak\hfil
		\penalty50\hskip1em\null\nobreak\hfil\squareforqed
		\parfillskip=0pt\finalhyphendemerits=0\endgraf}\fi}

\def\duzomniejsze{<\kern-.7mm<}
\def\duzowieksze{>\kern-.7mm>}

\def\textbf#1{{\bf #1}}
\def\beq{\begin{equation}}
\def\eeq{\end{equation}}
\def\be{\begin{equation}}
\def\ee{\end{equation}}

\def\bal{\begin{align}}
\def\eal{\end{align}}

\def\ben{\begin{eqnarray}}
\def\een{\end{eqnarray}}
\def\beqa{\begin{eqnarray}}
\def\eeqa{\end{eqnarray}}
\def\eea{\end{array}}
\def\bea{\begin{array}}

\newcommand{\bei}{\begin{itemize}}
	\newcommand{\eei}{\end{itemize}}
\newcommand{\bee}{\begin{enumerate}}
	\newcommand{\eee}{\end{enumerate}}
\newcommand{\nc}{\newcommand}
\nc{\1}{{\openone}}

\def\tr{{\rm Tr}}
\def\id{{\mathds{1}}}

\def\>{\rangle}
\def\<{\langle}

\newtheorem{axiom1}{Axiom}
\newtheorem{axiom2}{Axiom}
\newtheorem{lemma}{Lemma}
\newtheorem{corollary}{Corollary}
\newtheorem{proposition}{Proposition}
\newtheorem{theorem}{Theorem}
\newtheorem{definition}{Definition}
\newtheorem{observation}{Observation}

\newtheorem{rem}{Remark}

\newtheorem{notation}{Notation}
\newtheorem{remark}{Remark}
\newtheorem{property}{Property}
\theoremstyle{plain}

\newtheorem{ax1}{Axiom}

\newtheorem*{pr1}{Property 1}
\newtheorem*{pr2}{Property 2}

\def\bed{\begin{definition}}
	\def\eed{\end{definition}}
\def\bel{\begin{lemma}}
	\def\eel{\end{lemma}}
\def\bet{\begin{theorem}}
	\def\eet{\end{theorem}}
\def\bet{\begin{notation}}
	\def\eet{\end{notation}}
\def\bet{\begin{remark}}
	\def\eet{\end{remark}}
\def\bet{\begin{postulate}}
	\def\eet{\end{postulate}}

\usepackage{ragged2e}

\begin{document}

\title{Quantification of the energy consumption of entanglement distribution}% Force line breaks with \\

\author{Karol Horodecki$^{(1)}$, Marek Winczewski$^{(1,2)}$, Leonard Sikorski$^{(1)}$, Paweł Mazurek$^{(1)}$, Mikołaj Czechlewski$^{(1)}$, Raja Yehia$^{(3)}$ }

% \affiliation{$^1$Institute of Theoretical Physics and Astrophysics and National Quantum Information Centre in Gda\'nsk, University of Gda\'nsk, 80--952 Gda\'nsk, Poland}
% \affiliation{$^2$ International Centre for Theory of Quantum Technologies, University of Gda\'nsk, 80--952 Gda\'nsk, Poland}
% \affiliation{$^3$Institute of Informatics and National Quantum Information Centre in Gda\'nsk, Faculty of Mathematics, Physics and Informatics, University of Gda\'nsk, 80--952 Gda\'nsk, Poland}

\affiliation{$^{1}$ 
Institute of Informatics, National Quantum Information Centre, Faculty of Mathematics, Physics and Informatics, University of Gdańsk, Wita Stwosza 57, 80-308 Gdańsk, Poland
}
\affiliation{$^2$ Univ Lyon, Inria, ENS de Lyon, LIP, 46 Allee d’Italie, 69364 Lyon Cedex 07, France}
\affiliation{$^{3}$ 
ICFO - Institut de Ciencies Fotoniques, The Barcelona Institute of Science and Technology, Castelldefels, Spain
}

\begin{abstract}
Inspired by environmental sciences, we develop a framework to quantify the energy needed to generate quantum entanglement via noisy quantum channels, focusing on the hardware-independent, i.e. fundamental cost. Within this framework, we define a measure of the minimal fundamental energy consumption rate per distributed entanglement (expressed in Joule per ebit). We then derive a lower bound on the energy cost of distributing a maximally entangled state via a quantum channel, which yields a quantitative estimate of energy investment per entangled bit for future quantum networks. We thereby show that irreversibility in entanglement theory implies a non-zero energy cost in standard entanglement distribution protocols.

We further establish an upper bound on the fundamental energy consumption rate of entanglement distribution by determining the minimal energy required to implement quantum operations via classical control. To this end, we formulate the axioms for an energy cost measure and introduce a Hamiltonian model for classically-controlled quantum operations. The fundamental cost is then defined as the infimum energy over all such Hamiltonian protocols, with or without specific hardware constraints. The study of the energy cost of a quantum operation is general enough to be naturally applicable to quantum computing and is of independent interest.

Finally, we evaluate the energy demands of three entanglement distillation protocols for photonic polarization qubits, finding that, due to entanglement irreversibility, their required energy exceeds the fundamental lower bound by many orders of magnitude. The introduced paradigm can be applied to other quantum resources, with appropriate changes depending on their nature. 
\end{abstract}

\maketitle

\tableofcontents

%LSBIB\defaultbibliographystyle{apsrev} 
%LSBIB\defaultbibliography{References}
%LSBIB\begin{bibunit}

\section{Introduction}
Entanglement distillation \cite{bennett1996purification} is an important cornerstone of the era of technological advancement, in which the rise of quantum technologies represents both a groundbreaking opportunity and a significant challenge. Quantum communication \cite{QInt_Wehner2018}, quantum computing \cite{Feynman1982}, and quantum sensing \cite{Degen2017} promise to revolutionize industries, from cryptography to healthcare, by solving problems that are currently intractable for classical systems and enhancing current capabilities. However, the dual objective to the technological advancement is environmental sustainability and wider economic cost.

It is these dual objectives that motivate natural questions: how energetically costly entanglement distillation is in principle? How much can we improve implementation of the currently used protocols in given technologies with minimizing energy costs in mind? Finally, entanglement distillation is yet another instance of a task for which a general question can be asked: if it is discovered that a quantum technology solves a task 2 times faster than its classical counterpart but costs 3 times more energy, would we not prefer to build two instances of the classical device? (this effect has been already shown for the case of technological energy cost of QKD protocols in \cite{YehiaCentrone}). In this paper, we contribute towards addressing the above questions. 

Entanglement is undoubtedly one of the first recognized, well studied and widely applied quantum resource. It arouses curiosity of scientists in various areas of research from quantum and high energy physics \cite{HHHH09, Aad2024}, through cosmology \cite{capozziello2013}, to machine learning \cite{wang2021d2}.
It is the underpinning of many crucial protocols e.g. teleportation \cite{Teleport_PhysRevLett.70.1895}, superdense coding \cite{SuperDense_PhysRevLett.69.2881} or device dependent and independent QKD \cite{DDQKD_Bennett2014,DIQKD_PhysRevLett.98.230501}. Moreover, long distance entanglement created by entanglement swapping \cite{EntSwap_PhysRevLett.71.4287} is the building block of the quantum internet \cite{QInt_Wehner2018}. However, it comes with a pitch of salt--the phenomenon of irreversibility. Distant parties sharing entanglement can process it only by LOCC transformations. Generically, this implies that creation of a quantum shared state costs more in terms of entangled bits $E_C$ (entanglement cost) than the quantity $E_D$ (distillable entanglement) that the distant parties can obtain in useful form. As we will see, this fundamental irreversibility establishes a fundamental energetic lower bound on entanglement distillation, and is the basis of all the results, along with other fundamental principles like energy conservation and Landauer erasure principle.

\begin{figure}
    \includegraphics[scale=0.5]{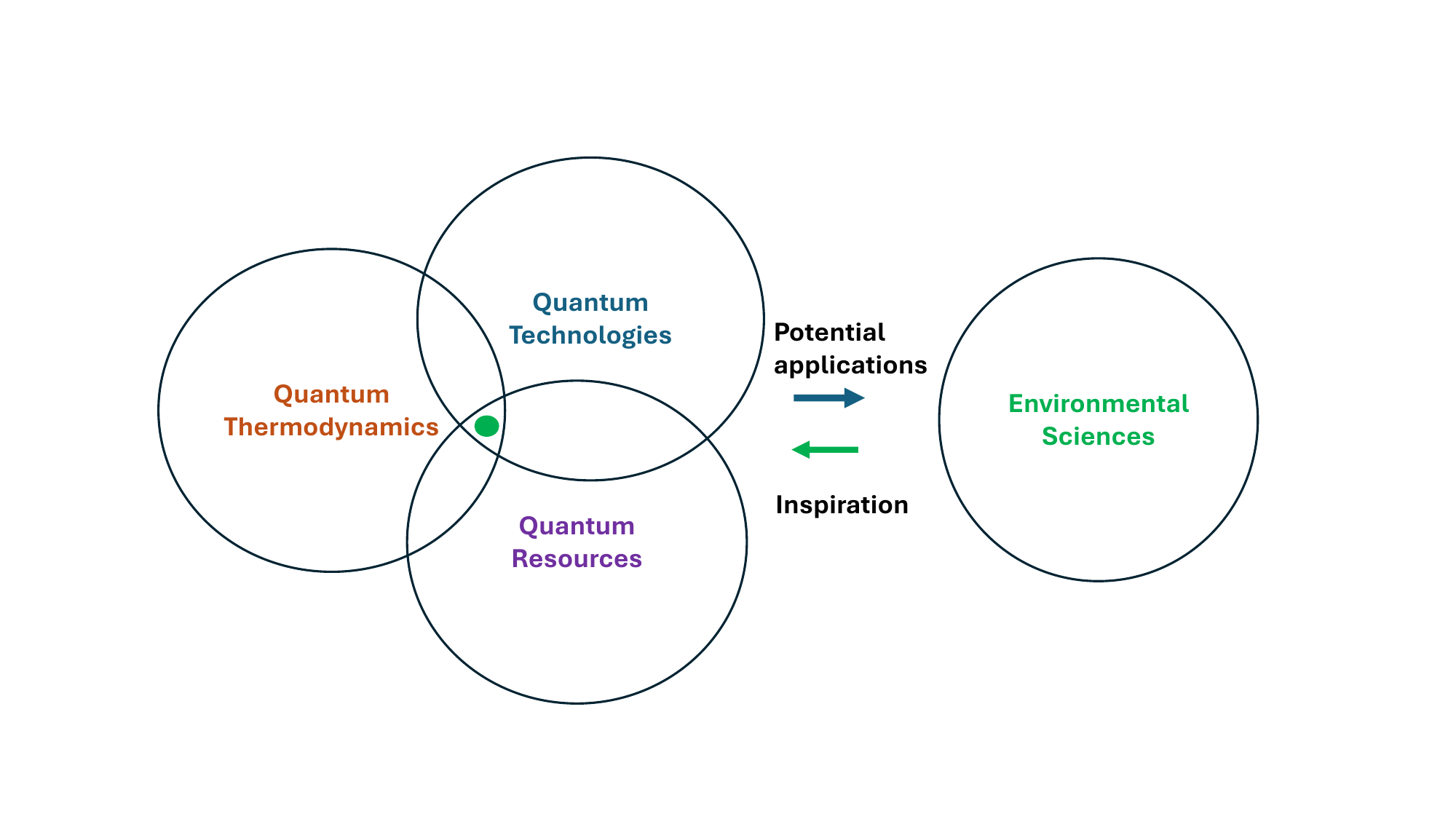}
    \caption{\justifying The quantification of the energy consumption rate of entanglement distribution in the landscape of scientific domains. It is based on the formalism of quantum resource theories, quantum thermodynamics and quantum technologies (where we focus on quantum channels). We emphasize the interplay between the cross-section of these domains and environmental sciences. While we are inspired by the latter, we see potential applications of the present work. }
    \label{fig:where_we_are}
\end{figure}

In the broader context, the problem of energetic costs of entanglement distillation falls into a cross-section of three established domains of quantum information theory: quantum resources, quantum thermodynamics
and quantum technologies, with a focus on quantum channels theory.
Additionally, we are inspired by environmental sciences. We focus not only on the work cost of the process, but we also take into account the heat pumped into a system so that a task, such as entanglement distribution, is achieved. We also expect potential applications of our findings in the field of environmental sciences. The landscape of the science's domain related to this article is summarized in Fig. \ref{fig:where_we_are}.

The results we present in this work provide the first insights to the fundamental energy consumption of entanglement distribution, taking into account entanglement creation, its transmission in a noisy quantum channel and its further distillation to a pure form. Our findings advance our understanding of the bridge between physical resources, such as energy and informational resource, such as entanglement, filling a critical gap in the current literature.

So far, particular attention has been put into assessing energy costs of quantum computation tasks \cite{Asiani_2019, Sagawa_2008, Faist_2018, PRXQuantum.2.040335}. In particular, in \cite{Asiani_2019} a broad paradigm of estimating energy efficiency of quantum processing has been put forward. While our approach is bares similarity to the Metric-Noise-Resource methodology (for discussion of the similarities and differences see Section \ref{ch:Lit}), it is conceptually closer to study of energy consumption of quantum communication tasks, such as letting two parties share an entangled state, as we propose to study here.
In the communication context, the analysis of the energy consumption of cryptographic network protocols was first introduced in Ref. \cite{YehiaCentrone}. There, Quantum Key Distribution (QKD) protocols are considered along with the definition of the energy efficiency in key-bit per Joule. The authors adopt a hardware dependent approach, to show the technological cost of photonic protocols operated on specific state of the art setups. Inspired by this study, yet diverging from it towards more fundamental statements, we develop a method to study the fundamental energy consumption of creation, sharing and distillation of entanglement. In  summary, our results differ and complement these  from~\cite{YehiaCentrone} in three ways: 

\begin{itemize}
\item We study the fundamental energy consumption, independent from any particular hardware realization. We then compare it with the  energy costs and performance of concrete entanglement distillation protocols. 

\item  We focus on the energy used in a given resource distillation process, rather than on the efficiency of a particular protocol. The main example of resource that we study in this article is entanglement, but analogous studies could be done for other quantum resources.

\item  We focus on the energy balance, \textit{i.e.} the invested energy minus the recovered one (in Joule). We then divide the balance by the number of generated resource-bit, the amount of entangled bit. 
\end{itemize}  

This paper is organized as follows: In Section \ref{ch:Results} we provide an overview of the model used, as well the energetic bounds for entanglement distillation. In Section \ref{ch:Lit} we discuss in more detail relation of the present study with other results in the field. In Section \ref{ch:Met}, we provide mathematically rigorous definitions and proofs of results of Section \ref{ch:Results}, to conclude with the Discussion in Section \ref{ch:Dis}.

\section{Results }\label{ch:Results}
This part in an accessible way introduces a reader to main results of the manuscript, exhibiting ideas standing behind the proposed paradigm. It presents the definition of the energy consumption rate, the definition of the energy cost of a quantum operation from axiomatic principles. Further, it demonstrates fundamental lower and upper bounds, linked to channel characteristics and protocol components. At the end it describes the relation between energy and entanglement generation.

\subsection{The energy consumption rate}
We start by describing our formalism for investigating the energy expense of entanglement distribution through a given quantum channel $\Phi$. In the scenario considered, we consider a bipartite scenario, and allow for the use of Local Operations and Classical Communication (LOCC) to distill entanglement from the output of the channel. The channel is assumed to be uncontrolled and passive, and a LOCC operation itself, so that it does not creates entanglement, but degrades it.
 The central question that we quantitatively address is the following: 

\begin{center}
\tcbox[  % Używamy polecenia \tcbox z pakietu tcolorbox
  colback=white, % Kolor tła ramki
  colframe=blue,  % Kolor obramowania
  boxrule=1pt,   % Grubość ramki
  rounded corners=all, % Zaokrąglenie narożników
  arc=5mm % Promień zaokrąglenia
]{%
  \begin{tabular}{c}
%\hline
     {\color{white}\tiny a}\\
     {\color{white}\Large A}\bf How much energy is fundamentally required to distribute a maximally entangled state {\color{white}\Large A}\\ {\color{white}\Large A}\bf through an entanglement degrading quantum channel  $\Phi$? {\color{white}\Large A}\\
     {\color{white}\tiny a}\\
%\hline
\end{tabular}
}
\end{center}

\begin{figure}[h!]
    \centering
    \includegraphics[width=0.8\linewidth]{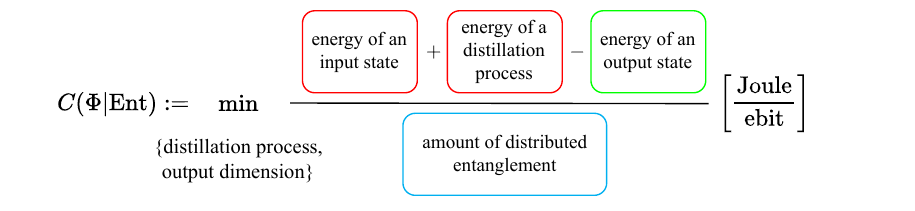}
    \caption{\justifying Visual represenation of the formal definition of energy consumption rate of entanglement distribution.}
    \label{fig:big-colorful-formulas}
\end{figure}

To answer this fundamental problem, we introduce the notion of the Energy Consumption Rate of Entanglement Distribution (ECRED), denoted as $C(\Phi|\mathrm{Ent})$. 
This quantity measures the minimal energy expenditure necessary to distribute entanglement via the channel $\Phi$.
%\begin{figure}[!ht]
%    \centering
%    \includegraphics[trim=0cm 6cm 0cm 0cm,width=0.8\linewidth]{corrected_prezentacja_idei.pdf}
%    \caption{The scheme of entanglement distribution via quantum channel $\Phi$.
%        The red highlights the components where the consumption of energy takes place, the green represents the gained energy in desired form and blue - the amount of distributed entanglement.}
%    \label{fig:scenario}
%\end{figure}
%\textcolor{red}{figure 3 bring here }
More formally, the ECRED is defined as the net energy invested, accounting for the input state energy plus the cost of entanglement distillation minus the output state energy, normalized by the amount of distributed entanglement. This amount is quantified by the logarithm in base 2 of the output dimension, when the output state approximates a maximally entangled state with local dimension $d$ (see Fig. \ref{fig:big-colorful-formulas} for the depiction of the scenario).

Following the development of the so-called beyond-iid quantum information theory \cite{4036001} we approach this problem rigorously in two steps. 
We first define the single-shot fundamental ECRED, $C^\epsilon(\Phi|\mathrm{Ent})$ (see Definition \ref{def:EnergyCost}). It captures the optimal ratio of energy used per entanglement generated, where the quality of the entanglement is accurate up to~$\epsilon$ in trace norm distance.  It involves three minimizations over different sets:

\begin{enumerate}\item {\bf Physical Realizations}  The set of all possible physical realizations (PR), which can be transmon qubits, ion traps, photons etc. (the list of PRs is not closed), as well as their specific setups (e.g. distances between the parties).
\item \textbf {Input state} The set of bipartite input states that are sent down the channel $\Phi$.
\item \textbf {Distillation protocols}
The set of LOCC distillation protocols that generates pure entanglement from the noisy output of the channel $\Phi$. 
\end{enumerate}

The ratio between the aforementioned net energy consumption and  the dimension of the output entanglement is minimized over these three sets.
The single shot quantity $C^{\epsilon}(\Phi,\mathrm{Ent})$ captures the energy consumption of a \textit {single} channel use, when one tolerates a fixed error $\epsilon$. More precisely, but still informally, it reads (in the spirit of Fig. \ref{fig:big-colorful-formulas}) as

\begin{equation}
C^{\epsilon}(\Phi|\mathrm{Ent}):=\inf_{\bf PR}\inf_{\rho_{in},\Lambda_{\mathrm{distill}} \in LOCC,} \left[\frac{E(\rho_{in})+E(\Lambda_{\mathrm{distill}},\rho_{int})-E(\rho_{out})}{\log d_{out}}\right] \left[\frac{Joule}{ebit}\right],
\end{equation}

where the infimum is taken over the operations $\Lambda_{\mathrm{ distill}}$ from the LOCC class that act on an intermediate state $\rho_{int}=\Phi(\rho_{in})$, which is the output of the channel $\Phi$. $\Lambda_{\mathrm{ distill}}$ outputs $\rho_{out}$ which is $\epsilon$-close in trace norm distance to a maximally entangled state $\psi^+_{d_{out}}=\sum_{i=0}^{d_{out}-1}\frac{1}{
\sqrt{d_{out}}}\ket{ii}$ i.e. the one having local dimension $\log d_{out}$ (and therefore that much of distillable entanglement)\footnote{We have omitted here the dependence of $\rho_{in}$ and $\Lambda$ on the physical realization $\bf PR$, and delegate the reader to the formal Definition \ref{def:ent-main} for details.}.

However, it is interesting and partially practical to consider a physical setup (channel) used many times. As a consequence, we quantify the asymptotic behavior of the single-shot ECRED with growing number of channel uses~$n$, while demanding the error $\epsilon$ to approach $0$ for large values of~$n$. This leads us to a definition of the asymptotic ECRED:
\begin{equation}
    C(\Phi | \mathrm{Ent}) = \sup_{\epsilon\in(0,1)} \liminf_{n\to \infty} C^{\epsilon}(\Phi^{\otimes n} | \mathrm{Ent}).
\end{equation}
It characterizes the minimal energy consumption rate in the asymptotic limit of many channel uses,
capturing the energy consumption rate per entanglement bit in the many-copy regime. It takes values in the range $[0,+\infty]$, where $0$ is reached for the identity channel, ideally transmitting the maximally entangled state e.g. $\psi^{+}_{d_{in}}:=\sum_{i=0}^{d_{in}-1} \frac{1}{\sqrt{d_{in}}}\ket{ii}$, while $+\infty$ is reached on any so-called binding entanglement channel \cite{HHHH09}.

While the above definition is fundamentally interesting, it is also incomputable due to the infimum over all physical realisations, including the ones that are even not yet known.

Enjoying the generality of the definition, we observe that entanglement is typically distributed using photons, rather other physical platform. We then propose two study platform-dependent definitions of $C(\Phi|\mathrm{Ent})$.
In particular, we focus on what we call {\it standard ECRED}, $C_{\mathrm{std}}(\Phi | \mathrm{Ent})$ (see Definition \ref{def:ecr_single_shot}), defined with respect to a canonical physical setup distributing entanglement. Namely, it includes: 
\begin{itemize}
\item \textbf{Standard Physical Realization}:
Entanglement encoded in polarized photons, 
\item \textbf {Standard Input state}: The input to the channel is the maximally entangled state: $\rho_{in} := \psi_{d_{in}}^+$, 
\item \textbf{Standardized protocol}: One of the typical ways to distribute entanglement is to create the entangled state in one location and send half of it to the other location. Effectively $\Phi:=\id_{A} \otimes \Phi_{A'\rightarrow B}$ \footnote{Other typical channels could be heralded entanglement which we do not consider here.}
\end{itemize}
By definition, the standard ECRED is an upper bound on the ECRED itself :
\begin{align}
   C(\Phi|\mathrm{Ent})&\leq C_{\mathrm{std}}(\Phi|Ent).
\label{eq:order} 
\end{align}

%%%%%%%%%%%%%%%%%%%%%%%%%%%%
\subsection{Definition of the energy cost of arbitrary quantum operation}

The ECRED and standard ECRED, depend on two different energy quantifiers: (i) the energy of a quantum state, which is (roughly) defined as $E(\rho) = \tr H_S\rho$, (ii) the energy cost of a quantum operation  $\Lambda$, denoted as $E(\Lambda, \rho)$ (in our case $E(\Lambda_{\mathrm{distill}},\rho)$. While we use only $\Lambda\in LOCC$ that distills entanglement, we provide a formal definition of $E(\Lambda,\rho)$ for any quantum operation $\Lambda$.  Indeed, the definition of energy cost of a quantum operation $E(\Lambda,\rho)$ is of independent interest. It can be used to quantify the energy expense of quantum computing platforms. In fact, our definition of the energy cost of a quantum operation is a formalization of concepts such as the {\it fundamental} and {\it on-cheap} energy costs, which has been put heuristically forward in the context of quantum computing \cite{Stevens_Grenoble}.

In principle, there are different ways of defining $E(\Lambda, \rho)$, depending on different purposes it has to serve. One can quantify the work expense \cite{Faist_2018,Oppenheim2002}, the heat pumped, or both, as well as their gained versions.
To clarify the way one can approach the problem, and capture tacit heuristic assumptions behind the approaches of \cite{YehiaCentrone,Stevens_Grenoble}, we propose an axiomatic framework for defining this cost. We start with two essential axioms:

\begin{axiom1} Subadditivity in time for sequential operations, i.e. $\forall {\Lambda_1:X\to Y},{\Lambda_2:Y\to Z},~f({\Lambda_2} \circ {\Lambda_1}) \le f(\Lambda_1)+f({\Lambda_2})$.\label{ax:subadd}
\end{axiom1}

\begin{axiom2}
Locality of energy: $\forall \Lambda_A,\,\, f(\Lambda_A\otimes\id)=f(\Lambda_A)$.
\label{ax:locofenergy}
\end{axiom2}

From the above, we also obtain a property of the definition of energy cost: subadditivity in space \textit{i.e.} subadditivity on tensor products of quantum operations:
$$\forall {\Lambda_1},{\Lambda_2},~ f({\Lambda_1} \otimes {\Lambda_2}) \leq f(\Lambda_1)+f({\Lambda_2}).$$
The above property will be crucial for deriving upper bounds on the energy of distillation protocols, since these protocols typically work in parallel on a tensor product of pairs of states. Thanks to this property, we can add the energy costs from each pair as an upper bound on the total energy cost.

We argue that any reasonable definition of $E(\Lambda, \rho)$ has to satisfy these axioms. In Supplementary information \ref{sec:axiomatic-apporach} we define $E(\Lambda, \rho)$ (see Definition \ref{def:EconsumFundamental}) and prove that it satisfies these two axioms. Our definition involves a physical model for realizing LOCC operations, rather than manipulating their Kraus representation as it is usual done in quantum information theory. This is where quantum thermodynamics comes explicitly into play.

To do so, we map LOCC operations onto a Hamiltonian formalism, so that the notions of the heat and work are well defined. This, in turn, allows us to study the energy balance (from the numerator of ECRED's definition). Our description of a quantum map $\Lambda$ involves three objects: 

\begin{itemize}
\item {\bf Quantum System:} A quantum system to process (e.g. polarized photons), initially in state $\rho_{int}=\id\otimes \Phi (\rho_{in})$, itself being the output of the channel on the input $\rho_{in}$.

\item {\bf Mesoscopic interface:} A mesoscopic device, \textit{i.e.} a physical system with distinguished relevant degrees of freedom, that is able to perform $\Lambda$ on $\rho$ (\textit{e.g.} an optical setup on an optical table). The device is small  enough to possess a few quantum degrees of freedom, yet bulky enough so that these degrees of freedom can be classically controlled.

\item {\bf Classical control: } A classical control system ensuring contact between the mesoscopic device and the relevant physical system, for example a heat bath, work reservoir and a CPU that runs it.
Its role is to exchange work and heat with the mesoscopic device so that the latter can perform gates and measurements on a quantum system $S$ (\textit{e.g}. to distill entanglement).
\end{itemize}

%The hamiltonian model of the mesoscopic interface performing $\Lambda$ is shown in Fig.~\ref{fig:made-by-pawel}.

%\begin{figure}
%    \centering
%    \includegraphics[width=0.5\linewidth]{image-from-presentation-made-by-pawel.png}
%    \caption{Graphical representation of the model we adopt to define the energy cost of quantum operations. It consists of a quantum system, a mesoscopic interface device and classical controls. To quantify the energy cost of an operation $\Lambda$ performed by the mesoscopic device on the system, we take into account heat and work currents directed from classical control to the mesoscopic device.}
%    \label{fig:made-by-pawel}
%\end{figure}

We then define the energy cost as the minimal energy of a protocol realizing a task (\textit{e.g.} entanglement distillation). The energy of a protocol equals the sum of heat pumped into the mesoscopic interface plus the amount of work performed by the classical control on the latter.
\begin{align}
    & E(\Lambda, \rho) = \inf_{\mathcal{P}}E(\Lambda_{\mathcal{P}}),\\
    & E(\Lambda_{\mathcal{P}}) = \int_0^{t^*}\operatorname{dt} \left(J_W^{\textrm{drive}\to}(t) + J_Q^{\textrm{drive}\to}(t) \right).
\end{align}
\\
\color{black}
The in-going work $J_W^{\textrm{drive}\to}(t)$ and in-going heat $J_Q^{\textrm{drive}\to}(t)$ are consequences of the classical control. The work contribution is related to the changes in the joint system Hamiltonian due to the {\it independent driving fields} that are described as an interaction with time-dependent classical potentials. The exchange of heat is the result of purely dissipative interactions with heat bath(s), which are also a part of classical drive, and can also be switched on and off. For the sake of simplicity, we refer to the latter type of interactions as the independent driving field. At a certain time $t$ of the protocol, both types of fluxes, from multiple control fields, can flow in and out of the quantum system.

In our approach we focus on the amount of energy that must be invested in order to perform a quantum task, \textit{e.g.}, entanglement distillation. Therefore, we only sum up the positive contributions from all the driving fields (denoted $\rightarrow$) for each field separately, only then transient work and heat fluxes are determined. Counting the contribution from all the driving fields separately allows us to determine how the {\it local operations} fill in the scenario. In the multipartite scenario, one can associate a se of driving fields with some agent (party) that uses them to execute the protocol. 

Within our framework {\it classical communication} can also be readily analyzed. The creation and detection of the signal is modeled by the action of the driving fields while its propagation is related to the free evolution of the auxiliary device. In this way, our framework is abundant enough to incorporate LOCC. Finally, we note that the cost of classical computing and signal processing is not taken into account in our considerations. 
\color{black}

%%%%%%%%%%%%%%%%%%%%%%%%
\subsection{Lower bounds on the standard energy consumption rate}

One of presented results is  fundamental lower bound on the standard ECRED (see Corollary \ref{cor:bound-by-ec-ed}). It is derived, under condition that
(nearly) pure entanglement is the only output of the entanglement distillation protocol. In that we follow a standard approach to entanglement distribution (see however \cite{SynakRadtke2005,Ganardi2025}). 

Under this assumption we firstly show, that entanglement degradation implies non-zero standard ECRED. To our knowledge, this is the first quantitative relation between irreversibility in entanglement theory and energy expense. It is known in entanglement theory that, for a generic mixed entangled state, the entanglement cost (the amount of maximally entangled states necessary to create many copies of a given bipartite state $\rho_{AB}$, denoted as $E_C(\rho_{AB})$) exceeds the distillable entanglement (the amount of maximally entangled states that can be distilled from many copies of $\rho_{AB}$, denoted as $E_D(\rho_{AB})$). With this notation irreversibility takes place if and only if

\begin{equation}
E_C(\rho_{AB})>E_D(\rho_{AB}).
\label{eq:irrev}
\end{equation}

By exhibiting a direct connection between irreversibility in entanglement manipulation and energy expense, we show that
\begin{equation}
C_{\mathrm{std}}(\Phi|\mathrm{Ent})\geq 2\hbar\omega \left ( \frac{E_C(\rho_\Phi)}{E_D(\rho_\Phi)} - 1\right ),
\label{eq:main_lb}
\end{equation}
where $\rho_{\Phi}$ is a bipartite state (known as the Choi-Jamio\l{}kowski state of the channel $\Phi$) of the form $\rho_\Phi:=\mathds{1}\otimes\Phi (\Psi^+_{d_{in}})$. We then derive an important corollary:
\begin{center}
\tcbox[  % Używamy polecenia \tcbox z pakietu tcolorbox
  colback=white, % Kolor tła ramki
  colframe=blue,  % Kolor obramowania
  boxrule=1pt,   % Grubość ramki
  rounded corners=all, % Zaokrąglenie narożników
  arc=5mm % Promień zaokrąglenia
]{%
  \begin{tabular}{c}
%\hline
     {\color{white}\Large A}\bf Entanglement irreversibility is a sufficient condition for non-zero  
       {\color{white}\Large A} {\color{white}\Large A}\\\bf  standard energy consumption of entanglement distribution.
     {\color{white}\tiny a}\\
%\hline
\end{tabular}
}
\end{center}

Indeed, if the strict inequality (\ref{eq:irrev}) holds for the state $\rho_{\Phi}$, then clearly the lower bound on the r.h.s. of the fundamental bound (\ref{eq:main_lb}) above, scales with the energy term $2\hbar \omega$ by a factor $\left(\frac{E_C(\rho_\Phi)}{E_D(\rho_\Phi)} -1\right)>0$. We note here that the factor $2$ appears in front, due to the fact that the state $\rho_{\Phi}$ is bipartite and we assume for simplicity that it has equal local dimensions.

Furthermore, for the non-zero energy consumption rate, it is in fact sufficient that the channel $\Phi$ under consideration {\it degrades} entanglement. Namely, it does not ideally (unitarily) transfer half of a maximally entangled states. If the input dimension exceeds the distillable entanglement of the channel's output state (\textit{i.e.}, $\log_2 \mathrm{dim}_{in}(\Phi) > E_D(\rho_\Phi)$), then the standard ECRED is strictly positive. Formally, we derive the following lower bound tighter than the above one:
\begin{equation}
    C_{\mathrm{std}}(\Phi | \mathrm{Ent}) \geq2\hbar\omega\left(\frac{\log_2 \mathrm{dim}_{in}(\Phi)}{E_D(\rho_\Phi)} - 1 \right).
\end{equation}
This establishes that entanglement distribution through such channels requires a strictly positive minimal energy investment. As mentioned above, to prove the result, we exploited the fact that in our considerations the energy that could potentially be restored from the target system is explicitly omitted. The bound then quantifies the price of non-resource aware approach to quantum entanglement distribution, and advocates for more energetically optimal approaches beyond standard paradigm.

%%%%%%%%%%%%%%%%%%%%%%
\subsection{Upper bound on the energy consumption rate of entanglement distribution}
We would like to stress that the proof of the above lower bounds relies on the non-negativity of the energy cost of entanglement distillation. It follows from the fact that, in the aforementioned task, a common approach is to focus on the specific task which only outputs the target states i.e. maximally entangled ones.
Hence, the lower bounds are obtained just by neglecting the term $E(\Lambda,\rho)$, the middle red box in the numerator in Fig. \ref{fig:big-colorful-formulas}. It is then both interesting, and important, to study how far is this lower bound of $C_{\mathrm{std}}(\Phi|\mathrm{Ent})$ from the energy of an actual distillation protocol. This leads us to the final part of this work : establishing upper bounds. 

Indeed, as the second main result, we provide upper bounds on the ECRED in the case of the so-called {\it depolarizing channel} \footnote{The depolarizing channel maps a quantum state to itself with probability $\lambda$ and, with probability $1-\lambda$, it outputs a maximally mixed state $\mathds{1}/d$.}.
We do so by using the fact that the ECRED is bounded from above by the standard ECRED (see Eq.\eqref{eq:order}), which is upper bounded by the energy consumption rate of a particular entanglement distillation protocol.

More precisely, we provide a numerical method to compute an upper bound on the standard ECRED for a class of entanglement distillation protocols. Let us exemplify this approach on the case of the Deutsch et al. entanglement distillation protocol \cite{deutsch1996quantum} (denoted as DEJMPS after names of the inventors). It consists of a repeated number of steps:

\begin{enumerate}
\item Divide the input states into pairs $\rho_{A_iB_i}\otimes \rho_{A_i'B_i'}$.
\item For each pair $i$, perform $CNOT_{A_i\rightarrow A_i'}$ and $CNOT_{B_i\rightarrow B_i'}$.
\item For each pair $i$, measure systems $A'_i$ and $B'_i$ in the computational basis and compare the results using a classical communication channel. If they match, save state of $A_iB_i$ system for the next round of iteration. Else, discard both systems $A_iB_i$ and $A_i'B_i'$.
\item Perform Hadamard gate on systems $A_i$ and $B_i$ selected in previous step for the next round of iteration.
\item Iterate steps $1$-$4$ until the selected state(s) have good enough fidelity with respect to the maximally entangled two-qubit state.
\end{enumerate}

Now, given the fact that $E(\Lambda,\rho)$ satisfies Axiom \ref{axiom2} (locality of energy), the energy consumption of joint operations of two entanglement distributing parties are upper bounded by the $i$th sum of the individual energy costs of the parties' actions. Furthermore, focusing on a single party, say $A$\footnote{The output of depolarizing channels is symmetric so the case of $B$ is analogous}, and given that $E(\Lambda, \rho)$ satisfies Axiom \ref{axiom1} (subadditivity of the energy cost in time), the energy cost of steps $1$-$4$ of the above DEJMPS protocol does not exceed the sum of the energy costs of its building blocks: energy of performing $2$ CNOT gates, $2$ measurements, $H$ gate and exchanging of $1$ classical bit (the measurement result).
While the detailed analysis of the energy cost is a bit lengthy, we give a hint of how it is computed. Using linear optics, the CNOT gate is fundamentally probabilistic, because photons do not interact easily. The best probability of performing it is $1/8$. Furthermore, the implementation of CNOT uses $2$ measurements \cite{Li_2021}. In this first estimation, we neglect the energy of classical communication and consider $H$ to be a passive gate. We focus on the upper bound on the energy of quantum measurements. Each such measurement needs a qubit from a quantum memory that  acquired $2\hbar \omega$ of energy and purified to a blank state $\ket{0}$ by erasure. While Landauer's principle sets a lower bound on such an erasure\footnote{Here $k_B$ is Boltzmann constant, $T$ is the temperature.}: $k_BT\ln 2$, we need to look for upper bounds instead.
We use results and techniques of Reeb et al. \cite{Reeb2014,Reeb_2018} to argue that multiplying the Landauer lower bound by a factor $2$ generates an upper bound.

In that manner, we compare the $3$ historically first distillation protocols: BBPSSW~\cite{bennett1996purification},  DEJMPS~\cite{deutsch1996quantum} and the recent P1-or-P2~\cite{miguel2018efficient}\footnote{We note that more recent protocols~\cite{Hiesmayr2024, Hiesmayr2025} have come out.} and find upper bounds of the form
\begin{equation}
    C(\mathrm{Depolarizing\,channel}|\mathrm{Ent})\leq C_{\mathrm{std}}(\mathrm{Depolarizing \,channel}|\mathrm{Ent})\leq O(10^{-12})\, \left[\frac{J}{ebit}\right],
\end{equation}

\begin{figure}[!h]
    \centering
    \includegraphics[width=0.5\linewidth]{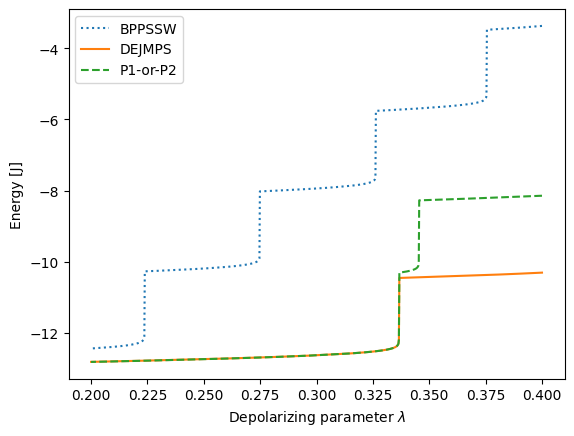}
    \caption{\justifying Upper bounds on the fundamental energy consumption required to obtain one quantum state with fidelity $0.9$ (with respect to maximally entangled state)  from a state $(1 - \lambda)\Phi^+ + \lambda \id/4$. In our analysis, we set the photons wavelength to 1550 nm, $p_{cnot} = 1/8$ and $T = 293K$. The energy axis is in the logarithmic scale.}
    \label{fig:drawing} 
\end{figure}

The comparison of performance in terms of energy consumed in distilling the two-qubit maximally entangled state $\psi^+_2$ is depicted in Fig. \ref{fig:drawing} for large enough parameter of depolarizing noise. This emphasizes the large gap between our lower bound and the actual, technological cost. Interestingly, we note that the protocol with the higher distillation rate is not the most energetically efficient. The DEJMPS protocol seems to be the most energy efficient (for large parameter of depolarizing noise) although it was the second to come out. 

%The results are schematized in Figure~\ref{fig:allresults},
%where 
%\textcolor{red}{pls copy past here the correct figure}

%By aggregating estimated energy costs of fundamental operations used in distillation protocols, such as probabilistic CNOT gates (in linear optics), measurement processes (bounded by Landauer's erasure limit), and classical communication, we provide a numerically upperbound $C(\Phi | \mathrm{Ent})$ for depolarizing channels.

%%%%%%%%%%%%%%%%%%%%%%

\subsection{Energy versus entanglement generation}
Finally, which is of separate interest, most of the results described above, can be presented on the plot of (properly rescaled) energy in [Joule] vs
entanglement in [ebit]. To give an example, we proceed as follows: (i) rewrite the RHS of the inequality \eqref{eq:main_lb}:
\begin{equation}
C_{\mathrm{std}}(\Phi|\mathrm{Ent})\leq 2\hbar\omega \left(\frac{E_C(\rho_\Phi)}{E_D(\rho_{\Phi})}-1\right) = 2\hbar\omega \left(\frac{E_C(\rho_\Phi)-E_D(\rho_\Phi)}{E_D(\rho_{\Phi})}\right),
\end{equation}

\begin{figure}[!ht]
    \centering
    \includegraphics[scale=.75]{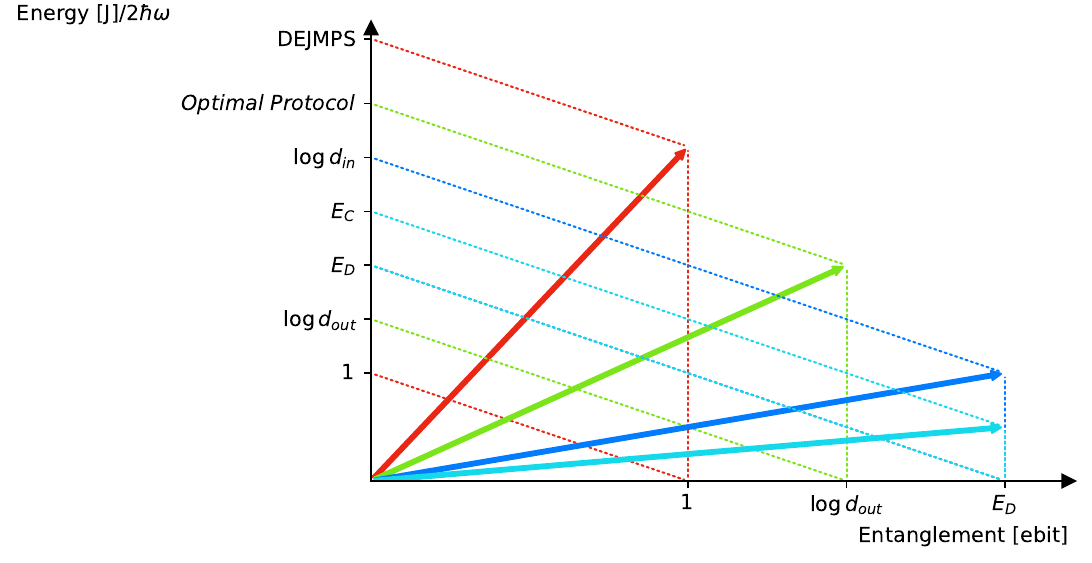}
    \caption{\justifying The energy consumption rate of entanglement distribution on a single plot, for a fixed channel $\Phi={\id}_A\otimes\Phi_{A'\rightarrow B}$. On the y-axis is the energy consumption rate (in Joule). On the x-axis is the rate of distilled entanglement, which is the amount of entanglement distilled by a corresponding protocol from the Choi state $\rho_\Phi$ of the channel $\Phi$.    
    %relation to Eq. \eqref{eq:energyConsRate} and Figure \ref{fig:EnergyVsForm} 
    %The $Env$ is modeled as a quantum channel $\Phi$ and the measure of form of energy $F$ is ebit
    The values of energy on the y-axis are normalized by a $2\hbar\omega$ factor. The green vector corresponds to $C_\mathrm{std}(\Phi|\mathrm{Ent})$ given in higher
    (\ref{def:ent-main}), which is the {\it standard} energy consumption rate involving the use of the optimal protocol distilling $\log_2 d_{out}$ ebits over the channel $\Phi$. The red vector corresponds to using the existing DEJMPS protocol \cite{deutsch1996quantum} to distill $1$ ebit, and the dark blue vector presents the lower bound given in Proposition \ref{prop:main_asymptotic} as a function of the distillable entanglement $E_{D}(\rho_\Phi)$. The light blue vector, whose coefficient equals $\frac{E_{C}(\rho_\Phi)}{E_{D}(\rho_\Phi)}-1$, reflects one of the main results (see Eq. \eqref{eq:bound-by-ec-ed}). Namely, that  irreversibility in entanglement theory is a sufficient condition for non-vanishing of the energy consumption rate.}
    \label{fig:all_inone_resuls_section}
\end{figure}

(ii) Put the denominator $E_D(\rho_\Phi)$ on the ebit axis (the $\mathrm{x}$-axis) and (iii) put the value of the numerator $E_C(\rho_\Phi)-E_D(\rho_\Phi)$, above the point marked in (ii), parallel to the energy axis (${\mathrm y}$-axis) (see dotted line on Fig.~\ref{fig:all_inone_resuls_section}). (iv) Connect the origin of the graph and the end of dotted line.

In this way we obtain the light blue vector on Fig. \ref{fig:all_inone_resuls_section}, which has a slope equal to the ratio $ \left(\frac{E_C(\rho_\Phi)-E_D(\rho_\Phi)}{E_D(\rho_{\Phi})}\right)$. Hence, its slope represents the lower bound on $C_{\mathrm{std}}(\Phi|\mathrm{Ent})$, according to Eq. \eqref{eq:main_lb}. Similarly, the slope of the red vector represents the performance of the DEJMPS protocol \textit{i.e.} the energy balance when performing a protocol generating 1 ebit. Our quantity of interest, the standard consumption rate of entanglement distribution for an energy-optimal protocol, is represented on the slope of the green vector, which is located somewhere between the blue and red. 

%\textcolor{red}{Mikołaj - pls put here the drawing of all results from above and describe it in the text syaing that all the %main results can be shown in the plot of Joules vs ebits}

\section{Related literature}\label{ch:Lit}
The first work pertaining to the energetic cost of quantum communication protocols has been done in~\cite{YehiaCentrone}, as mentioned in the introduction. There, they focus on the \textit{technological cost} and define different figure of merits to study the energy cost of performing basic communications protocol such as quantum key distribution and conference key agreement. For example, they study the cost of generating 1Gbit of secret key using a hardware-dependent approach and listing commonly used equipment along with their power consumption and their initialization cost. Their findings show that, depending on the situation and the task, it might be more efficient to work at different wavelength or to sacrifice some time using less efficient but less energy consuming detectors that do not require cryogenics. They also study multipartite networking and present some energetic bottleneck in multipartite entanglement generation.\\
It is also worth mentioning that classical capacity of quantum channel per unit cost, which can be associated with energy, was studied in \cite{Jarzyna_2021}, showing that classical capacity of noisy bosonic Gaussian channels can be asymptotically achieved using a simple modulation and detection scheme in the low energy limit. As our aim is to  describe energy costs of distillation protocols, and compare them according to their energy efficiency, here we will adopt Joule per ebit as a unit.

Research caring for the energetic cost of quantum technologies has been triggered by the publication of the Metric-Noise-Resource (MNR) methodology \cite{PRXQuantum.3.020101}. There, one defines a metric which is a measure of success of the experiment  
%and study the optimization of the noise given a fixed resources or the converse.
and minimise the resource required to reach a targeted, fixed, metric of success, or the reverse, however in the less general approach, in some earlier paper of the same authors \cite{PRXQuantum.2.040335}.

For example, for a given noise model, one can maximize the ratio of the metric divided by the resource to obtain the efficiency of a given experiment.
%For example, for a given level of noise one needs to maximize the ratio of metric divided by resource to obtain the efficiency of a given experiment (usually a quantum computation on a specific platform~\cite{Asiani_2019}). 
While, in general, our approach can be seen as following the MNR methodology, the studied quantity and idea behind differ from~\cite{Asiani_2019}. We detail the similarities and differences between the two approaches in Table~\ref{tab:compMarco} below.

\begin{table}[!h]
    \centering
    \begin{tabular}{|c|c|c|}
    \hline
    MNR methodology's entities \cite{Asiani_2019} &Task: Entanglement distribution [this work] & Task: Cracking n-bit RSA\cite{Asiani_2019}\\ \hline
      Parameters &
        \begin{tabular}{c}
         Physical realization (PR), \\ input state $\rho_{in}$,\\distillation operation $\Lambda\in LOCC$
      \end{tabular} &
    \begin{tabular}{c}
    Temperature $T$,\\
    attenuation $A$,\\
    code concatenation level $k$
    \end{tabular}\\ \hline
        Metric  & $(1-\epsilon)$ where $\epsilon>0$ satisfies $\rho_{out}\,\approx_\epsilon\,\psi^+_{d_{out}}$ & Probability of correct factoring \\\hline
         Noise& 
         \begin{tabular}{c}
              Channel $\mathds{1}\otimes\Phi$\\
              acting on $\rho_{in}$,\\
              fixed i.e. independent from parameters
         \end{tabular}& 
         \begin{tabular}{c}
              Pauli channel, which parameter\\
              depends on temperature and\\
              attenuation acting on\\
              physical qubits
         \end{tabular}\\ \hline
         (Consumed) Resource & Energy balance: $
         E_{\mathrm{input }}+E_{\mathrm{distillation}}-E_{\mathrm{output}}$ &Invested power: $
         P_{\mathrm{computation}}$\\\hline
         Gained Resource [this work]& $\log_2 d_{out}$ of distributed entanglement  & Size of the factorized number $n$ \\\hline
         Optimization& $\min{\bigg(
         \frac{\text{Consumed \,Resource}}
         %E_{\mathrm{input }}+E_{\mathrm{distillation}}-E_{\mathrm{output}}}
         {\text{Gained Resource}}
         \bigg)}$
         & $\max{\big(\frac{\text{Metric}}{\text{(Consumed) Resource}}\big)}$ \\\hline
    \end{tabular}
    \caption{\justifying Comparison between the Metric-Noise-Resource methodology of \cite{Asiani_2019}, and
    the paradigm of the energy consumption rate of entanglement distribution introduced in this work. We compare our analysis with the one of the energy cost of factoring $n$-bit number using Shor's algorithm, given in \cite{Asiani_2019}. The ``Gained resource'' which, in our case, is the amount of distributed entanglement during the energetically optimal entanglement distribution protocol is explicitly introduced here, while only implicitly used in \cite{Asiani_2019}. We propose to call the Resource in MNR methodology as Consumed Resource, to differentiate from the Gained Resource. The paradigm proposed in this work can be phrased in a general  as minimizing the ratio of a consumed vs gained resource. In this manuscript these are energy and entanglement respectively, but in general the pair of resources can be arbitrary.
    } 
    \label{tab:compMarco}
\end{table}
The similarities are the following (i) we fix some parameters over which desired quantity is optimized (however they are no longer numbers), (ii) we  fix a metric which quantifies the quality of a considered functionality (in our case of entanglement distillation, and in case of \cite{Asiani_2019} the correctness of the output of the quantum computing) (iii) we consider noise model. 
Here comes the first technical difference: in our case the noise does not depend on parameters chosen in point (i). Instead of varying noise depending on parameters we minimize over, we characterize a fixed model of noise -- a channel by the energy consumption of achieving the task that involves channel's usage. 
The crucial conceptual difference  between ours and the approach of Ref. \cite{Asiani_2019} is twofold: first the MNR technique suggests variation over a ``resource'' with the meaning of consumed valuable, while we simply consider it to be energy\footnote{Note, that our approach is however general enough to consider as consumed resource any quantum resource other than energy, such as e.g. coherence \cite{RevModPhys.89.041003}, purity \cite{Oppenheim2002}, non-stablizerness \cite{https://doi.org/10.48550/arxiv.quant-ph/0402171,PhysRevA.71.022316} etc.}. We propose to refer to it as to consumed' resource, to differentiate from what we dab ``gained''  resource. Second, MNR technique focuses on Metric, while we focus on the aforementioned Gained Resource, namely the {\it quantum} resource -- be it entanglement \cite{Bennet_1996} , non-locality \cite{RevModPhys.86.419}, non-stabilizerness \cite{https://doi.org/10.48550/arxiv.quant-ph/0402171,PhysRevA.71.022316} etc. (see \cite{Gourbook} for other prominent examples) which, as above, do not need to be a natural metric in general. This gained resource, in our case entanglement, has not been used explicitly in the MNR methodology. Here we consider it explicitly as a center of the proposed paradigm.

Another difference is that we propose to optimize the ratio of the energy balance to the gained resource. In Ref. \cite{Asiani_2019} the resource (e.g. the power) is solely optimized and not in the form of a balance. The gained power, corresponding to our $-E({\mathrm{\rho_{out}}})$, is negligible in the case of the practical cost of quantum computing.

%in the case of quantum computing.

%We can map (however only partially) our approach to this general paradigm. 
%In the entanglement distillation case studied in this work, the noise comes from a channel $\Phi$ used to generate the quantum resource (e.g. entanglement). In the MNR paradigm, the energy consumed, or power, corresponds to the resource while the amount of quantum resource generated corresponds 
%to the metric. 
%to a parameter.
%To be more precise we minimize the resource generation cost, in Joule per ebit, for a given noise represented by a parameter $p$ of a depolarizing channel $\Phi(\rho)= p(\rho)+(1-p)\frac{I}{4}$. 

The authors of previously mentioned papers are active members of the Quantum Energy Initiative~\cite{QEIwebsite}, a community of researchers caring for the physical cost of emerging quantum technologies. In the same community, O. Ezratty has contributed to high-level vision of the current landscape of quantum technologies, which include some energetic aspects~\cite{ezratty2022mitigating,ezratty2024understanding}.

Work on the fundamental energy consumption of quantum technologies are scarce. The analysis \cite{Bassman_Oftelie_2024} gives insights into some fundamental bounds of cooling quantum system, therefore distilling $\ket{0}$ states, for quantum computing algorithms. A follow up~\cite{oftelie20az25measurementworkstatisticsopen}  reports on the measurement of the work statistics of open quantum systems. In the thesis~\cite{YoannThese} a subsection dedicated to the fundamental energy cost of performing CV-QKD shows a first attempt at quantifying a lower bound on the energy consumption of communication protocols.

Fundamental energy bounds for some of the quantum processes can be casted in the thermodynamic language \cite{Faist_2018}. In this setting,  free operations are represented as ones that preserve a certain state and do not increase the trace. This however does not apply directly to distillation protocols, which allow for LOCC operations, and lack a notion of the specific state to be preserved. The problem of energetic requirements for entanglement distillation can also be posed within the framework of generalized resource theories. It may seem at first glance that, since we study two different resource, energy and entanglement,
the construction of multi-resource theories applies \cite{Sparaciari_2020}.
However, we do not fall into this class, since we allow for operations from entanglement theory, that in principle may generate energy for free, thus are not free in theory of quantum thermodynamics.
Our paradigm is preserved because we consider infimum over such operations. Thus, our considerations cannot be directly seen as a particular case of a multi-resource theory \cite{Sparaciari_2020}.
A different, but technically closer, approach has been put forward in \cite{Arqand2023}. There, the energetically constrained quantum channel capacity has been considered, assisted by LOCC. More precisely, the energy of the input state is by assumption bounded from above. The authors of \cite{Arqand2023} consider also energetically constrained entanglement measures of the channel.
%one can formulate the distillation problem corresponding  resource theory as the intersection between the resource theory of energy, and that of (approximate) singlet states which can be obtained through a distillation process. So far however, already a simpler theory of local control under energetic restrictions is  lacking characterization of interconversion of resources, apart from a subclass of states \cite{Sparaciari_2020}.

\section{Methods}\label{ch:Met}
In this section, we present detailed definitions and proofs of the concepts introduced in the previous section.
\color{black}

\subsection{Fundamental energy consumption of entanglement distribution - the scenario}
Let us begin with the notation used in this section. The acronym LOCC stands for the set of Local Operations and Classical Communication \cite{Bennet_1996,Chitambar2014}. While all the presented results, including the main definitions of ECRED and standard ECRED, can be considered as involving only LOCC with one-way classical communication, we will focus in what follows on LOCC with two-way communication. This is because typical, practical entanglement distillation schemes \cite{deutsch1996quantum,bennett1996purification,miguel2018efficient,Hiesmayr2024} involve $2$-way classical communication.

The LOCC channel $\Phi$ is a completely positive trace preserving (CPTP) map which admits LOCC Kraus representation \textit{i.e.} corresponds to an operation from the set of LOCC \cite{Chitambar2014}.
Furthermore, $F(\rho,\sigma)= ||\sqrt{\rho}\sqrt{\sigma}||^2$ is the fidelity function \cite{Uhlmann1976}. The Choi state of the channel $\Phi$ is denoted $\rho_{\Phi}$ and
$||A||_{1}=\text{Tr}\sqrt{A^{\dagger}A}$. 
We denote a maximally entangled state of dimension $d$ by $\psi_{d}^+ = \frac{1}{\sqrt{d}}\sum_{i=0}^{d-1}\ket{ii}$. We are interested in the task of generating and sharing $\psi_{d}^+$ for some dimension, between two parties Alice and Bob (or A and B) at different spatial locations, using a quantum channel $\Phi$. In what follows we also make use of so-called {\it Choi state} of a channel $\Phi$, which is defined as $\rho_{\Phi} := (\id \otimes \Phi)(\psi_d^+)$.

\begin{figure}[!ht]
    \centering
    \includegraphics[trim=0cm 6cm 0cm 0cm,width=0.8\linewidth]{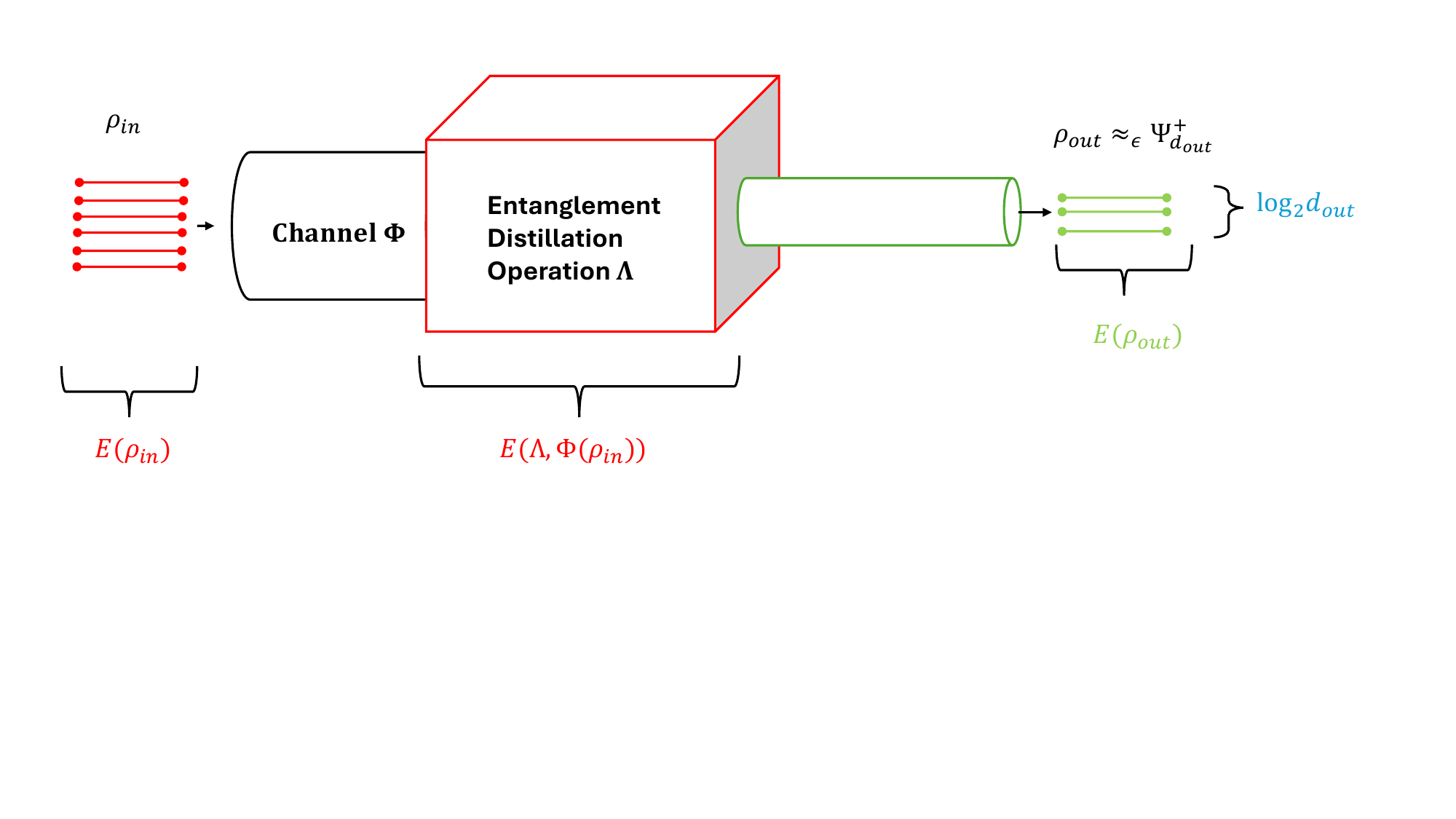}
    \caption{\justifying The scheme of entanglement distillation. It is assumed that the channel $\Phi$ is passive, i.e. it does not increase the distillable entanglement of the input states $\rho_{in}$. The red components highlight where the consumption of energy takes place: during the acquisition of the local input state $E(\rho_{in})$ and during the distillation operation $\Lambda$. The green color shows the gained energy while distributing $\log_2 d_{out}$ of entanglement between distant parties (represented by $\rho_{out}$ which approximates the ideal maximally entangled sate $\psi^+_{d_{out}}$ of local dimension $d_{out}$).}
    \label{fig:scenario}
\end{figure}

In the simplest scenario, one of the parties  produces an input state $\rho^{(in)}_{A''B''}$ on systems $A''B''$.  The system $B''$ is then used as input to the channel $\Phi:A''B''\rightarrow A'B'$. The channel is assumed to be passive in the sense that it cannot increase the distillable entanglement of the input state. The resulting state is called intermediate and denoted as $\rho^{int}_{A'B'}:=\Phi(\rho^{(in)}_{A''B''})$, where we identify $A''$ with $A'$ to indicate a new phase of the scenario although system $A''$ undergoes an identity operation when becoming $A'$. Next, the distillation operation $\Lambda^{\epsilon} :A'B'\rightarrow AB$ belonging to the class of Local Operations and Classical Communication (LOCC) is applied by the parties. The output $\rho_{out} = \Lambda^{\epsilon}(\rho^{int})$ is assumed to be $\epsilon$ close in trace norm distance to the singlet state $\Psi_{d_{out}}^+$. This scenario, shown schematically in Fig~\ref{fig:scenario}, effectively generates entanglement, in the form of the singlet \textcolor{black}{state} $\Psi_{d_{out}}^+$, between Alice and Bob. \textcolor{black}{This scenario} has variants, coming as \textcolor{black}{its} natural extensions, such as the scenario  \textcolor{black}{where a third party in the middle distributes entangled photons to parties A and B.}

We are now ready to introduce our figure of merit to study the performance of the distillation scenario shown in Fig.~\ref{fig:scenario}. It is given by the {\it ratio of the energy balance per the rate of distilled entanglement i.e. $\log_2 d_{out}$:}

% {\mw Karol said we should think about (and over) the base of log in the below def.. We need to point out that it should be the same in nominator and denominator and then nothing depends on it, so we can take it to be 2.}
\begin{definition}[Energy Consumption rate] \label{def:EnergyCost}
    For a quantum channel $\Phi:A''B''\rightarrow A'B'$ and $\epsilon >0$ the {\bf single shot fundamental energy consumption rate per ebit generated via channel $\Phi$} with input $\rho_{A''B''}^{(in)}$ and output $\rho_{AB}^{(out)}$ of dimension $d_{out}$ is 
    \begin{align}
C^{\epsilon}(\Phi|\mathrm{Ent}):=\inf_{\mathrm{PR}}\,\inf_{\substack{d_{out}({\mathrm{PR}})\in \mathrm{N}_+ \,\\\Lambda({\mathrm{PR}}) \in LOCC, \,\\\rho_{A''B''}^{(in)}({\mathrm{PR}})}} \,
    &\left[\frac{ E(\rho_{A''B''}^{(in)})+ E(\Lambda,\rho^{(int)}_{A'B'}) -E(\rho_{AB}^{(out)})}{\log_2 d_{out}}\,:\rho^{out}_{AB}:=
    \Lambda(\Phi (\rho_{A''B''}^{(in)}(\mathrm{PR})))\approx_\epsilon \psi^+_{d_{out}}\right],
    \end{align}
    where the intermediate state $\rho^{int}_{A'B'}$ is $\rho^{int}_{A'B'}:=\Phi(\rho^{(in)}_{A''B''}(\mathrm{PR}))$, $\mathrm{PR}$ means Physical Realizations, LOCC stems for Local Operations and Classical Communication, $\Lambda : A'B'\rightarrow AB$ are LOCC applied after the channel $\Phi$ and $\sigma_1 \approx_\epsilon \sigma_2 \iff 1/2\| \sigma_1 - \sigma_2 \|_1 \leq \epsilon$. \\
    Its regularized version is 
    \begin{equation}
C(\Phi|\mathrm{Ent}):=\sup_{\epsilon \in (0,1)} \liminf_{n\rightarrow \infty} C^{\epsilon}(\Phi^{\otimes n}|\mathrm{Ent}).
    \end{equation}
    \label{def:ent-main}
\end{definition}

\noindent 
The energy balance takes into account two energetic factors: $E(\Lambda, \rho^{(int)}_{A'B'})$ and  $E(\rho^{(in)}_{A''B''}) - E(\rho_{AB}^{(out)})$. The former expresses the energy needed to apply $\Lambda$ on state $\rho^{(int)}_{A'B'}$, while the latter quantifies energy losses in the physical system, which is the carrier of the considered resource. These losses can happen both in $\Phi$ channel or during the application of $\Lambda$ operation. Moreover, since we do not know \textit{a priori} which dimension of the output implies the lowest energy consumption per generated ebit, we take the infimum of the considered ratio over the choice of the output dimension. 
%The energy balance takes into account the energy of the input $E(\rho^{(in)}_{A''B''})$  and $E(\Lambda^\epsilon, \rho^{int}_{A'B'})$ which is the energy needed to apply $\Lambda^\epsilon$ on state $\rho^{int}_{A'B'}$, and then subtract the energy of the output state $E(\rho_{out})$. 
To justify the second of the considered factors, i.e., $E(\rho^{(in)}_{A''B''})-E(\rho^{(out)}_{AB})$, we note that the first term of this factor is equal to the "static" investment, i.e., energy in form of the input state. The second term is subtracted, because it corresponds to the energy of output which is in desired form of entangled state and as such is not yet used.
Indeed, since we have $\rho_{out}$ at our disposal, the corresponding energy survives the channel action and can be used in whatever way the output state is further used (e.g. for quantum  teleportation). 
%Since we do not know \textit{a priori} which dimension of the output implies the lowest energy consumption per generated ebit, we take the infimum of the considered ratio over the choice of the output dimension. 

Furthermore, the energy consumed by a given quantum channel, may depend on its physical
realization. Processing entanglement encoded in photons may be cheaper than processing it when encoded in ions in an trap, due to a different physical carrier of this resource. We thus also take the infimum over the choice of the physical realization to obtain a quantity which is independent of this choice. We end up with a quantity that is solely a function of the channel and the resource (in this case entanglement).

We propose a general definition, since we believe that no matter what is the physical realization of the qubits, possessing a physical system realization of a qubit inevitably costs energy. It is thus interesting which physical platform is energetically the cheapest realization of resource recovery. {This consumption may heavily depend on the physical platform, be it quantum dots, photons, transmon qubits or any other quantum states. Here, we propose a fundamental quantity with the infimum over energy consumption over all possible physical platforms. }Naturally some physical context uses imply a natural choice for a given platform, therefore the energy consumption for a fixed platform is of great importance. Indeed, future quantum internet will require photonic platform instead of electric circuits like it was the case for the telegraph. Thus enjoying a fundamental - physical realization independent definition of energy consumption we propose to study also independently the minimal energy consumption for a given platform.
\color{black}

In the definition \ref{def:EnergyCost} we decided to use trace norm distance as a quantifier of distance between quantum states, i.e. $\sigma_1 \approx_\epsilon \sigma_2 \iff 1/2\| \sigma_1 - \sigma_2 \|_1 \leq \epsilon$. However, note that using fidelity, i.e. $\sigma_1 \approx_\epsilon \sigma_2 \iff F(\sigma_1, \sigma_2) \geq 1 - \epsilon$ would lead to the equivalent definition.

Let us note that the quantity defined in Def~\ref{def:EnergyCost} has units of Joule per ebits 
\begin{equation}
\left[\frac{J}{ebit}\right]
\end{equation}
Furthermore, contrary to many asymptotic quantities the (such e.g., as entanglement measures) $C$ does not involve a factor $\frac1n$ which is needed for the so called regularization of the quantity \cite{Gourbook}.
This is because this quantity is a ratio of two extensive quantities, and hence intensive. Indeed, in the case of non-interactive particles which we focus on here, the energy balance is additive hence extensive quantity. Furthermore, $\log_2 d_{out}$ is also extensive, if only $\Phi$ has non-zero 
quantum capacity.

Moreover, it is important to note that whenever $\Phi$ is not a binding entanglement channel \cite{HHHH09}, i.e., the channel Choi matrix of which has $E_D(\rho_\Phi)=0$, the quantity $C^\epsilon(\Phi|\mathrm{Ent})$ is finite, as a quotient of naturally finite quantities. On the other hand, for all channels the Choi state of which has zero distillable entanglement,  the above defined energy consumption is equal to to $\infty$. It is reflects the fact, that their is infinite energy cost of distributing pure entanglement by them, since no matter what energy one uses up,
one cannot distill even a single ebit by any LOCC from its Choi state, either because the state is separable i.e., not entangled, or because it is bound entangled \cite{bound}.

% It is important to comment on the choice of the family of channels $\Phi$. Naturally we consider such that do not produce entanglement. However, the most common choice of channels used to produce bipartite entanglement are (i) $\Phi = \id_{A''} \otimes \tilde{\Phi}_{B''}$, which corresponds 
% to the creation of a singlet in one location and the distribution of half of it through a noisy channel $\tilde{\Phi}$ to another location,
% or (ii) $\Phi = \Phi^{(1)}_{A''} \otimes \Phi^{(2)}_{B''}$, that corresponds to the distillation of entanglement in a middle node and its sharing though two channels to two distant parties. Other viable options can be more abstract.

\begin{rem}
    The protocol that is the most energetically efficient one is not necessarily the one that is the optimal protocol in the standard sense. Consider entanglement distillation protocols as an example. The optimal protocol here is the one that transforms minimal number of input states into maximal number of maximally entangled states. However, both input states and quantum operations have their energetic costs. Therefore, if operations constituting the optimal protocol are energetically expensive it might be energetically favorable to use the protocol that is not optimal in the standard sense. The protocol that uses least energy per production of one maximally entangled state would be the most energetically efficient one in that case.
\end{rem}

\subsection{Definitions of energetic quantities }

\color{black}

In the following paragraphs, we informally describe a general framework for estimating the fundamental energy cost of performing a quantum task. The detailed and formal description is given in Appendix~\ref{sec:axiomatic-apporach}. The methodology described herein serves as a minimal self-consistent framework capable of non-trivial quantification of energy required for performing a quantum task. In the forthcoming considerations, we assume a fixed, yet arbitrary, physical realization of the task, i.e., the physical features of target quantum system $S$ are all well-specified. We remark that our approach does not take into account any energetic costs of classical computing required during a protocol realizing a desired quantum task.

From a physicist’s perspective, energy is a conserved quantity; however, in engineering terms, devices performing coherent quantum tasks inevitably consume power—typically supplied by classical control systems. Our goal is to bridge these perspectives by rigorously defining the \emph{minimum} energy that a classical device must invest to implement a quantum operation.

\begin{definition}[Energy of a system]
    We call $E(\rho):=\inf_{ \rho_S|\rho}\Tr [\mathrm{H}_{S}\rho_S]$ the energy of a system $S$ holding state $\rho$. Here, $\mathrm{H}_S$ is the Hamiltonian of the system $S$, dimension of which is the minimal one that allows to encode logical state $\rho$ in the physical state $\rho_S$ belonging to the Hilbert space associated with $\mathrm{H}_S$.
\end{definition}

\begin{definition}[Energy of a distillation operation]
    We call $E(\Lambda,\rho)$ the energy used to obtain the output state $\Lambda(\rho)$ from an input state $\rho$ by means of the distillation process $\Lambda$. 
\end{definition}

We formalize this with the concept of \emph{fundamental energy cost}, denoted $E(\Lambda, \rho)$, where $\Lambda$ represents the quantum operation (a completely positive, trace-preserving map) acting on the input state $\rho$. This cost quantifies the minimal energy investment required to achieve the desired output state $\Lambda(\rho)$. However, there are plenty of ways to define a function $E(\Lambda, \rho)$, each of them focusing of different aspects of energy expenditure. Nevertheless, we claim that each ''valid'' energy cost function should satisfy the following two, physically motivated axioms:
\begin{ax1}[Sub-additivity on compositions]\label{axiom1}
    $\forall \left\{{\Lambda_1:X\to Y},{\Lambda_2:Y\to Z}\right\}~f({\Lambda_2} \circ {\Lambda_1}) \le f(\Lambda_1)+f({\Lambda_2})$.
\end{ax1}

\begin{ax1}[Locality of energy]\label{axiom2}
 % Locality of energy: $\forall\left\{\Lambda_A\otimes\id \equiv (\tilde{\Lambda}_A \otimes \id_B,\rho_{AB},\sigma_{AB}),{\Lambda_A}\equiv (\tilde{\Lambda}_A,\Tr_B[\rho_{AB}],\Tr_B[\sigma_{AB})]\right\}~f(\Lambda_A\otimes\id)=f(\Lambda_A)$.
 $\forall \Lambda_A,\,\, f(\Lambda_A\otimes\id)=f(\Lambda_A)$.
\end{ax1}

In this spirit, we introduce the model that is grounded in a Hamiltonian formalism, wherein a \emph{mesoscopic device}—an interface between the quantum system and its classical controller—mediates the energy transfer. The mesoscopic device is designed to be large enough for reliable classical driving yet small enough to preserve essential quantum degrees of freedom. This allows us to describe the \emph{Hamiltoanian model protocol} $\mathcal{P}$ (see Definition \ref{def:HamiltonianProtocol}) implementing $\Lambda$ via a time-dependent Liouvillian $\mathcal{L}_{SA}(t)$ that captures both the coherent and dissipative, classically-driven dynamics of the system $S$ and device $A$. The classical control acts through this time-dependent drive and dissipator, providing a transparent link between theoretical energy definitions and practical implementation. With a chosen Liouvillian $\mathcal{L}_{SA}(t)$, we identify operators describing independent driving fields $\hat{A}_i$, interaction between system $S$ and a device $A$ $H_{SA}^{(i)}$, and dissipators $D_{A, \beta_i}^{(i)}$ describing interaction with a heat bath at inverse temperature $\beta_i$. With the use of the first two type of operators, we define work currents $J_{W}^{\mathrm{drive}\stackrel{\raisebox{-0.5ex}{$\rightarrow$}}{\raisebox{-0.5ex}{$\leftarrow$}}}(t)$, while using dissipators, we define heat currents $J_Q^{\mathrm{drive}\stackrel{\raisebox{-0.5ex}{$\rightarrow$}}{\raisebox{-0.5ex}{$\leftarrow$}}}(t)$ (see Definition \ref{def:positive-and-negative-currents} for details). These two quantities allow us to define energy current in the following way

\begin{equation}
        J_E^{\mathrm{drive}\stackrel{\raisebox{-0.5ex}{$\rightarrow$}}{\raisebox{-0.5ex}{$\leftarrow$}}}(t)  = J_Q^{\mathrm{drive}\stackrel{\raisebox{-0.5ex}{$\rightarrow$}}{\raisebox{-0.5ex}{$\leftarrow$}}}(t)  + J_{W}^{\mathrm{drive}\stackrel{\raisebox{-0.5ex}{$\rightarrow$}}{\raisebox{-0.5ex}{$\leftarrow$}}}(t),
\end{equation}
where the direction of the arrow in superscript determines if the current is ingoing ($\rightarrow$), or outgoing ($\leftarrow$). Furthermore, we first determine the in(out)-going currents due to all independent driving fields separately. Only in the end, all positive contributions (ingoing currents) sum up to total work and hear current. Using as an example work current (mutatis mutandis  for heat current), we have
\begin{align}
    & J_{A_i}^{\mathrm{drive}}(t) = \tr \left[ \frac{\partial V_i(t)}{\partial t} \hat{A}_i \rho(t)\right] +\mathrm{c.c.},~~J_{H_i}^{\mathrm{drive}}(t)  = \tr \left[ \frac{\partial H_{SA}^{(i)}(t)}{\partial t}\rho(t) \right],\\
    & J_{A_i}^{\mathrm{drive}\stackrel{\raisebox{-0.5ex}{$\rightarrow$}}{\raisebox{-0.5ex}{$\leftarrow$}}}(t)  = \eta\left( \pm J_{A_i}^{\mathrm{drive}}(t) \right) \cdot J_{A_i}^{\mathrm{drive}}(t), ~~J_{H_i}^{\mathrm{drive}\stackrel{\raisebox{-0.5ex}{$\rightarrow$}}{\raisebox{-0.5ex}{$\leftarrow$}}}(t)  = \eta \left( \pm J_{H_i}^{\mathrm{drive}}(t) \right) \cdot J_{H_i}^{\mathrm{drive}}(t),\\
    & J_{W}^{\mathrm{drive}\stackrel{\raisebox{-0.5ex}{$\rightarrow$}}{\raisebox{-0.5ex}{$\leftarrow$}}}(t)  = 
        \sum_i J_{H_i}^{\mathrm{drive}\stackrel{\raisebox{-0.5ex}{$\rightarrow$}}{\raisebox{-0.5ex}{$\leftarrow$}}}(t)  + \sum_i J_{A_i}^{\mathrm{drive}\stackrel{\raisebox{-0.5ex}{$\rightarrow$}}{\raisebox{-0.5ex}{$\leftarrow$}}}(t),
\end{align}
where $\eta(\cdot)$ is the Heaviside step function, $\mathrm{c.c}$ stands for complex conjugate, and scalar functions $V_i(t)$ determine time-dependent coupling strengths. Considering the contributions from all driving fields ensures consistency of the description, while at the same time the framework is general enough to properly describe multipartite scenarios.

Now, having defined energy current, we define the \emph{energy cost of execution of a protocol} $\mathcal{P}$ (see Definition \ref{def:EconsumP}) as 
\begin{align}
    E(\Lambda_\mathcal{P}) := \int_0^{t^*}dt  \left(J_W^{\mathrm{drive}\rightarrow}(t)+J_Q^{\mathrm{drive}\rightarrow}(t)\right).
\end{align}
Finally, since our aim was to define $E(\Lambda, \rho)$ as a fundamental quantity, we take infimum over all protocols $\mathcal{P}$ implementing $\Lambda$ and in this way we obtain the desired definition (see Definition \ref{def:EconsumFundamental}) of the considered quantity
\begin{align}  
E(\Lambda,\rho):=\inf_{\mathcal{P}}E(\Lambda_\mathcal{P}).
\end{align}
As a conclusion to our derivations on the definition of energy cost function, we note that in our model, $E(\Lambda, \rho)$ satisfies axioms 1, 2. Moreover, it also has the following two properties (see Proposition \ref{prop:SatisfyAxioms})
\begin{property} The trivial task, i.e., $\Lambda = \mathds{1}$, does not possess energy cost: 
    $f(\mathds{1})=0$, (zero element exists). 
    \label{property:1}
\end{property}
\begin{property}The energy cost is non-negative: 
    $\forall \Lambda,~f(\Lambda) \ge 0$, (non-negativity).
    \label{property:2}
\end{property}
which allow us to interpret $E(\Lambda, \rho)$ as a quantifier of inevitable energy expanse. 

Our Hamiltonian framework establishes a rigorous, technology-agnostic benchmark for assessing the energy efficiency of quantum technologies, guiding future efforts to design sustainable and low-power quantum devices. Finally, we remark that the framework developed here can be modified, generalized, and developed in many directions. One of the possibilities is to include classical registers and gates in order to incorporate the expense of classical computing within the model. Another possibility is to treat time of the execution of the protocol as a resource or to incorporate entire spatial and temporal structure of the setup.

\subsection{Lower bound on the standard fundamental energy consumption rate stemming from the irreversibility phenomenon in entanglement theory}
In this section we introduce a quantity called (single shot) {\it standard} ECRED. It is a simplified version of the single shot ECRED (in fact an upper bound), where the choices 
of the physical realization, entanglement encoding and input states are standard for the case of entanglement distribution. 
Recall, that this standard approach is encapsulated by the following choices:
\begin{itemize}
    \item {\bf Standard Physical Realization}:
Entanglement encoded by means of polarized photons, 
\item {\bf Standard Input state}: Typically the input to the channel is maximally entangled state: $\rho_{in} := \psi_{d_{in}}^+$, 
\item {\bf Standarized protocol}: One of typical ways of distributing entanglement is sending half of the input state down the channel. Effectively $\Phi:=\id_{A} \otimes \Phi_{A'\rightarrow B}$.
\end{itemize}
Indeed, current standard protocols use singlet states as inputs to noisy quantum channels, and further distill entanglement from their outputs. We propose the following definition of the {\it standard fundamental energy consumption rate of  entanglement distribution}.
By definition we have:
\begin{equation}
C_{\mathrm{std}}^\epsilon(\Phi|\mathrm{Ent})\geq C^\epsilon(\Phi|\mathrm{Ent}),
\end{equation}
due to the fact that $C_{\mathrm{std}}^\epsilon(\Phi|\mathrm{Ent})$ reflects the commonly experienced fundamental energy consumption per ebit generation according to standard approach, we study it on its own right in this section. In particular we will place a lower bound on it, exhibiting limits of standard approach, focused on a task rather than its energy cost. Note that, while the definition is written in the case where party A generates and sends the singlet state $\psi^+_{d_{in}}$ to another party B, it is straightforward to write it for the case where a middle party generates and sends $\psi^+_{d_{in}}$ to both A and B. 

The remark below motivates substitution of energy of maximally entangled state in place of the not necessarily equal to the latter output.

\begin{rem} Entanglement is usually encoded
in non-energetic degree of freedom e.g. such as polarization of photons. 
Thus, $E(\rho_{AB}^{(out)})=E(\psi^+_{d_{out}})$ (all polarization states have the same energy). Later in Section \ref{sec:upper_bounds}, considering the case when
$E(\rho_{out})\neq E(\psi^+_{d_{out}})$ 
%we consider RHS of the Eq (\ref{eq:en_dissip_photon}), however, without the additive term $\epsilon E_{max}$. In such a case, 
we will derive continuity for energy, expressed in Lemma \ref{lem:1orMoreparticleEnegyContinuityB}, resulting in additional term
$\epsilon E_{max}$, which does not appear below.
\end{rem}

\begin{definition}\label{def:ecr_single_shot}
The single shot standard fundamental energy consumption per ebit distribution through channel $\Phi$
\begin{equation}
C_{\mathrm{std}}^\epsilon(\Phi|\mathrm{Ent}):=
\inf_{d_{out}\in \mathbb{N}_+, \,\newline\Lambda \in LOCC \,\newline}
    \left[\frac{ E(\psi^+_{d_{in}})+ E(\Lambda,\id\otimes \Phi(\psi_{d_{in}}^+)) -E(\psi^+_{d_{out}})}{\log_2 d_{out}} \,:
    \Lambda(\id \otimes \Phi (\psi^+_{d_{in}}))\approx_\epsilon \psi^+_{d_{out}}\right],    \label{eq:def_en_dissip_photon}
\end{equation}
where $\epsilon \in (0,1)$ and $\sigma_1 \approx_\epsilon \sigma_2 \iff 1/2\| \sigma_1 - \sigma_2 \|_1 \leq \epsilon$. 
Its regularized version is
\begin{equation}
C_{\mathrm{std}}(\Phi|\mathrm{Ent}):=\sup_{\epsilon\in(0,1)}\liminf_{n\rightarrow \infty} C_{\mathrm{std}}^\epsilon(\Phi^{\otimes n}|\mathrm{Ent}).
\end{equation}
\end{definition}
 %{\mw this sentence confronts fundamental aspect}. 
In what follows, we derive a lower bound to this quantity, which estimates in turn the inevitable, or fundamental energy dissipation quantum networks distributing entanglement though distillation of singlet states.

 Following notation of \cite{Gourbook}, we define the single shot distillable entanglement~\cite{BuscemiDatta} as 
\begin{definition} Given a bipartite state $\rho_{AB}$ and a parameter $\epsilon \in (0,1)$, the single shot distillable entanglement is defined as
\begin{equation}
E_D^{\epsilon}(\rho_{AB}):= \max_{d \in \mathbb{N}} \left\{ \log_2 d: \frac{1}{2}\min_{\Lambda_{d}}||\Lambda_{d}(\rho_{AB})- \psi^{+}_{d}||_{1}\leq \epsilon\right\}
\label{eq:ss_ed}
\end{equation}
where $\Lambda_{d}\in LOCC({\cal H}_{AB},{\cal H}_{A'B'})$.
\end{definition}

In what follows, it appears natural that some LOCC based entanglement distillation protocols have a lower rate than the optimal one. An LOCC operation  $\Lambda^\epsilon$ acting on $\rho_{AB}$,
is called a single-shot distillation protocol if $\Lambda^\epsilon(\rho_{AB})=\rho_{A'B'}$ and
$||\rho_{A'B'}-\psi^+_{d}||_1\leq \epsilon$. The {\it achievable rate} of a protocol $\Lambda^\epsilon$, denoted as   $R(\Lambda^\epsilon,\rho_{AB})$ is  $R(\Lambda^\epsilon,\rho_{AB}):= \log_2 d$. 

We also recall that the asymptotic, (two way) distillable entanglement is \cite{Gourbook}
\begin{equation}
E_D(\rho) := \inf_{\epsilon \in (0,1)}\limsup_{n\rightarrow \infty}\frac1n E_D^\epsilon (\rho^{\otimes n}).
\end{equation}

We now have all the elements to derive the lower bound on the energy consumption during resource distribution. It holds for input states represented by non-interacting particles, e.g. photons. First note that:
%It consists of two terms. The first is proportional to the irreversibility in entanglement generation  (\rj{what does this mean?}), while the second is the energy consumed by the distillation operation per generated ebit.

% when we cout energy then we consider
%effective hamiltonian
% \begin{observation}
% When the maximally entangled state $\psi_d^+$ of local dimension $d\geq 2$ is represented by the non-interacting physical subsystems (e.g. photons), each representing a qubit with energy $E$, the density of energy $E_{dens}(\psi_{d}^+)$ equals $E$.

% \label{obs:density}
% \end{observation}
% \begin{proof}
% The bipartite state $\psi_{d}^+$ of dimension $d^2$ can be optimally encoded in $k=2\lceil\log_2 d\rceil$ qubits. We also have that $E_{dens}(\psi_d^+)=\frac{E(\psi_d^+)}{2\lceil\log_2 d\rceil}$. We further note that, since the bipartite  state resides on $k$ qubits, represented by non-interacting particles, its energy equals $k\times E$. Hence the assertion follows. 
% \end{proof}
%NAPOZNIEJ\textcolor{red}{From the above we get two independent lower bounds that hold for non-interacting particles. }
\begin{rem}In what follows, we assume that the distillation operation $\Lambda$ does not alter the form, nor the parameters of the local Hamiltonian of the system. Specifically:
\begin{enumerate}
\item Physical realization of entanglement: the input entanglement and output entanglement are encoded on the same physical carriers (e.g. photons). %\sout{The same will be assumed in the context of other resources than entanglement.} 
Therefore, the form of the system Hamiltonian remains unchanged.

\item For simplicity of presentation, we assume that the distillation operation $\Lambda$ is passive, in the sense that it does not alter the energy of a particle representing single qubit. For example, in the MNR setting~\cite{Asiani_2019}, the final field configuration is the same as initially, so parameters in the Hamiltonian are not modified. Therefore, the input and output are quantified in the same units of energy.  
\end{enumerate}
\end{rem}

\begin{theorem}
Let $\epsilon >0$. For an LOCC channel $\Phi$ with input dimension $d_{in}$ such that the input to this channel $\psi_{d_{in}}^+$ is represented by non-interacting particles (e.g. photons), each representing a logical qubit with energy $E$, we have that:
\begin{equation}
   C_{\mathrm{std}}^{\epsilon}(\Phi|\mathrm{Ent}) \geq  2E \times \left[ \frac{\lceil \log_2 d_{in}\rceil}{\lceil E_D^\epsilon(\rho_\Phi)\rceil} -1\right]
   \label{eq:lower_bound_ceil}
\end{equation}
\label{thm:single_shot_lb}
\end{theorem}

\begin{proof}
Let $R(\Lambda^\epsilon,\rho_\Phi)$ be the rate of entanglement distillation $\log_2 d_{out}$ of a protocol, that minimize the RHS in (\ref{eq:def_en_dissip_photon}) (in Definition \ref{def:ecr_single_shot}). We then note that it does not need to be an optimal entanglement distillation rate achieving $E_D^{
\epsilon}(\rho_\Phi)$, as it focuses on minimizing energy expense rather than maximizing the distilled entanglement. That is, we have

\begin{equation}
E_D^{\epsilon}(\rho_\Phi)\geq R(\Lambda^\epsilon,\rho_\Phi).
\label{eq:non-optimal}
\end{equation}

Moreover, by definition of $R(\Lambda^\epsilon,\rho_\Phi)$ the map $\Lambda^{\epsilon}$ outputs a state close to a singlet of local dimension $d_{out}$. We then have 

\begin{equation}
C_{\mathrm{std}}^\epsilon(\Phi|\mathrm{Ent})=
\frac{ E(\psi^+_{d_{in}})+ E(\Lambda^{opt}_\epsilon,\rho_{\Phi}) -E(\psi^+_{d_{out}})}{R(\Lambda^{opt}_\epsilon,\rho_\Phi)} 
\label{eq:smarater}
\end{equation}
%where $\Lambda^{opt}_\epsilon$ is an optimal the optimal 
%protocol, i.e. that minimizes RHS in (\ref{eq:en_dissip_photon}). \rj{something is missing to arrive to (28).}
%\sout{From definition of energy density given in Eq.(\ref{eq:en_density})
%there is $E(\psi_{d_{in}}^+)= E_{dens}(\psi_{d_{in}}^+)
%\times 2 \lceil\log_{2} d_{in}\rceil$. Now, by Observation \ref{obs:density} we have
%that $E_{dens}(\psi_{d_{in}}^+)=E$. Hence $E(\psi_{d_{in}}^+)=E\times 2\lceil \log_2 d_{in}\rceil$.}
\textcolor{black}{By assumption, the maximally entangled state is encoded on non-interacting particles. Hence, we have $E(\psi_{d_{in}}^+)=E\times 2\lceil \log_2 d_{in}\rceil$. Here, $E$ is the free energy of a single carrier subsystem (particle), and $2\lceil \log_2 d_{in}\rceil$ is the minimal number of carriers required to encode state $\psi_{d_{in}}^+$.} Similarly $E(\psi_{d_{out}}^+)=E\times 2\lceil R(\Lambda^{opt}_\epsilon,\rho_\Phi)\rceil$ because $\log_2 d_{out} =R(\Lambda^{opt}_\epsilon,\rho_\Phi)$ is the rate of ebits distilled by protocol $\Lambda^{opt}_\epsilon$ from the state $\rho_\Phi$, %\sout{and by Observation (\ref{obs:density}) we have $E_{dens}(\psi_{d_{out}}^+)=E$}.
We substitute these identities, obtaining

\begin{equation}
C_{\mathrm{std}}^\epsilon(\Phi|\mathrm{Ent})=
\frac{ 2E\times\lceil \log_2 d_{in}\rceil +E(\Lambda_\epsilon^{opt},\rho_\Phi) -2E\times \lceil R(\Lambda^{opt}_\epsilon,\rho_\Phi)\rceil}{R(\Lambda^{opt}_\epsilon,\rho_\Phi)}
\label{eq:fraction_bound_0}
\end{equation}

We now use the fact that, by assumption, the energy of entanglement distillation satisfies Axiom 1 \textit{i.e.} is a non-negative function . This implies that $E(\Lambda_\epsilon^{opt},\rho_\Phi)\geq 0$. We can then drop it, obtaining much lower but simpler to estimate bound:

\begin{equation}
C_{\mathrm{std}}^\epsilon(\Phi|\mathrm{Ent}) \geq
\frac{ 2E\times\lceil \log_2 d_{in}\rceil -2E\times \lceil R(\Lambda^{opt}_\epsilon,\rho_\Phi)\rceil}{R(\Lambda^{opt}_\epsilon,\rho_\Phi)}
\label{eq:fraction_bound}
\end{equation}

We further  notice that
\begin{equation}
    \lceil \log_2 d_{in}\rceil \geq \lceil R(\Lambda_\epsilon^{opt},\rho_\Phi)\rceil
\end{equation}
which is due to the fact that distillable entanglement is non-increasing property under a channel $\Phi$~\cite{BDSW96}.
Indeed, the channel $\Phi$ is, by assumption, LOCC, which do not increase distillable entanglement (initially equal to $\log_2 d_{in}$). Hence, the numerator in Eq.~\eqref{eq:fraction_bound} is non-negative and we can lower bound the total fraction by replacing $R(\Lambda_\epsilon^{opt},\rho_\Phi)$ in the denominator by an upper bound, namely (due to Eq.~\eqref{eq:non-optimal}) $\lceil E_D^\epsilon(\rho_\Phi)\rceil$.
Using again the fact that the single shot distillable entanglement upper bounds $R(\Lambda^{opt}_\epsilon(\rho_\Phi))$, and owing to the fact that $E\geq 0$, we can replace the latter by $\lceil E^\epsilon_D(\rho_\Phi))\rceil$ in numerator and arrive at the lower bound
\begin{equation}
C_{\mathrm{std}}^\epsilon(\Phi|\mathrm{Ent})\geq
2E
\times \left(\frac{ \lceil \log_2 d_{in}\rceil}{\lceil E_D^\epsilon(\rho_\Phi)\rceil} - 1\right)
\end{equation}
\end{proof}

While we can consider better lower bounds,
the ones obtained under this assumption separate energy cost and irreversibility factor from each other in appealing way. %\cite{Future}.

\begin{rem}
We note that the choice of basis of the logarithm in the numerator of the Eq. \eqref{eq:fraction_bound} is arbitrary as long as the base of the logarithm in the definition of $E_D^\epsilon$ (in \eqref{eq:ss_ed}) has the same basis. 
%In case of other resources than entanglement, such as magic, it is natural to choose the basis such that the single-shot distillable resource of a target state is $1$. In case of magic, when we distill e.g. the so called Norren or Strange states, the natural base would be $3$, since these are qutrit states \cite{}.
\end{rem}

Following from the above theorem, we come to an important result which states the lower bound for the fundamental energy consumption of an asymptotic entanglement distillation protocol. It requires the following mathematical fact, proved in Observation \ref{obs:liminf_limsup} of the Supplemental Material: for a bounded sequence of non-negative numbers $\{a_n\}_n$, we have:
\begin{equation}
\liminf_{n\rightarrow \infty} \frac{1}{a_n} =
\frac{ 1}
{\limsup_{n\rightarrow \infty} a_n}.
\end{equation}

We can now show the asymptotic version of theorem \ref{thm:single_shot_lb}. It yields a non-trivial lower bound already for a single qubit channel, \textit{i.e.} in the case where the bound \eqref{eq:lower_bound_ceil} is not useful, equal either to infinity in case of channels having non-distillable Choi-Jamiłkowski states or zero (indeed, in the case of a 2-qubit distillable state,
$E_D^\epsilon(\rho_\Phi)\leq 1$ by definition).

\begin{proposition}
For a channel $\Phi$ with input dimension $d_{in}$ such that the input to this channel $\psi_{d_{in}}^+$ is encoded in non-interacting particles (e.g. photons) 
and $E_D(\rho_\Phi) >0$, where $\rho_{\Phi}$ is the Choi-Jamiołkowski state of the channel $\Phi$, we have
\begin{equation}
C_{\mathrm{std}}(\Phi|\mathrm{Ent})\geq
2E\times\left(\frac{ \log_2 d_{in}}{E_D(\rho_\Phi)} - 1\right).\label{eq:main_asymptotic}
\end{equation}
\label{prop:main_asymptotic}
\end{proposition}

\begin{proof}
By Theorem \ref{thm:single_shot_lb} we know that
\begin{equation}
\sup_{\epsilon \in(0,1)}\liminf_{n\rightarrow \infty} C_{\mathrm{std}}^\epsilon(\Phi^{\otimes n}|\mathrm{Ent})\geq 
\sup_{\epsilon \in(0,1)}\liminf_{n\rightarrow \infty}
\left[2E\times\left(\frac{ \lceil\log_2 d_{in}^n\rceil}{\lceil E_D^\epsilon(\rho_\Phi^{\otimes n})\rceil} - 1\right)\right]
\end{equation}
Indeed, 
noticing that the dimension of the input to the $\Phi^{\otimes n}$ is $d_{in}^n$, we have
\begin{equation}
C_{\mathrm{std}}^\epsilon(\Phi|\mathrm{Ent})\geq
2E\times\left(\frac{ \lceil  \log_2 d_{in}^n\rceil}{\lceil E_D^\epsilon(\rho_\Phi^{\otimes n})\rceil}  - 1\right)
\label{eq:current}
\end{equation}
and we take $\sup_{\epsilon\in(0,1)}\liminf_{n\rightarrow \infty}$ on both sides.
We further lower bound the fraction on the RHS in a sequence of (in)equalities which we comment below.
\begin{align}
\sup_{\epsilon\in(0,1)}\liminf_{n\rightarrow \infty}\left[ \frac{\lceil\log_2 d_{in}^n\rceil}{\lceil E_D^\epsilon(\rho_\Phi^{\otimes n})\rceil}\right]  =\\
\sup_{\epsilon\in(0,1)}\liminf_{n\rightarrow \infty}\left[ \frac{\frac1n\lceil\log_2 d_{in}^n\rceil}{\frac1n\lceil E_D^\epsilon(\rho_\Phi^{\otimes n})\rceil}\right] \geq 
\label{eq:numerator}
\\
\sup_{\epsilon\in(0,1)}\liminf_{n\rightarrow \infty}\left[ \frac{\log_2 d_{in}}{\frac1n\lceil E_D^\epsilon(\rho_\Phi^{\otimes n})\rceil}\right]  =
\label{eq:liminf_limsup_down}\\
%\inf_{\epsilon>0}\left[ \frac{\liminf_{n\rightarrow \infty}\frac1n\lceil\log_2 d_{in}^n\rceil}{\limsup_{n\rightarrow \infty}\frac1n\lceil E_D^\epsilon(\rho_\Phi^{\otimes 
%n})\rceil}\right] \geq \label{eq:numerator} \\
\sup_{\epsilon\in(0,1)}\left[ \frac{\log_2 d_{in}}{\limsup_{n\rightarrow \infty}\frac1n\lceil E_D^\epsilon(\rho_\Phi^{\otimes n})\rceil}\right] = 
\label{eq:infsup}\\
 \frac{\log_2 d_{in}}{\inf_{\epsilon\in(0,1)}\limsup_{n\rightarrow \infty}\frac1n\lceil E_D^\epsilon(\rho_\Phi^{\otimes n})\rceil} \geq
\label{eq:fraction}\\
 \frac{\log_2 d_{in}}{\inf_{\epsilon\in(0,1)}\limsup_{n\rightarrow \infty}\left(\frac1n E_D^\epsilon(\rho_\Phi^{\otimes n}) +\frac1n
 \right)}  \geq\\
 \frac{\log_2 d_{in}}{\inf_{\epsilon\in(0,1)}\left[\limsup_{n\rightarrow \infty}\left(\frac1n E_D^\epsilon(\rho_\Phi^{\otimes n}) +\limsup_{n\rightarrow \infty}\frac1n\right)\right]
 }=\\
 \frac{\log_2 d_{in}}{E_D(\rho_\Phi)}.
\end{align}
For the first equality, we multiply numerator and denominator of the fraction in the above equation by $\frac1n$. We then use the fact that, for a real positive $x$, we have 
\begin{equation}
\frac1n \lceil n x\rceil \geq \frac1n \times nx = x .
\end{equation}
Hence, we can lower bound the  numerator (and the whole expression) in (\ref{eq:numerator}) by $\log_2d_{in}$.
The inequality (\ref{eq:liminf_limsup_down}) follows from Observation \ref{obs:liminf_limsup}.
We note here that the sequence $\frac1n E_D^\epsilon(\rho_\Phi^{\otimes n})$ is bounded since,
by definition of single-shot distillable entanglement (see Eq.~\eqref{eq:ss_ed}), for any $\epsilon\in(0,1)$ there is
$\log_2 d_{in}=\frac1n\log_2 d_{in}^n \geq \frac1n E_D^\epsilon(\rho_\Phi^{\otimes n})\geq 0$. Indeed, the highest value of $E_D^\epsilon(\sigma)$ is for $\sigma =\psi^+_{d_{in}}$ and equals $\log_2 d_{in}$.
In the inequality (\ref{eq:infsup}) we use the fact that for non-negative function $c(\epsilon)$ of $\epsilon\in(0,1)$ there is
\begin{equation}
\sup_{\epsilon \in(0,1)} \frac{1}{c(\epsilon)} =
\frac{1}{\inf_{\epsilon\in(0,1)} c(\epsilon)}.
\end{equation}
Further in (\ref{eq:fraction}) we use 
the inequality (for a non-negative $y$):
\begin{equation}
\frac1n \lceil y\rceil \leq \frac1n y +\frac1n.
\end{equation}
In the last equality, we use definition of $E_D(\rho_\Phi)$ and the fact that, for two sequences $\{a_n\}_n$ and $\{b_n\}$, there is 
\begin{equation}
\limsup_{n\rightarrow \infty} (a_n+b_n)\leq
\limsup_{n\rightarrow \infty} (a_n)+
\limsup_{n\rightarrow \infty} (b_n)
\end{equation}
to get that the denominator is upper bounded by
$E_D(\rho_\Phi)+ \inf_{\epsilon\in(0,1)}\limsup_{n\rightarrow\infty}\frac1n=E_D(\rho_\Phi)$, and the assertion 

\begin{equation}
C_{\mathrm{std}}(\Phi|\mathrm{Ent})\geq
2E\times\left(\frac{\log_2 d_{in}}{ E_D(\rho_\Phi) }  - 1\right)
\end{equation}
follows.
\end{proof}

%we drop the last term of \ref{eq:main_lower_bound} as it is non-negative. We obtain regularization noticing that $E_C(\rho^{\otimes n})=nE_C(\rho^{\otimes n})$ and similarly for $E_D$, hence we can
%    regularize the LHS keeping the lower bound.

As a comment to the above proposition, let us note, that the lower bound amounts to infinity for non-distillable states including separable and bound entangled states, since for these states $E_D=0$. This fact reflects the situation that no matter how much energy we spend, if  the channel is binding entanglement \cite{Horodecki2000} or entanglement breaking \cite{Horodecki2003}, there is no way of producing even a single ebit.
%\textcolor{red}{@Marek - pls add citations of ppl who introduced the quantities}
\subsection{Entanglement irreversibility as a  sufficient condition for positive standard energy consumption rate }
We now show two corollaries which provides appealing lower bounds that relate our quantity to the notions of entanglement irreversibility and channel capacity respectively. To state the first one, we recall the definitions of two well known entanglement measures: the entanglement of formation and the entanglement cost.
\begin{definition}
    Let $\rho_{AB}$ be a bipartite quantum state. The entanglement of formation $E_F$ is defined as follows
    \begin{equation}
        E_F(\rho_{AB}) := \inf \sum_k p_k S_A(\psi_{AB}),
    \end{equation}
    where the infimum is taken over all decompositions of a state $\rho_{AB}$ into mixtures of pure states $\rho_{AB} = \sum_k p_k |\psi_k \> \< \psi_k|$ and $S_A(\psi_{AB})$ is the Von Neumann entropy of the subsystem A of a pure state $\psi_{AB}$. 
\end{definition}

\begin{definition}
    The asymptotic entanglement cost $E_{C}(\rho)$ and one-shot entanglement cost $E_{C}^{\epsilon}(\rho)$ of a bipartite quantum state $\rho_{AB}$ are defined as 
    \begin{align}
        E_{C}(\rho) & := \sup_{\epsilon \in (0,1)} \limsup_{n\rightarrow\infty} \frac{1}{n} E_{C}^{\epsilon}(\rho^{\otimes n}) , \\
        E_{C}^{\epsilon}(\rho) & := \inf_{\substack{\Lambda \in \mathrm{LOCC},\\ d \in \mathbb{N}_+}}
        \left\{
        \begin{array}{c}
            \log_2 d : \\
            \frac{1}{2} \|\Lambda(\psi_d^+)- \rho \|_1\leq \epsilon
        \end{array}\label{eq:def-one-shot-entanglement-cost}
        \right\},
    \end{align}
    where the infimum in \eqref{eq:def-one-shot-entanglement-cost} is taken over every LOCC channel $\Lambda$ and every positive integer $d$.
\end{definition}

\begin{corollary}
\label{col:bound-by-ec-ed}
For a channel $\Phi$ with input dimension $d_{in}$ such that the input to this channel $\psi_{d_{in}}^+$ is encoded in non-interacting particles (e.g. photons), each representing a (logical) qubit of energy $E$, and the Choi state of $\Phi$ is $\rho_\Phi$, there is 
\begin{equation}
    \label{eq:bound-by-ec-ed}
   C_{\mathrm{std}}(\Phi|\mathrm{Ent}) \geq  2E\times\left[ \frac{E_C(\rho_\Phi)}{E_D(\rho_\Phi)} -1\right]
\end{equation}
\label{cor:bound-by-ec-ed}
\end{corollary}

\begin{proof}
It follows from the fact that for a bipartite state $\rho $ of local dimension $d_{in}$, there is $E_C(\rho)=E_F^\infty(\rho)\leq E_F(\rho) \leq \log_2 d_{in}$, where $E_C(\rho)$ and  $E_F(\rho)$ stand for entanglement cost and entanglement of formation, respectively
\cite{Hayden_2001}.
\end{proof}

%We have thus a quantitative statement:
% On the other hand, from Proposition \ref{prop:main_asymptotic} it is clear that LOCC irreversibility is not a necessary condition for non-zero energy expense in entanglement distribution it is non-zero even for channels which transform maximally entangled states into pure states with biased Schmidt coefficients. Namely we obtain:
%\begin{equation}
%C_{\mathrm{std}}(\Phi|\mathrm{Ent})\geq 2E\times \left (\frac{\log_2 d_{in}}{S_A(\phi)} -1 \right )
%\end{equation}
%where $S_A(\rho_\Phi)=E_D(\phi)$ is the von-Neumann entropy $-Tr\rho\log_2\rho$ of the subsystem A of a %pure state $\phi$ that is the output of the channel \cite{Bennet_1996, HHHH09}. Clearly the above bound %is zero only for the identity channel.
% \begin{rem}
% Note that LHS is J/ebit  and so is with the RHS, while the factor $(E_C/E_D -1)$ seems to be dimensionless.
% The upshot is that $E_C$ is a dimensionless number (there is no protocol done which has $E_C$ rat 
%  {\pmaz ?}), while $E_D$ in numerator which
% truly happens in the distillation process has
% dimension $ebit/second$. {\pmaz why per second?} Similarly
% the $-1$ term is not dimensionless,
% it is counted in $1/(ebit)$.
% \end{rem}

Before showing the other corollary, let us recall the definition of the two-way quantum capacity \cite{MarkWildeBook} : 

\begin{definition} The two-way quantum capacity $Q^{\leftrightarrow}$ of the channel $\Phi$ is defined as:
\begin{equation}Q^{\leftrightarrow}(\Phi):= \sup \left\{\log_2  D:\lim_{n\rightarrow\infty}\inf_{\phi_n}[1-F(\phi_n,\mathcal{R}_n(\phi_n))]=0\right\}
\end{equation}
where $F$ is fidelity, ${\cal R}_n$ is the resultant operation after $n$ uses of the channel incorporating all adaptive processes involving two way classical communication, and $\phi_n$ is of dimension $D$. 
\end{definition}

\begin{corollary}
For a channel $\Phi$ with input dimension $d_{in}$ such that the input to this channel $\psi_{d_{in}}^+$ is represented by non-interacting particles (e.g. photons) each representing a (logical) qubit of energy $E$, there is \begin{equation}
   C_{\mathrm{std}}(\Phi|\mathrm{Ent}) \geq  2E\times \left[ \frac{\log_2 d_{in}}{Q^{\leftrightarrow}(\Phi)} -1\right]
\end{equation}
\label{cor:lb-by-channel}
\end{corollary}

\begin{proof}
It follows from Propositon~\ref{prop:main_asymptotic} in an analogous way to the previous corollary, however keeping the $\log_2 d_{in}$ while upper bounding the denominator: $Q^{\leftrightarrow}(\Phi)\geq E_{D}({\rho_\Phi})$ \cite{Bennet_1996} (see in this context \cite{Pirandola2017}). To recall the proof of the latter inequality note, that sending a half of the singlet state, distilling entanglement by two-way classical communication and performing quantum teleportation is a viable, but possibly suboptimal protocol for sending qubits via $\Phi$. This potentially suboptimal protocol has rate of faithfully sent qubits equal to $E_D(\rho_\Phi)$. 
\end{proof}
\begin{remark}
For certain channels like erasure and dephasing channels the corollaries \ref{cor:lb-by-channel} the bound from Eq. (\ref{eq:main_asymptotic}) in Proposition \ref{prop:main_asymptotic} coincide, because $Q^{\leftrightarrow}=E_D^{\leftrightarrow}$ for them \cite{BDSW96,Pirandola2017}. However, they are not equivalent in general \cite{MarkWildeBook}.
\end{remark}

Let us note that the bounds given in Proposition \ref{prop:main_asymptotic} and Corollaries \ref{cor:bound-by-ec-ed} and \ref{cor:lb-by-channel}, although non-zero in non-trivial cases of mixed states, are not easily computable. The following corollary gives a computable bound. 

\begin{corollary}
\label{cor:text-lb-by-logneg}
For a channel $\Phi$ with input dimension $d_{in}$ such that the input to this channel $\psi_{d_{in}}^+$ is represented by non-interacting particles (e.g. photons) each representing a (logical) qubit of energy $E$, and the Choi state of $\Phi$ is $\rho_\Phi$, there is 
\begin{equation}
   C_{\mathrm{std}}(\Phi|\mathrm{Ent}) \geq  2E\times \left[ \frac{\log_2 d_{in}}{\log_2 ||\rho_{\Phi}^{\Gamma}||_{1}\label{cor:lb-by-logneg}
   } -1\right]
\end{equation}
\end{corollary}
\begin{proof}
    It is known that NPT-entanglement is bounded from above by the Rains' bound, namely for a Choi state $\rho_{\Phi}$
    \begin{equation}
        E_{D}(\rho_{\Phi})\leq \min_{\sigma}\{D(\rho_{\Phi}||\sigma)+\log_2 ||\sigma^{\Gamma}||_{1}\}\ ,
    \end{equation}
where $\sigma$ is a state from the set of all bipartite mixed quantum states. For $\sigma:=\rho_{\Phi}$ the above inequality simplifies to 
    \begin{equation}
        E_{D}(\rho_{\Phi})\leq \log_2 ||\rho_{\Phi}^{\Gamma}||_{1}\label{eq:distUB_neg}\ ,
    \end{equation}
where the right hand is logarithmic negativity of state $\rho_{\Phi}$. Hence, from the Eqs.\eqref{eq:distUB_neg} and \eqref{eq:main_asymptotic} it is straightforward that
\begin{equation}
C_{\mathrm{std}}(\Phi|\mathrm{Ent})\geq
2E\times\left(\frac{ \log_2 d_{in}}{E_D(\rho_\Phi)} - 1\right)\geq 2E\times\left(\frac{ \log_2 d_{in}}{||\rho_{\Phi}^{\Gamma}||_{1}} - 1\right)\ .
\end{equation}
\end{proof}

In fact, for any upper bound $U(\rho_\Phi)\geq E_D(\rho_\Phi)$ such as the Rains' bound, distillable key, relative entropy of entanglement
and squashed entanglement (see definitions along with original literature in \cite{Gourbook}), we obtain a lower bound of the form
\begin{equation}
    C_{\mathrm{std}}(\Phi|\mathrm{Ent})\geq 2 E\times\left[\frac{\log_2 d_{in}}{U(\rho_\Phi)} -1\right].
\end{equation}

We have singled out the log-negativity to give an example of an easily computable bound. Indeed, the other lower bounds to the standard fundamental energy consumption per ebit distributed mentioned in this work are challenging to estimate.

%%%%%%%%%%%%%%%%%%%%%%%%
\subsection{Computing the lower bound on the energy consumption rate for a few examples of quantum channels}
\subsubsection{Quantum erasure channel}
As an example, let us have a look at the quantum erasure channel (QEC) $\Phi_{\rm eras}^p$ which, with probability p, replaces the incoming qubit with an erasure state, orthogonal to both $\ket{0}$ and $\ket{1}$, thereby both erasing the qubit and informing the receiver that it has been erased. The two-way capacity of the QEC is given by~\cite{Bennett_QEC}:
\begin{equation}
\label{eq:2waycapacQEC}
    Q^{\leftrightarrow}(\Phi_{\rm eras}^p) = 1-p .
\end{equation}

Let us now imagine that Alice prepares photonic two-qubits singlet states $\ket{\psi^+}$ and sends single photons to Bob through $\Phi_{\rm eras}^p$. The previous corollary shows that the optimal protocol distilling entanglement will have a fundamental cost lower bounded by 

\begin{equation}
\label{eq:boundQEC}
    C_{\mathrm{std}}(\Phi_{\rm eras}^p|\mathrm{Ent}) \geq  2\hbar\omega\times \left[ \frac{p}{1-p} \right]
\end{equation}.

\begin{figure}[!ht]
    \centering
    \includegraphics[width=0.8\linewidth]{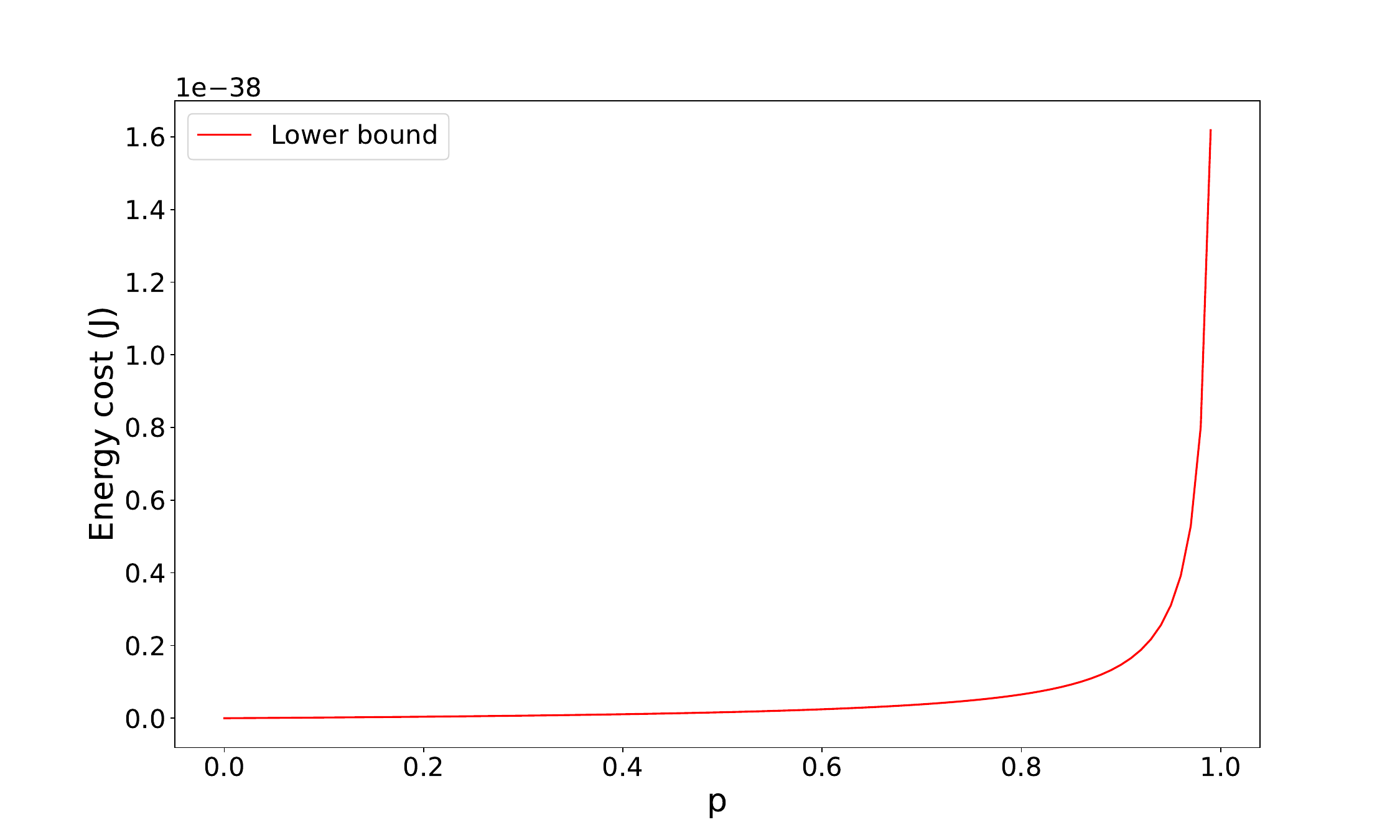}
    \caption{\justifying Lower bound on the standard fundamental energy consumption of a distillation protocol using the quantum erasure channel $\Phi_{\rm eras}^p$ at telecom wavelength ($ 1550$ nm) as a function of the erasure probability $p$.}
    \label{fig:lowerboundQEC}
\end{figure}

This bound, shown in Fig.~\ref{fig:lowerboundQEC}, blows up to infinity as $p\rightarrow 1$, as expected since all the photons are then lost. Another known bound is the bosonic dephasing channel $\Phi_{\rm deph}^p$, studied in~\cite{Lami_2023} and given by
\begin{equation}
\Phi_{\rm deph}^p(\rho):=\int_{-\pi}^\pi d \phi \, p(\phi) e^{-i a^{\dagger} a \phi} \rho e^{i a^{\dagger} a \phi},
\end{equation}

where $p$ is a probability density function on the interval $[-\pi,\pi]$ and $a^\dagger a$ is the photon number operator. The two-way quantum capacity is given by 

\begin{equation}
\label{eq:2waycapacBDC}
    Q^{\leftrightarrow}(\Phi_{\rm deph}^p) = \log_2(2\pi) - h(p) .
\end{equation}

where
\begin{equation}
h(p):=-\int d \phi p(\phi) \log_2(p(\phi))
\end{equation}

is the differential entropy of the probability density p.

Unfortunately, very few other channels have an explicit formula for their two-way quantum capacity.   This makes the lower bound on the energy cost of distilling entanglement on these channels hard to compute.  

%%%%%%%%%%%
\subsubsection{Quantum depolarizing and amplitude damping channels}
\begin{figure}[!ht]
    \centering
    \includegraphics[width=0.8\linewidth]{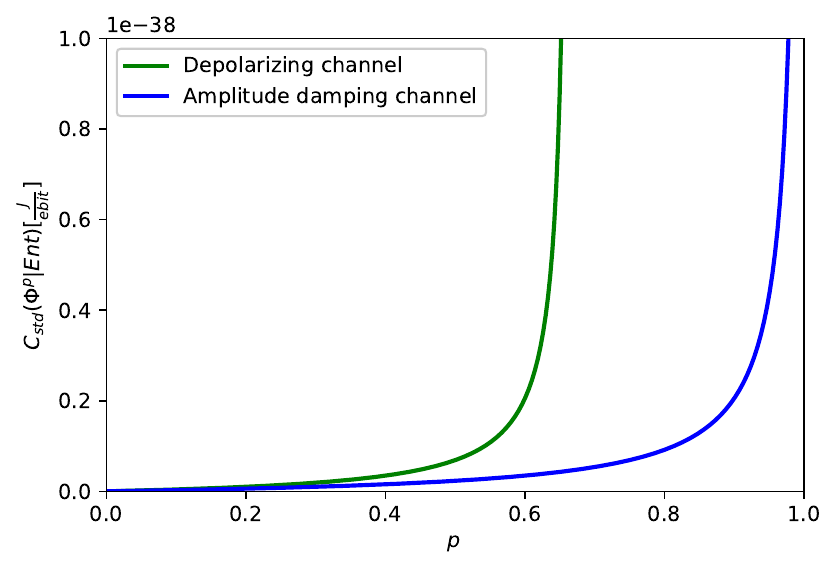}
    \caption{\justifying Lower bounds on the standard fundamental energy consumption rate of distillation protocols, for the quantum depolarizing and amplitude damping channel at telecom wavelength ($1550$ nm) as a function of the depolarizing (damping) probability $p$.}
    \label{fig:lowerboundQDepAndAdCHanel}
\end{figure}

Since commonly used noise models for photons are depolarizing and amplitude damping channels it is natural to calculate their lower bounds of energy consumption rate (Eq. \eqref{eq:bound-by-ec-ed}). Although,  value of distillable entanglement $E_D$  for these channels is not known its upper bound already is (Corollary \ref{cor:text-lb-by-logneg}). Their Choi's states are the following
\begin{align}
    \rho_{depol}&=(1-p)\ketbra{\Phi^{+}}{\Phi^{+}}+\frac{p}{4}\id \label{eq:depolChoi}\ ,\\
    \rho_{ad}&=\frac{1}{2}(\ketbra{00}{00}+(1-p)\ketbra{11}{11}+\sqrt{(1-p)}(\ketbra{00}{11}+\ketbra{11}{00})+p\ketbra{10}{10})\ ,\label{eq:adChoi}
\end{align}
where $p$ is the probability of depolarization \eqref{eq:depolChoi}  or damping \eqref{eq:adChoi} of a channel.
Equation \eqref{cor:lb-by-logneg} shows that it is enough to derive lower bounds for quantum depolarizing and amplitude damping channels, respectively
\begin{align}
    C_{\mathrm{std}}(\Phi_{\rm depol}^p|\mathrm{Ent}) &\geq  2\hbar\omega\times \left[ \frac{1}{\log_{2}(2-\frac{3p}{2})}-1 \right]\ ,\label{eq:boundQDepCH}\\
    C_{\mathrm{std}}(\Phi_{\rm ad}^p|\mathrm{Ent})& \geq  2\hbar\omega\times \left[ \frac{1}{\log_{2}(2-p)}-1 \right]\ ,\label{eq:boundQAdCH}
\end{align}
where expressions in the denominators stands for the logarithmic negativity of the Choi's state for the given channel. The above relations are depicted in Fig. \ref{fig:lowerboundQDepAndAdCHanel}. 
%\begin{figure}[!ht]
%    \centering
%    \includegraphics[width=0.8\linewidth]{LBAdChannel.eps}
%    \caption{Lower bound on the standarad fundamental energy %consumption of a distillation protocol using the quantum %amplitude damping channel $\Phi_{\rm adamp}^p$ at telecom %wavelength (\textit{i.e.} $\omega = 1550$ nm ) as a function %of the amplitude damping probability $p$.}
%    \label{fig:lowerboundQEC}
%\end{figure}

\subsection{Upper bound on fundamental energy consumption rate of entanglement distribution}
\label{section:UB}

Let us expand on the problem of computing the value of $C^{\epsilon}(\Phi|\mathrm{Ent})$ for a certain channel $\Phi$ (see Definition ~\ref{def:EnergyCost}). This task seems untractable for at least three reasons, which we want to emphasize now. 
\begin{align}
    & C^{\epsilon}(\Phi|\mathrm{Ent}):= \nonumber\\
    &\underbrace{\inf_{\mathrm{PR}}}_{\text{first problem}}\,\underbrace{\inf_{\substack{d_{out}({\textrm{PR}})\in \mathrm{N}_+ \,\\\Lambda({\mathrm{PR}}) \in LOCC, \,\\\rho_{A''B''}^{(in)}({\mathrm{PR}})}}}_{\text{second problem}} \,
    &\left[\frac{ E(\rho_{A''B''}^{(in)})+ \overbrace{E(\Lambda,\rho^{(int)}_{A'B'})}^{\text{third problem}} -E(\rho_{AB}^{(out)})}{\log_2 d_{out}}\,:\rho^{out}_{AB}:=
    \Lambda(\Phi (\rho_{A''B''}^{(in)}(\mathrm{PR})))\approx_\epsilon \psi^+_{d_{out}}\right],
\end{align}

\begin{enumerate}
    \item First of all, one has to consider an infimum over all physical realizations (PR), which seems to be a demanding task given the fact that even science-fiction literature does not cover all of them. The hardness of this task is that one has to take into account all possible, even unimaginable, physical platforms.
    \item Secondly, in the formula there is also an infimum over output dimensions $d_{out}({\mathrm{PR}})\in \mathrm{N}_+$, distillation protocols $\Lambda({\mathrm{PR}}) \in LOCC$ and input states $\rho_{A''B''}^{(in)}({\mathrm{PR}})$. Those objects are mathematically well defined, but the problem that one has to overcome here is that the choice of the parameters that is optimal in terms of distillation rate may not be optimal in terms of energy consumption. Thus, it is not clear, which choice of the parameters ensures the lowest energy consumption.
    \item Thirdly, the task of computing $E(\Lambda,\rho^{(int)}_{A'B'})$ seems to be highly non-trivial (see Section \ref{sec:axiomatic-apporach}). Thus we have to find a suitable way for dealing with this task.
\end{enumerate}

Since the task of computing $C^{\epsilon}(\Phi|\mathrm{Ent})$ precisely is untractable, we restrict ourselves to the estimation of an upper bound on this quantity. In this section, we develop a numerical framework for achieving this task. Then, we use it to estimate the upper bound on $C^{\epsilon}(\Phi|\mathrm{Ent})$ for a range of depolarizing channels $\Phi(\rho) = (1-\lambda)\rho + \lambda \id/d$ parametrized by a depolarizing parameter $\lambda$.

First, we have to overcome the three problems mentioned above. Observe, that the infimum over physical realizations (problem 1) can be upper bounded by a choice of specific physical realization. We choose linear optics, working at room temperature, and using light at telecom wavelength ($1550$ nm). The same logic applies to the infimum over $d_{out}({\mathrm{PR}})\in \mathrm{N}_+$, $\Lambda({\mathrm{PR}}) \in LOCC$ and $\rho_{A''B''}^{(in)}({\mathrm{PR}})$ (problem 2). In our analysis, we set $d_{out}({\mathrm{PR}}) = 1$ and $\rho_{A''B''}^{(in)}({\mathrm{PR}}) = \psi_+^{\otimes n}$. Moreover, we consider three different entanglement distillation protocols, namely BBPSSW \cite{bennett1996purification}, DEJMPS \cite{deutsch1996quantum} and P1-or-P2 \cite{miguel2018efficient}, as a choice of $\Lambda({\mathrm{PR}}) \in LOCC$. Those protocols are summarized in the tables below. We acknowledge that tables describing BBPSSW and DEJMPS protocols are taken from \cite{rozpkedek2018optimizing}.

\newcommand{\proj}[1]{|#1\rangle\!\langle#1|}
\begin{algorithm}[H]
\caption{BBPSSW protocol (description from \cite{rozpkedek2018optimizing})}\label{algorithm:BBPSSW-protocol}
\begin{algorithmic}[1]
	\State Depolarize the two available copies of the state to the isotropic state form:
	$\tau = p \proj{\Phi^+} + (1-p) \frac{\id}{4},$ with fidelity $F = (3p+1)/4$.
	\State Apply bi-local CNOT gates between the two copies.
	\State Measure the target qubits and communicate the results.
	\If{The measured flags are 00 or 11 (this occurs with probability $p_{\rm succ} =F^2+2F(1-F)/3+5[(1-F)/3]^2$) }
	 	\State The source (first) copy becomes more entangled than before (the fidelity to $\ket{\Phi^+}$ increases). We obtain a Bell diagonal state with fidelity $F'$ such that
	$$F' = \frac{F^2 + [(1-F)/3]^2}{p_{\rm succ}}.$$
	\EndIf
\State \textbf{return}
\end{algorithmic}
\end{algorithm}

\begin{algorithm}[H]
\caption{DEJMPS protocol (description from \cite{rozpkedek2018optimizing})}\label{algorithm:DEJMPS-protocol}
\begin{algorithmic}[1]
    \If{The input state $\rho_{AB}$ is not a Bell diagonal state}
	   \State Twirl the two available copies of the state to the Bell diagonal state using LOCC
    \EndIf
	\State Perform local rotations on both Alice's and Bob's qubits so that the two copies are in the form
	$$\tau = p_1 \proj{\Phi^+} + p_2 \proj{\Psi^+} + p_3 \proj{\Phi^-} +p_4 \proj{\Psi^-},$$
	with $p_1 >0.5$, $p_1 > p_2 \ge p_3 \ge p_4$ and $p_1 + p_2 + p_3 + p_4 = 1$. This ordering of the Bell coefficients allows to achieve the highest fidelity \cite{Dehaene2003}.
	\State Perform additional rotations: rotate both qubits on Alice's side by $\pi/2$ around $X$-axis and by $-\pi/2$ on Bob's side.
	\State Apply bi-local CNOT gates between the two copies.
	\State Measure the target qubits and communicate the results.
	\If{The measured flags are 00 or 11 (this occurs with probability $p_{\rm succ} =(p_1 + p_4)^2 + (p_2 + p_3)^2$) }
	 	\State The source (first) copy becomes more entangled than before (fidelity to $\ket{\Phi^+}$ increases). We obtain a state:
	$$\eta = p'_1 \proj{\Phi^+} + p'_2 \proj{\Psi^+} + p'_3 \proj{\Psi^-} +p'_4 \proj{\Phi^-},$$
	with $p'_1 = (p_1^2 + p_4^2)/p_{\rm succ}, \, p'_2 = (p_2^2 + p_3^2)/p_{\rm succ}, \, p'_3 = 2p_2 p_3/p_{\rm succ}, \, p'_4 = 2p_1 p_4/p_{\rm succ}.$
	\EndIf
\State \textbf{return}
\end{algorithmic}
\end{algorithm}

\begin{algorithm}[H]
\caption{P1-or-P2 protocol}\label{algorithm:P1-or-P2-protocol}
\begin{algorithmic}[1]
\If{The input state $\rho_{AB}$ is not a Bell diagonal state}
    \State Twirl the two available copies of the input state to the Bell diagonal state $\rho_{AB} = \sum_{k,j,=0}^{1}\alpha_{kj} |\psi_{kj}\>\<\psi_{kj}|$, where $|\psi_{mn}\> = 1/\sqrt{2}\sum_{r=0}^1(-1)^{m\cdot r} |r\>\otimes |r\ominus n\>$ using LOCC
\EndIf
\If{$\alpha_{0, 0} + \alpha_{1, 0} \leq \alpha_{0, 0} + \alpha_{0, 1}$}
    \State {\bf (P1 routine)} Apply bi-local CNOT gates between the two copies.
    \State Measure the target qubits and communicate the result.
    \If{The measured flags are $00$ or $11$ (this occurs with probability $p_{succ}=\sum_{k_1,k_2,j_1=0}^1 \alpha_{k_1j_1}\alpha_{k_2j_1}$)}
        \State The source (consisting of control qubits) copy becomes more entangled than before. We obtain a state:
        $$\rho_{control} = \sum_{k,j_1=0}^1 \tilde{\alpha}_{kj_1}|\psi_{kj_1}\>\<\psi_{kj_1}|,$$ with $\tilde{\alpha}_{kj_1} = \sum_{\{(k_1, k_2) : k_1\oplus k_2 = k\}}\alpha_{k_1j_1}\alpha_{k_2j_1}/p_{succ}$.
    \EndIf
\Else
    \State {\bf (P2 routine)} Apply local Quantum Fourier Transform on each qubit.
    \State Apply bi-local CNOT gates between the two copies.
    \State Apply local Quantum Fourier Transform on control qubits.
    \State Measure the target qubits and communicate the result.
    \If{The measured flags are $00$ or $11$ (this occurs with probability $p_{succ}=\sum_{k_1,j_1,j_2=0}^1 \alpha_{k_1j_1}\alpha_{k_1j_2}$)}
        \State The source (consisting of control qubits) copy becomes more entangled than before. We obtain a state:
        $$\rho_{control} = \sum_{k_1,j=0}^1 \tilde{\alpha}_{k_1j}|\psi_{k_1j}\>\<\psi_{k_1j}|,$$
        with $\tilde{\alpha}_{k_1j} = \sum_{\{(j_1, j_2) : j_1\oplus j_2 = j\}}\alpha_{k_1j_1}\alpha_{k_1j_2}/p_{succ}$.
    \EndIf
\EndIf

\State \textbf{return}
\end{algorithmic}
\end{algorithm}

Lastly, we want to overcome the problem of computing $E(\Lambda,\rho^{(int)}_{A'B'})$ (problem 3). However, it requires much deeper analysis then the former problems. Fortunately, $E(\Lambda,\rho^{(int)}_{A'B'})$, as a quantifier of the fundamental energy consumption of a quantum task $\Lambda$ (see Definition \ref{def:EconsumFundamental}) obeys Axiom 1 (sub-additivity on compositions) and Axiom 2 (sub-additivity on tensor products). 
We can split the task $\Lambda$ both in terms of subsequent (compositions) and parallel (tensor products) applications of quantum sub-operations and, therefore, we can upper bound $E(\Lambda,\rho^{(int)}_{A'B'})$ by the sum of the energies of each sub-operation separately.
Considered herein entanglement distillation protocols can be split into the following building blocks operations: 
depolarization, twirling, local unitaries, bi-local CNOT gates, measurements, classical communication, state post-selection (see Algorithms \ref{algorithm:BBPSSW-protocol}, \ref{algorithm:DEJMPS-protocol}, \ref{algorithm:P1-or-P2-protocol}). Thus, we will upper bound $E(\Lambda,\rho^{(int)}_{A'B'})$ by the sum of energies required to perform each building block separately times the number of its applications during the distillation protocol.

\subsection{Estimation of the fundamental energy cost of entanglement  distillation}
\label{subsec:energy-of-entanglement-distillation}

In this section, we show how to estimate the energy consumption of performing building blocks of a task $\Lambda$. Let $\mathcal{L}$ denotes one of those building blocks. Observe that, by the the Definition \ref{def:EconsumFundamental}, $E(\mathcal{L}, \rho) = \inf_{\mathcal{P}} E(\mathcal{L}_\mathcal{P})$, where the infimum is taken over all possible protocols $\mathcal{P}$ (having a Hamiltonian model) realizing the operation $\mathcal{L}$. We can always upper bound this infimum by the particular choice of a protocol $\mathcal{P}$. Unfortunately, directly applying Definition \ref{def:EconsumP} for $E(\mathcal{L}_\mathcal{P})$ seems to be a formidable task.
Hence, we make a sequence of upper bounds on $E(\mathcal{L}_\mathcal{P})$ for particular choices of $\mathcal{P}$, preceded by few auxiliary lemmas. In what follows, to remain coherent with the notation from cited papers, we calculate the Shannon entropy $H(\cdot)$ and von Neumann entropy $S(\cdot)$ in natural units of information (nats).\\

\begin{lemma}\label{lem:imporved-landauer}
    Let $\Lambda$ denote a quantum task of information erasure on system $S$, i.e. realizing the following transformation $\rho_S \mapsto \rho_S' = |0\> \< 0|_S$. For each $\epsilon > 0$, there exist a protocol $\mathcal{P}$ realizing $\Lambda_{\mathcal{P}}$ for which $E(\Lambda_\mathcal{P}) \leq (\Delta S - \ln (1-\epsilon))/\beta$, where $\Delta S = S(\rho_S) - S(\rho_S')$, $\beta = 1/k_B T$ and $E(\Lambda_\mathcal{P})$ is defined as in Definition \ref{def:EconsumP}.
\end{lemma}

\begin{proof}
    Let $\epsilon > 0$. At first, observe that the transformation $\Lambda$ realizing $\rho_S \mapsto |0\> \< 0|_S$ does not increase the rank of state. Hence, we can make use of the protocol (parametrized by $\epsilon$) proposed in Supplemental Material D of \cite{Reeb2014}. It has the following structure
    \begin{equation}
        \rho_S \otimes \rho_R \xrightarrow[]{U_{SR}^{\epsilon}} \rho_{SR}',
    \end{equation}
    where $\rho_R$ is infinite-dimensional reservoir, initially in a thermal state with inverse temperature $\beta$. Herein, we set $\rho_S' = |0\> \<0|_S$, which implies that $\rho_{SR}' = |0\> \<0|_S \otimes \rho_R'$. The heat dissipated in this protocol satisfies
    
    \begin{equation}\label{eq:upper bound-on-relative-enrtopy-reeb2014}
        \Delta Q = \Delta S + D(\rho_R' || \rho_R) \leq \Delta S - \ln (1-\epsilon),
    \end{equation}
    
     where $\Delta S = S(\rho_S) - S(\rho_S')$ and $D( \cdot ||\cdot)$ is a quantum relative entropy.
    Moreover, in this erasure protocol, heat flows only in one direction. We can then ignore the Heaviside function used in the definition of $J_W^{\mathrm{drive}\rightarrow}(t)$ and $J_Q^{\mathrm{drive}\rightarrow}(t)$ (see Definition \ref{def:EconsumP}) and simply upper bound sum of those quantities by $(\Delta S - \ln (1-\epsilon))/\beta$. Hence, with appropriate time parameter scaling we obtain $E(\Lambda_{\mathcal{P}}) \leq (\Delta S - \ln(1-\epsilon))/\beta$, which ends the proof.
\end{proof}

\begin{lemma}\label{lem:reeb-measurement}
    Let $\Lambda$ denotes a quantum task of performing a quantum measurement on system $S$.
    For each $\epsilon > 0$ there exist a protocol $\mathcal{P}$ realizing quantum task $\Lambda$ for which $E(\Lambda_\mathcal{P}) \leq \Delta E_S + \left( -\ln (1-\epsilon) + H(\{p_k\}) \right)/\beta$, where $H(\{p_k\})$ is the Shannon entropy of the measurement outcomes probability distribution, $\beta = 1/k_B T$, $\Delta E_S$ is the energy change in the measured system $S$ and $E(\Lambda_\mathcal{P})$ is defined as in Definition \ref{def:EconsumP}.
\end{lemma}

\begin{proof}
    Let us recall the protocol for quantum measurement proposed in \cite{Reeb_2018}. It is divided in two steps: measurement and memory resetting, and makes use of three systems: the measured system $S$, the memory system $M$ and the thermal bath system $B$. Each system is modeled by an appropriate Hilbert space, equipped with a Hamiltonian $H_S, H_M$ and $H_B$ respectively. Initially, states of these quantum systems are described by $\rho_S, \rho_M$ and $\rho_B' = \exp(-\beta H_B)/\tr \exp(-\beta H_B)$. A graphical representation of this protocol is depicted in Fig. \ref{fig:reeb-measurement}.
    
    \begin{figure} [!h]
        \centering
        \includegraphics[width=0.5\linewidth]{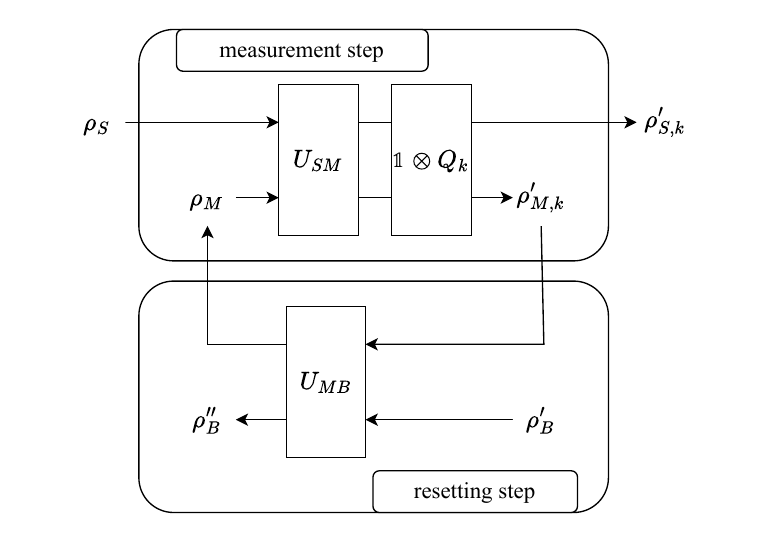}
        \caption{\justifying A protocol realizing quantum measurement proposed in \cite{Reeb_2018}. $\rho_S$, $\rho_M$ and $\rho_B$ states describe measured system, memory system and thermal reservoir system respectively. The protocol consists of two steps - measurement and resetting, each in a separate box. For detail description see \cite{Reeb_2018}.}
        \label{fig:reeb-measurement}
    \end{figure}
    
    The energy cost of realizing the protocol is given by
    \begin{equation}
        E_{cost} = \Delta E_\mathcal{R} + \Delta E_\mathcal{M},
    \end{equation}
    where $\Delta E_\mathcal{R}$ and $\Delta E_\mathcal{M}$ denotes the energy required for the memory reset step and the measurement step respectively. These quantities are defined in the following way
    \begin{align}
        & \beta\Delta E_\mathcal{R} = -\beta \Delta F_{M} + \beta \Delta F_B + \mathcal{I}_{MB},\\
        & \beta\Delta E_\mathcal{M} = \beta\Delta E_S + \beta\Delta  F_M + S(\rho_M') - S(\rho_M),
    \end{align}
    where $\Delta F_X$ is a change of free energy in a system $X$ and $\mathcal{I}_{MB} = \mathcal{I}(M:B)_{\rho_{MB}''}$ is a mutual information between the memory and heat bath system after the resetting step. Summing all the terms, we obtain 
    \begin{align}
        & \beta\Delta E_\mathcal{R} + \beta\Delta E_\mathcal{M} = -\beta \Delta F_{M} + \beta \Delta F_B + \mathcal{I}_{MB} + \beta\Delta E_S + \beta \Delta F_M +  S(\rho_M') - S(\rho_M) = \\
        & \qquad \qquad \beta \Delta F_B + \mathcal{I}_{MB} + \beta\Delta E_S + S(\rho_M') - S(\rho_M)
    \end{align}
    Since $\beta \Delta F_B = D(\rho_R' || \rho_R)$, by Lemma \ref{lem:imporved-landauer} for every $\epsilon > 0$ we can construct a memory resetting protocol for which $\beta \Delta F_B \leq -\ln(1-\epsilon)$. Moreover, since $\rho_{MB}''$ is a state of system $MB$ after the resetting step, satisfying $\tr_B \rho_{MB}'' = \rho_M$, we can write $\mathcal{I}_{MB} = \mathcal{I}(M:B)_{\rho_{MB}''}\leq S(\rho_M)$.
    Lastly, since we are constructing a particular protocol realizing quantum measurement, without loss of generality we can assume that $\rho_M$ is pure. Thus, following \cite{Reeb_2018}, i.e. that $S(\rho_M') = S(\rho_M) + H(\{p_k\})$, where $\{p_k\}$ is a measurement outcome probability distribution, we can write $S(\rho_M') = H(\{p_k\})$. Finally, we arrive at
    \begin{align}
        & E_{cost} = \Delta E_\mathcal{R} + \Delta E_\mathcal{M} \leq \Delta F_B + \Delta E_S + (\mathcal{I}_{MB} + S(\rho_M') - S(\rho_M))/\beta\\
        & \qquad \qquad \Delta E_S + \left( -\ln (1-\epsilon) + S(\rho_M) + S(\rho_M') - S(\rho_M) \right)/\beta = \Delta E_S + \left( -\ln (1-\epsilon) + H(\{p_k\}) \right)/\beta.
    \end{align}
    In this particular protocol $\mathcal{P}$, realizing quantum measurement $\Lambda$, heat flows only in one direction, so we can ignore the Heaviside function used in the definition of $J_W^{\mathrm{drive}\rightarrow}(t)$ and $J_Q^{\mathrm{drive}\rightarrow}(t)$ (see Definition \ref{def:EconsumP}) and simply upper bound sum of those quantities by $\Delta E_S + \left( -\ln(1-\epsilon) + H(\{p_k\}) \right)/\beta$. Hence, with appropriate time parameter scaling we obtain $E(\Lambda_{\mathcal{P}}) \leq \Delta E_S + \left( -\ln(1-\epsilon) + H(\{p_k\}) \right)/\beta$, which ends the proof.
\end{proof}

Now, we can list all the remaining bounds on fundamental energy consumption.

\begin{enumerate}
\item The Landauer's principle \cite{landauer1961irreversibility} states that erasing one bit of information requires a minimum energy dissipation of $k_BT\ln2$ into the environment, where $T$ is an ambient temperature. Unfortunately, it is a lower bound on inevitable energetic cost of information erasure, while our aim is to upper bound this quantity. Thus, we make use of Lemma \ref{lem:imporved-landauer}, which states that for every $\epsilon > 0$, the fundamental energy consumption is bounded by $(\Delta S - \ln (1-\epsilon))/\beta$. We bound $\Delta S$ by $1$ and set $\epsilon = 1/2$ to conclude that fundamental energy consumption of one bit information erasure is upper-bounded by $E_{\mathrm{Landauer,ub}} = 2k_BT\ln2$.

\item The fundamental energy consumption of adding an auxiliary system in the state $\ket{0}$ consists in two parts: physical carrier and information erasure. The first one is quantified by $\hbar \omega$, while the second one can be upper-bounded by already the defined $E_{\mathrm{Landauer,ub}}$. Thus we can upper bound this quantity by $E_{\textrm{aux}} = \hbar \omega + E_{\mathrm{Landauer,ub}}$.

\item Communication of one classical bit can be realized by encoding it on the polarization degree of freedom of a photon. Besides $\hbar \omega$, the energy necessary to prepare a pure state $\ket{0}$, given by $E_{\mathrm{Landauer,ub}}$, also has to be taken into account. Overall it gives us $E_{CC} = E_{\mathrm{Landauer,ub}} + \hbar\omega$ for one bit of classical communication.

\item To upper-bound the fundamental energy of a qubit measurement with classical (macroscopic) outcome, we make use of Lemma \ref{lem:reeb-measurement}, which states that for each $\epsilon > 0$, it is not greater than $\Delta E_S + \left( -\ln (1-\epsilon) + H(\{p_k\}) \right)/\beta$. Since we use the polarization degree of freedom, we have $\Delta E_S = 0$. Moreover, $H(\{p_k\}) \leq 1$. Finally, by setting $\epsilon = 1/2$ we obtain the desired upper bound $E_{\mathrm{measurement}} = 2k_B T \ln 2$.

\item Random bits can be generated by quantum measurements. To obtain one random bit, it is enough to perform a single measurements on an auxiliary qubit system. Thus, we set $E_{\mathrm{randomness}} = E_{\mathrm{measurement}} + E_{\mathrm{aux}}$.

\item Averaging over applications of bilateral local unitaries from the set $G$ can be realized as random application of a unitary from this set. Random choice of an element from $G$ can be performed with a usage of $\lceil \log_2 |G| \rceil$ bits of randomness, which by our estimations introduces $E_{\mathrm{randomness}}\cdot \lceil \log_2 |G| \rceil$ to the energetic cost. Then this random choice of unitary has to be communicated via classical communication, which introduces $\lceil\log_2 |G|\rceil E_{CC}$ to the energetic cost. So, we count the total energetic cost of this operation as $\lceil \log_2 |G| \rceil E_{\mathrm{randomness}} + \lceil \log_2 |G| \rceil E_{CC} = \lceil \log_2 |G|\rceil (E_{\mathrm{randomness}} + E_{CC})$.

\item Twirling a bipartite qubit-qubit state into the Bell diagonal state can therefore be performed as averaging over bilateral local single-qubit Clifford group, which consists of $24$ elements. We set $E_{\mathrm{twirling}} = \lceil \log_2 24 \rceil(E_{\mathrm{randomness}} + E_{CC})$.

\item Depolarization of a Bell diagonal state into the isotropic state can be performed as averaging over bilateral local unitaries from the set $\{B_x, B_y, B_z, \id \}$. Thus, in our estimation we take $E_{\mathrm{depolarization}} = \lceil \log_2 4 \rceil (E_{\mathrm{randomness}} + E_{CC})$.

\item One-qubit local unitaries in linear optics are energetically free from the fundamental point of view, since they are implemented by passive elements.

\item Unitary transformations are for free, but in optical settings with polarization, CNOTs gates are always probabilistic and hard to implement.
In more detail, the non-demolition realization of CNOT gates is necessary for consecutive usage of its outputs in later stages of distillation protocol. Therefore, we will focus here on the heralded implementation of CNOTs in optical settings \cite{Li_2021}, \cite{knill2002quantum}, in contrast to those based on destructive measurements and post-selection \cite{Pittman_2022,  OBrien_2003, Langford_2005,Okamoto_2005} (therefore, these can be used only as last gates in the circuit). While schemes for execution of control-flip gate (equivalent to CNOT gate) by exploiting photonic orbital angular momentum degrees of freedom are already available \cite{liu_2024}, herein we restrict ourselves, for the sake of upper bounding the infimum over physical protocols, to the polarization degrees of freedom.

In our estimations, we choose an implementation of controlled not gate proposed in \cite{Li_2021}. It can be used to implement a CNOT gate that operates with success probability $1/8$ and requires two auxiliary systems and two qubit measurements. 
Thus we put $E_{CNOT} = 2E_{\mathrm{aux}} + 2E_{\mathrm{meaurement}}$.
\item From the fundamental perspective, post-selection operation does not carry any energetic cost, apart from the one associated with processing of classical information. Thus, we do not count it, assuming it is negligible. 

\end{enumerate}

To summarize, in our estimation of the energy consumption of entanglement distillation $E(\Lambda, \rho_{A'B'}^{(int)})$, only energy required to implement CNOT gates, depolarizing and measurements contribute positively to $E(\Lambda, \rho_{A'B'}^{(int)})$. 
Thus we want to estimate, how much of them suffices to distill a state $\approx_{\epsilon}\psi_+$ with a protocol $\Lambda$ acting on the input state $\Phi(\psi_+)^{\otimes n}$. To achieve this, we introduce a ''forward-backward'' numerical method.

%%%%%%%%%%%%%%%
\subsubsection{Numerical approach for estimation of the upper bound on the energy consumption rate}
Notice that the considered entanglement distillation protocols are iterative and probabilistic, which means that they consist of many steps, which can succeed with some probability. After each successful step, the fidelity with the maximally entangled state is increased and this procedure is repeated until the desired fidelity is achieved. Note that the definition of $C^{\epsilon}(\Phi | \mathrm{Ent})$ also requires from the result to be $\epsilon-$close to the maximally entangled state in the trace norm distance. We can assure that by imposing that the result has a fidelity at least $1-\epsilon$ with the maximally entangled state. The aim of the ''forward-backward'' method is to numerically estimate the sufficient number of quantum operations and input states having some initial fidelity $F_{in} = F(\Phi(\psi_+), \psi_+)$ that allows to distill, with an iterative and probabilistic protocol $\Lambda$, at least one quantum state with fidelity $F_{des} = 1-\epsilon$.

The ''forward-backward'' method as its name point out, consists of two phases - forward and backward.
In the forward phase we analyze how each subsequent successful application of a chosen entanglement distillation protocol increases the fidelity of a state and what is the success probability of each subsequent step. After the forward analysis, we have the following array of pairs
\begin{equation}
\label{eq:fidelity-probability-path}
    (F_1 := F_{in}, p_1), (F_2, p_2),\ldots ,(F_{K_{steps}-1}, p_{K_{steps}-1}),(F_{K_{steps}} := F_{target}, p_{K_{steps}}),
\end{equation}
where $F_k$ is a fidelity of state before $k$-th application of an entanglement distillation protocol, $p_k$ is a probability that $k$-th application ends up with success and $F_{target}$ is the smallest fidelity that can be reached from $F_{in}$ such that $F_{target} > F_{des}$. 

These protocols are probabilistic, thus they can fail in some iteration. If such a failure occurs, we set the result state to $|00\> \< 00|$. Thus, the result state in our approach is
\begin{equation}
    p_{succ} \cdot \rho_{F_{target}} + (1-p_{succ}) \cdot |00\> \< 00|,
\end{equation}
where $p_{succ}$ is a probability that the whole procedure succeed. As can be easily checked, the above state has fidelity $p_{succ} \cdot F_{taget} + 1/2(1-p_{succ})$. The aim of backward phase in our analysis is to find the smallest number of states with initial fidelity $F_{in}$ such that $p_{succ}$ is high enough to guarantee $p_{succ} \cdot F_{taget} + 1/2(1-p_{succ}) \geq F_{des}$. Suppose now that $F_{target}$ can be reached from $F_{in}$ in $K_{steps}$ steps. Let $\eta = 1 - \sqrt[\uproot{2} K_{steps}]{\frac{F_{des} - 1/2}{F_{target} - 1/2}}$.
If we assure that failure probability of each iteration is lower than $\eta$ then 
\begin{align}
    & p_{succ} \geq \prod_{k=1}^{K_{stes}} (1 - \eta) = (1 - \eta)^{K_{steps}} =  \frac{F_{des} - 1/2}{F_{target} - 1/2} \\
    & \Rightarrow p_{succ}(F_{target} - 1/2) \geq F_{des} - 1/2 \\
    & \Rightarrow p_{succ} \cdot F_{taget} + 1/2(1-p_{succ}) \geq F_{des}.
    \label{eq:goodenough}
\end{align}
which is the required condition. 
Now we aim to estimate minimal value $n_1$ of copies of initial state, sufficient to assure that failure probability of each iteration is lower than $\eta$. In our analysis, we assign four parameters $(n_k, F_k, p_k, S_k)$ to each iteration, which have the following meaning: 
\begin{itemize}
    \item $n_k$ - number of states at the begining of that iteration, 
    \item $F_k$ - fidelity of states used in that iteration, which is calculated during forward phase of our analysis,
    \item $p_k$ - success probability of each
    pair in $k$-th steps.
    It is calculated during forward phase of our analysis,
    \item $S_k$ - random variable that counts number of successful applications of te investigated distillation protocol in $k$-th iteration. It has a binomial distribution with $n_k/2$ trials and success probability $p_k$.
\end{itemize}
Using the Chernoff bound (see Theorem 4.5 in \cite{upfal2005probability}), we upper bound the failure probability of $k$-th iteration, $k\in\{1,\ldots,K_{steps}\}$, as follows
\begin{equation}
    P(S_{k} < n_{k+1}) \leq P(S_{k} \leq n_{k+1}) = P(S_{k} \leq (1-\delta_{k})\mu_{k}) \leq e^{-\delta_{k}^2\mu_{k}/2},
\end{equation}
where $\mu_{k} = n_{k}/2\cdot p_{k}$ and $\delta_{k} = 1 - n_{k+1}/\mu_{k}$. In the above, the condition $S_k < n_{k+1}$ expresses the failure, i.e.
that in the $k$-th step there
was smaller number of successes than the value of the variable $n_{k+1}$ expressing the demanded number of states  in the $k+1$-th step.
Notice that in order to use this bound, one has to ensure that $\delta_k \in (0, 1)$. We also put $n_{K_{steps} + 1} = 1$, because for the purpose of the upper bound, we analyze protocols obtaining one two-qubit output state of desired fidelity. As it was stated earlier, we want to impose that the failure probability of each iteration is lower than $\eta$, thus we need to solve
\begin{equation}
    e^{-\delta_{k}^2\mu_{k}/2} \leq \eta =: e^{-t_{\eta}}.
\end{equation}
Which is equivalent to the following quadratic inequality 
\begin{equation}\label{eq:energy-estimate-quadratic-inequality}
    \left(1 - \frac{n_{k+1}}{\frac{n_k}{2} \cdot p_k}\right)^2\frac{n_k}{2} \cdot p_k \geq 2t_\eta \Rightarrow n_k^2 \left( \frac{p_k}{2}\right)^2 - n_kp_k(n_{k+1} + t_{\eta}) + n_{k+1}^2 \geq 0,
\end{equation}
where we want to solve it for $n_k$. More precisely, we want to find the lowest integer $n_k$, such that above inequality is satisfied and such that $\delta_k \in (0, 1)$. This recursive relation can be computed effectively, starting from $k=K_{steps}$ (proved in Eq. \eqref{eq:goodenough}) to be sufficient for achieving fidelity not smaller than $F_{des}$)  and iterating backwards. Thus finally we will obtain $n_1$, which was the desired, but unknown parameter. Summing up $n_k/2$ from each iteration, we calculate the total number of performed distillation protocol. Energy required for each run of the protocol is equal to $E(\Lambda)$. Thus the total energy used is 
\begin{equation}
    \underbrace{2n_1 \cdot \hbar\omega}_{\text{energy of input physical carries}} + \underbrace{E(\Lambda)}_{\text{energy of a single protocol run}}\times \underbrace{\sum_{k=1}^{K_{steps}}n_k/2}_{\text{number of single protocol runs}}  - \underbrace{2\hbar\omega}_{\text{output energy}},
\end{equation}
where $\hbar$ is a reduced Planck constant and $\omega$ is a wave frequency of used photons. In table \ref{tab:energy-of-distillation} one can find a summary with formulas for different entanglement distillation protocols.
\begin{table}[h!]
    \centering
    \begin{tabular}{| c | c | c | c |}
    \hline
    \multirow{2}{*}{Protocol} & \multicolumn{2}{c}{Energy of a single ebit generation} &\\ 
    % \end{tabular}
    % \begin{tabular}{| c | c | c | c |}
    \cline{2-4}
     & $E(\Lambda, \rho^{(int)})$ & $E(\rho^{(in)})$ & $E(\rho^{(out)})$\\ 
    \hline
    % BPPSSW & $(2E_{CNOT} + 2E_{measurement} + 2\underbrace{k T \ln 12 + \text{physical carrier}}_{\text{random twirling}})\cdot \sum_{k=1}^{K_{steps}}n_k/2$ & $2n_1 \cdot \hbar\omega$ & $2\hbar\omega$\\ 
    BPPSSW & $(2E_{\mathrm{depolarization}} + 2E_{CNOT} + 2E_{\mathrm{measurement}} + 2E_{CC})\cdot \sum_{k=1}^{K_{steps}}n_k^{BPPSSW}/2$ & $2n_1^{BPPSSW} \cdot \hbar\omega$ & $2\hbar\omega$\\  
    \hline
    DEJMPS & $(2E_{CNOT} + 2E_{\mathrm{measurement}} + 2E_{CC})\cdot \sum_{k=1}^{K_{steps}}n_k^{DEJMPS}/2$ & $2n_1^{DEJMPS} \cdot \hbar\omega$ & $2\hbar\omega$\\
    \hline
    P1-or-P2 & $(2E_{CNOT} + 2E_{\mathrm{measurement}} + 2E_{CC})\cdot \sum_{k=1}^{K_{steps}}n_k^{P1orP2}/2$ & $2n_1^{P1orP2} \cdot \hbar\omega$ & $2\hbar\omega$\\
    \hline
\end{tabular}
    \caption{\justifying Upper-bounds on fundamental energy consumption required to perform various entanglement distillation protocols. $E_{CNOT}$, $E_{\mathrm{measurement}}$, $E_{\mathrm{depolarization}}$ and $E_{CC}$ refer to energy required to perform a CNOT gate, a projective measurement, depolarization and communicate one classical bit respectively.}
    \label{tab:energy-of-distillation}
\end{table}

Lastly, let us remind that in this section we aim to numerically upper bound energy consumption rate (see Definition \ref{def:EnergyCost}) of entanglement distillation for depolarizing channel. But considered distillation protocols are probabilistic in their nature, hence it may happen that the recursive procedure succeeded more that once, producing many copies of states closed to maximally entangled states. Thus, it may be confusing what one should take as the output dimension. To overcome this issue, we assume that all systems but one are trace out after the distillation protocol. This allows us to write $d_{out} = 2$, which implies that the denominator in the upper bounded formula (see Definition \ref{def:EnergyCost}) is equal to $1$.

\begin{figure}[t!]
    \centering
    \begin{subfigure}[t]{0.5\textwidth}
        \centering
        \includegraphics[width=1.0\linewidth]{entanglement_distill_2.png}
        \caption{Results for three different entanglement distillation protocols: BPPSSW, DEJMPS and P1-or-P2 in no-noise regime.}
    \label{fig:j-per-ebit-3-protocols}
    \end{subfigure}%
    ~ 
    \begin{subfigure}[t]{0.5\textwidth}
        \centering
        \includegraphics[width=1.0\linewidth]{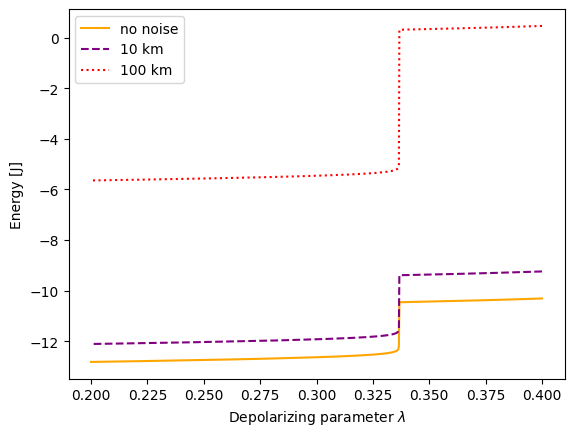}
        \caption{Results for DEJMPS protocol. Here, the effect of probabilistic photon loss in memory fiber for different distances between parties (0km, 10km and 100km) has been taken into account.}
    \label{fig:j-per-ebit-with-noise}
    \end{subfigure}
    \caption{\justifying Upper bounds on the fundamental energy consumption required to obtain one quantum state with fidelity $0.9$ (with respect to maximally entangled state) starting from $(1 - \lambda)\Phi_+ + \lambda \id/4$. In our analysis, we chose photons of the wavelength 1550 nm, $p_{cnot} = 1/8$ and $T = 293K$. Energy axis is in the logarithmic scale.}
\end{figure}

In any real experimental scenario noise, is an inevitable part of quantum information processing. Thus, as a final remark to this section, we consider the effect of noise on the quantum entanglement distillation. For the sake of simplicity, we consider the simplistic, but yet very useful, model of noise, namely probabilistic photon loss in optical fiber. The probability of photon loss in an optical fiber of length $L$ and attenuation parameter $\eta$ is commonly modeled by the following formula
% \begin{equation}
%     P(\text{photon loss}) = 1 - \exp(-\lambda\cdot L),
% \end{equation}
\begin{equation}
    P(\text{photon loss}) = 1 - 10^{\frac{-\eta \cdot L}{10}},
\end{equation}

where a typical value of fiber attenuation is $0.18$ $\mathrm{dB}/\mathrm{km}$ at telecom wavelength ($1550$ nm). We aim to incorporate this noise model into our numerical estimations. Each of considered here entanglement distillation protocols consist of many rounds, between which, classical communication is exchanged. In a typical entanglement distribution scenario, parties $A$ and $B$ are spatially separated by some distance $d$, thus during the exchange of classical communication, all necessary quantum states has to be stored in memory, which can be implemented as a fiber loop of length $d$ stored locally. The memory will be affected by probabilistic photon loss noise. Effectively, this noise contribute to the decrease of success probability $p_k$ (see Eq. \eqref{eq:fidelity-probability-path}) in each entanglement distillation step. More precisely, each success probability $p_k$ will be multiplied by a factor $P(\text{photon loss}) = 1 - 10^{\frac{-\eta \cdot L}{10}}$. Numerical results obtained for DEJMPS distillation protocol including this noise model for quantum memory for various distances between parties are depicted on Fig. \ref{fig:j-per-ebit-with-noise}.

%%%%%%%%%%%%%%%%%%%%
\section{Discussion and outlook}\label{ch:Dis}
In this manuscript we have quantified the fundamental energy needed for entanglement distribution over long distances via quantum channels. We thus contributed to the development of resource aware quantum information processing. The main contribution of this work is the first formal definition of the fundamental energy cost of entanglement distillation. We focus mostly on entanglement distillation, as it is a building block of quantum networking. Other examples of quantum resource are  magic states for computation, non-locality or contextuality. Our work opens the path to many energy related studies, in particular the study of the distillation of these other resources.

We show that the fundamental energy cost associated to distilling entanglement using a fixed quantum channel is lower bounded by the so-called \textit{distillable} entanglement of the channel.  We then reformulate this lower bound using the entanglement cost. While this bound is a slightly weaker bound, it has the appealing property to connect the energy cost with the concept of irreversibility from entanglement theory. Another reformulation connects the fundamental cost of entanglement distillation with the two-ways quantum capacity of the channel used. Since none of these bounds are easily computable, we also derive a lower bound based in the log negativity.

These lower bounds are of the order of $10^{-34}$ J, which is far from any realistic implementation cost. After showing a generic upper bound based on the singlet, we look at 3 historically relevant distillation protocols: BBPSSW~\cite{bennett1996purification},  DEJMPS~\cite{deutsch1996quantum} and the recent P1-P2~\cite{miguel2018efficient} and find upper bounds of the order of $10^{-12}$ J. This emphasizes the large gap between fundamental cost and technological cost. Interestingly, we note that the protocol with the higher distillation rate is not the most energetically efficient. The DEJMPS protocol seems to be the most energy efficient although it was the second to come out. We explained in detail how to quantify the energy of concrete protocols by dividing the protocol in building blocks, such as operations and measurement, and using fundamental energy concepts such as the Landauer bound. The results are schematized in Fig.~\ref{fig:all_inone_resuls_section}.

The presented approach paves the way for answering a number of important open questions. They show that taking the energetic cost as a benchmark gives a different perspective than the usual vision based solely on the entanglement distillation rate. Sustainable quantum technologies might favor a protocol taking a longer time but more efficient energetically. Although this is very fundamental and cannot be achieved by any hardware, these results can prove to be very important when scaling up systems. Indeed entanglement, or ebits are used as building blocks in quantum network. However, many ebits might be needed for long distance communication, as envisioned in current quantum repeater schemes \cite{repeaterNV,avis2022requirements,coopmans2021netsquid}. Therefore a quantum internet will have a much more stringent scaling on resources than the classical internet. Moreover, it appears that the lower bound Eq. \eqref{eq:lower_bound_ceil} holds also in the multipartite case of GHZ distribution. In general similar results are expected in case of other resource theories, when the target state to be distributed is pure, and the physical platform is photonic. However, in the case of other resource theories and physical platforms, depending on the type and on the properties of a given resource, the formula for the energy consumption will differ from the one from this work. For example, when working with a non-photonic realization of a quantum computer, we do not expect loss of qubits unless a major damage happens. Hence the energy of input and output will cancel out. 

The proposed approach lays also the foundations for an estimation of energy consumption of (parts of) algorithms using quantum entanglement as an elementary step \textit{e.g.} for teleportation, or certain delegated computation schemes. In such a case, it is natural to use, as the upper bound, the proposed energy of distillation process using the particular architecture of Def.\ref{def:EconsumArchitecture}, based on a particular gate set. Other quantum resources can also be studied using the same method. Magic state distillation and contextuality are example of necessary resources for any quantum advantage. As building blocks of quantum information technologies, studying their fundamental energy consumption is of high interest. By making the appropriate changes in the numerator in our definition of the fundamental energy consumption (Definition~\ref{def:ent-main}), one could expand our framework to these resources. The study of the energy cost of a quantum operation presented in this work is general enough to be applicable in other contexts, as well as to study the technological cost of quantum protocols.

Tighter upper bounds on the energy consumption of entanglement distillation protocols are welcome. In particular, it will be crucial to compare the recently proposed entanglement distillation protocols in Refs. \cite{PoppHiesmayr2024a,PoppHiesmayr2024b} including those using non-Clifford operations \cite{DurPurif2024}.

In this manuscript we took a resource aware approach, quantifying all consumed energy that is invested in form of the input state and leaked due to channel and distillation. It would be interesting, however, from the point of view of technology, to separately study the work expense using a quantum thermodynamical approach, in a similar way to~\cite{thompson2025} in the case of agents.

Finally, let us emphasize, that the energy of measurement, and in general of quantum operations, is a topic of recent debate~\cite{Faist2015,Cimini2020,FellousAsiani2021,Deffner2021,Chiribella2021,Stevens2022,Stevens2025}. The definition of energy expense of entanglement (an other resource) distillation protocols should be updated accordingly in future works.

\section*{Acknowledgments}
The authors thank Siddhartha Das, Marco Fellous-Asiani, Karol Ławniczak, Monika Rosicka, Mark M. Wilde and Marek \.Zukowski for fruitful comments and discussions.
KH, MW, LS and MC acknowledge National Science Centre, Poland, grant Opus 25, 2023/49/B/ST2/02468. This work benefited from a government grant managed by the Agence Nationale de la Recherche under the Plan France 2030 with the reference ANR-22-PETQ-0006.

RY acknowledges funding from the European Union (ERC, ASC-Q, 101040624) and the Horizon Europe program (QUCATS).  It is also supported by the LUBOS project, the Government of Spain (Severo Ochoa CEX2019-000910-S and FUNQIP), Fundació Cellex, Fundació Mir-Puig, Generalitat de Catalunya (CERCA program).

% \putbib[References.bib]
%LSBIB\putbib
%LSBIB\end{bibunit}
%@@@
%LSBIB\newpage
%LSBIB\begin{bibunit}
\section{Supplementary information}

\subsection{Axiomatic approach to energy cost in quantum tasks via Hamiltonian models \label{sec:axiomatic-apporach}}

%{- bridging the gap between mathematics, physics and intuition} \rj{This title is too long}
From the perspective of quantum physicist, energy is a well-defined and fundamentally conserved quantity. However, from the perspective of an engineer, any type of device that coherently performs some task consumes power, usually provided by a voltage source available. In the following paragraphs we want to bridge this gap by formally defining the fundamentally minimal energy required for a classical device to perform quantum operation. \textcolor{black}{The quantity introduced here, namely the fundamental energy cost \textcolor{blue}(and related quantities), is to be interpreted as the optimal energy investment. That is, we focus on minimizing the energy that must be supplied to carry out an operation on a quantum system, disregarding any energy outgoing as a byproduct of the process. }

Consider a quantum task $\Lambda$ which transforms the state of a quantum system from its initial state $\rho$ into output state $\sigma$ with a completely positive and trace-preserving (CPTP) map $\tilde{\Lambda}$, i.e., $\sigma = \tilde{\Lambda}(\rho)$. Herein, the bridge between the reversible, unitary
quantum world and the irreversible classical world is established by introducing a mesoscopic device (or instrument) that serves as an interface between the two realms. Examples of such interface devices are  photodetectors and the electromagnetic field— both mediate energy transfer between the quantum system and the classical environment. Such a device can be defined as a setup consisting of many instruments, possibly at different spatial locations, used sequentially or in parallel, yet we refer to them as a single object for the sake of simplicity. This device is imagined to be small enough to possess controllable quantum degrees of freedom, yet bulky enough to allow for classical driving (at least for the sake of description). We call these distinguished degrees of freedom the relevant ones, as these are the ones that couple to the quantum system and allow to perform the task $\Lambda$. The remaining degrees of freedom of interface device, if existent, must be decoupled both from the relevant ones and the quantum system of interest. Only then we can consider them to be a part of the environment. The classical drive is modeled with a time-dependent Louvillian acting\footnote{Strictly speaking, the Louvillian acts both on mesoscopic interface and the quantum device. However, only the interface is subjected to the classical driving, while the interaction between the interface and the quantum system is fully quantum.} on the relevant degrees of freedom of the mesoscopic device, thereby allowing for an unambiguous definitions of power and heat injected into the system, with energy and entropy treated as fundamental thermodynamic resources\footnote{\textcolor{black}{The energy transferred between the mesoscopic system, during execution of the task, and its environment does not depend on the ambient temperature, but only on the energy difference between the joint system $SA$ initial and final states. We note that the energy cost quantifiers developed herein do not depend explicitly on the ambient temperature. Nevertheless, they still depend on the protocol pathway, which may involve contact with heat baths at arbitrary temperatures, under the constraint that the mesoscopic device returns to its designated initial state. 
% Whether the fundamental energy cost depends on the ambient temperature therefore remains an open question~(cf. Ref.~\cite{thompson2025}).
}}.  We formalize the above consideration in the following definition.

% \footnote{Indeed, stabilizing the temperature is crucial for monitoring heat flows. However, the final energy consumption, calculated using a fundamental Hamiltonian model, should be independent of the ambient temperature. This is because the protocols are designed to reset the mesoscopic device to its initial state at the end of the protocol. The energy transferred between the mesoscopic system and its environment relies not on the ambient temperature, but solely on the difference in energy between the initial and final states of the system. Consequently, the fundamental energetic cost should also be independent of temperature, although it may still depend on the optimal pathway, consistent with the assumed constraints, by which the task is accomplished.} 

\begin{definition}\textcolor{black}{[Of Hamiltonian model protocol]} Let the triple $\Lambda \equiv (\tilde{\Lambda},\rho_S,\sigma_S)$ be a quantum task transforming the input state $\rho_S$ of system $S$ into the output state $\sigma_S$ via the completely positive and trace-preserving (CPTP) map $\tilde{\Lambda}$ (i.e., $\sigma_S=\tilde{\Lambda}(\rho_S)$). Given a fixed physical realization (platform and setup) performing the task $\Lambda$ on system $S$, we call a tuple \textcolor{black}{$\mathcal{P} \equiv (\mathcal{L}_{SA}(t),t^*,A,\tau_A,\rho_S,\sigma_S)$} a Hamiltonian model protocol of performing the task $\Lambda$, where $A$ is an auxiliary system, $\tau_A$ is the initial and the final state of the auxiliary system, and $\mathcal{L}_{SA}(t)$ is a classically-driven (on system $A$) time-dependent \textcolor{black}{Liouvillian} of the total system $S+A$, that performs the task $\Lambda$ in the time interval~$[0,t^*]$.
% \textcolor{blue}{The spatial partitions of system $SA$ are denoted $SA\equiv S_1A_1 : S_2A_2 : \ldots : S_2A_2 $, where colon (":") indicates the spatial separation.}
\label{def:HamiltonianProtocol}
\end{definition}

\begin{definition}
    We define a composition of protocols $\mathcal{P}_1 = (\mathcal{L}^{(1)}_{SA}(t),t_1^*,A,\tau_{A},\rho_{S}^{(1)},\sigma_{S}^{(1)})$ and $\mathcal{P}_2 = (\mathcal{L}^{(2)}_{SA}(t),t_2^*,A,\tau_{A},\sigma_{S}^{(1)},\sigma_{S}^{(2)})$ as
    \begin{equation}
        \mathcal{P}_2\circ\mathcal{P}_1 \equiv (\mathcal{L}_{SA}(t),t_1^* + t_2^*,A,\tau_A,\rho_{S}^{(1)},\sigma_{S}^{(2)}),
    \end{equation}
    where
    \begin{equation}
        \mathcal{L}_{SA}(t) = \left\{ \begin{array}{ll}
            \mathcal{L}^{(1)}_{SA}(t) &  ,t\in [0, t_1^*]\\
            \mathcal{L}^{(2)}_{SA}(t-t_1^*) & ,t\in (t_1^*, t_1^* + t_2^*]
        \end{array} \right.
    \end{equation}
\end{definition}

\begin{definition} We define the tensor product of protocols, both starting at $t=0$, in the following way.
\begin{align}
    &\mathcal{P}_1 \otimes \mathcal{P}_2 \equiv  (\mathcal{L}^{(1)}_{S_1A_1}(t),t_1^*,A_1,\tau^{(1)}_{A_1},\rho_{S_1}^{(1)},\sigma_{S_1}^{(1)}) \otimes (\mathcal{L}^{(2)}_{S_2A_2}(t),t_2^*,A_2,\tau^{(2)}_{A_2},\rho_{S_2}^{(2)},\sigma_{S_2}^{(2)}) \\ &:= \left( \mathcal{L}^{(1)}_{S_1A_1}(t) \otimes \mathds{1}_{S_2A_2} + \mathds{1}_{S_1A_1} \otimes \mathcal{L}^{(2)}_{S_2A_2}(t), \max\{t_1^*,t_2^*\}, A_1A_2, \tau^{(1)}_{A_1} \otimes \tau^{(2)}_{A_2},\rho_{S_1}^{(1)} \otimes \rho_{S_2}^{(2)},\sigma_{S_1}^{(1)} \otimes \sigma_{S_1}^{(2)} \right).
\end{align}
Furthermore, protocols starting at different time points can be always composed by immersing them in protocols starting at the same time, however dynamics of which is trivial for the extended time interval.
\label{def:ten1}
\end{definition}

% \begin{definition} Let $\mathcal{P} \equiv (\mathcal{L}_{SA}(t),t^*,A,\tau_A,\rho_S,\sigma_S)$ be a protocol realized by a $N$-partite device $SA \equiv S_1A_1:S_2A_2:\cdots : S_NA_N$, \sout{where the colon punctuation mark, i.e., ":" denotes spatial separation,} and let $\mathcal{L}_{SA}(t) = \sum_{i=1}^{N} \mathcal{L}^{(i)}_{S_iA_i}(t)$ be a multipartite local Louvillian\footnote{\textcolor{blue}{For example, given $N=2$ we have $\mathcal{L}_{SA}(t) = \mathcal{L}^{(1)}_{S_1A_1}(t) \otimes \mathds{1} _{S_2A_2} + \mathds{1}_{S_1A_1} \otimes \mathcal{L}^{(2)}_{S_2A_2}(t)$.}}. We call the following object
% \begin{align}
%     \mathcal{P}_i \equiv (\mathcal{L}^{(i)}_{S_iA_i}(t),t^*,A_i,\Tr_{\neq i}[\tau_A],\Tr_{\neq i}[\rho_S],\Tr_{\neq i}[\sigma_S]),
% \end{align}
% the reduced protocol of the $i$-th party.
% \end{definition}
In this place, we comment on the above definition. According to it, the reduced protocol of the $i$-th party is defined only if its dynamics is independent from the other parties. This is not true when parties exchange classical communication. In that case the reduced dynamics of one subsystem depends on the state of the other(s) subsystem.

\color{black}

\begin{definition} We define the tensor product of quantum tasks in the following way
\begin{align}
    \Lambda_1 \otimes \Lambda_2 \equiv (\tilde{\Lambda}_1,\rho_1,\sigma_1) \otimes (\tilde{\Lambda}_2,\rho_2,\sigma_2) := 
    (\tilde{\Lambda}_1 \otimes \tilde{\Lambda}_2,\rho_1 \otimes \rho_2,\sigma_1 \otimes \sigma_2).
\end{align}
\label{def:ten2}
\end{definition}
In the above definitions, the auxiliary system(s) corresponds to the mesoscopic device(s) considered in the text preceding the definitions. \textcolor{black}{Furthermore, the dimension of system $S$ at the input and output of the task $\Lambda$ may differ. However, we assume that $\sigma$ is the unique quantum state at the output of $\Lambda$. The dimension reduction is achieved by discarding (or destroying) any unnecessary components (such as particles) of system $S$ (mutandis mutatis increase in dimension).}

\begin{remark}
    Notice that, upon a proper re-scaling of the \textcolor{black}{ Liouvillian (energy scale), and execution time $t^*$ of a protocol} by a constant factor, we can obtain $t^*=1$ as a standardized execution time of every protocol.  
\end{remark}

\begin{remark} The above definition of a Hamiltonian model protocol for a physical realization of operation $\Lambda$ goes along the lines with the idea of Stinespring dilation theorem
\color{black}
\begin{align}
    &\mathcal{L}_{SA}(t):~~i\partial_t U_{SAE}(t,0)=H_{SAE}(t)U_{SAE}(t,0),\\
    &\partial_t \rho_{SA}(t)=\mathcal{L}_{SA}(t)[\rho_{SA}] =\partial_t\tr_{E} \left[U_{SAE}(t,0)\rho_{SA}\otimes \chi_{E} U_{SAE}^\dagger(t,0)\right], ~~\sigma_{SA} = \rho_{SA}(t^*),
\end{align}
% where in our model $A$ is the mesoscopic auxiliary system (instrument) that interacts with system $S$, i.e., system of interest. Here, system $E$ in state $\chi_E$ is the environment that drives the mesoscopic interface $A$ and therefore exchanges energy with it. 
where, in our model, $A$ denotes the mesoscopic auxiliary system (or 'instrument') that interacts with the primary system of interest, $S$. Meanwhile, $E$ represents the environment, initially in the state $\chi_E$, which drives the mesoscopic interface $A$ and thereby exchanges energy with it.
\end{remark}
\color{black}

\begin{notation}
    We introduce shortened notation in which $\Lambda_\mathcal{P}$ has the meaning of operation $\Lambda$ performed by means of protocol $\mathcal{P}$.
\end{notation}

Having the above setup in mind, we want to propose a Hamiltonian model to calculate the energy cost of a protocol $\mathcal{P}$ of a classical drive that allows us to perform a quantum operation $\Lambda$ by manipulating the relevant degrees of freedom of the mesoscopic device. To achieve this we first need to specify what we mean by the energy cost, a quantity that is usually well understood intuitively, yet requires a mathematical description. We chose an axiomatic approach in which we postulate the axioms which any function $f$ that quantifies the energy cost of operation $\Lambda$ should satisfy

\begin{axiom1}
    Sub-additivity on compositions:
    $\forall \left\{{\Lambda_1:X\to Y},{\Lambda_2:Y\to Z}\right\}~f({\Lambda_2} \circ {\Lambda_1}) \le f(\Lambda_1)+f({\Lambda_2})$.
\end{axiom1}
\begin{axiom2}
     Locality of energy: $\forall\left\{\Lambda_A\otimes\id \equiv (\tilde{\Lambda}_A \otimes \id_B,\rho_{AB},\sigma_{AB}),{\Lambda_A}\equiv (\tilde{\Lambda}_A,\Tr_B[\rho_{AB}],\Tr_B[\sigma_{AB})]\right\}~f(\Lambda_A\otimes\id)=f(\Lambda_A)$.
\end{axiom2}
\begin{observation}[Sub-additivity on independent tasks]
\label{obs:axiom-sub-additivity-in-space}
    Let $\Lambda_{AB}\equiv (\tilde{\Lambda}_A \otimes \tilde{\Omega}_B$, $\rho_{AB},\sigma_{AB}),{\Lambda_A}\equiv (\tilde{\Lambda}_A,\Tr_B[\rho_{AB}],\Tr_B[\sigma_{AB}])$, ${\Omega}_B\equiv (\tilde{\Omega}_B,\Tr_A[\rho_{AB}],\Tr_A[\sigma_{AB}])$ be quantum tasks. Then each function $f$ satisfying axioms 1 and 2 also satisfies 
    \begin{equation}
        f(\Lambda_{AB}) \leq f(\Lambda_A)+f(\Omega_B).
    \end{equation}
\end{observation}
\begin{proof}
    Observe that for each quantum channels $\tilde\Lambda_A, \tilde\Omega_B$ we have $\tilde\Lambda_A\otimes\tilde\Omega_B = (\tilde\Lambda_A\otimes \id_B)\circ (\id_A\otimes \tilde\Omega_B)$. Let us introduce the following quantum tasks 
    $\Lambda_{AB}\equiv (\tilde{\Lambda}_A \otimes \tilde{\Omega}_B,\rho_{AB},\sigma_{AB})$,
    $\Lambda_{A}\otimes \id \equiv (\tilde\Lambda_{A}\otimes \id, \rho_{AB}, \tilde\Lambda_{A}\otimes \id(\rho_{AB}))$,
    $ \id \otimes \Omega_B\equiv (\id\otimes\tilde\Omega_B, \rho_{AB}, \id\otimes\tilde\Omega_B(\rho_{AB}))$.
    Now, by using Axiom 1 we have
    \begin{equation}
        f(\Lambda_{AB}) \leq f(\Lambda_A\otimes\id) + f(\id\otimes\Omega_B).
    \end{equation}
    By introducing ${\Lambda_A}\equiv (\tilde{\Lambda}_A,\Tr_B[\rho_{AB}],\Tr_B[\sigma_{AB}]=\Tr_B[\tilde\Lambda_{A}\otimes \id(\rho_{AB})])$,
    ${\Omega}_B\equiv (\tilde{\Omega}_B,\Tr_A[\rho_{AB}],\Tr_A[\sigma_{AB}]=\Tr_A[\id\otimes\tilde\Omega_B(\rho_{AB})])$ and using Axiom 2 we obtain
    \begin{equation}
        f(\Lambda_A\otimes\id) + f(\id\otimes\Omega_B) = f(\Lambda_A) + f(\Omega_B),
    \end{equation}
    which ends the proof.
\end{proof}
% \begin{ax2}
%     Additivity on independent tasks $\forall {\Lambda_1},{\Lambda_2},~ f({\Lambda_1} \otimes {\Lambda_2}) = f(\Lambda_1)+f({\Lambda_2})$.
% \end{ax2}
% \color{blue}
% \begin{ax3}
%      Locality of energy $\forall\Lambda_{AB}\equiv (\tilde{\Lambda}_A \otimes \tilde{\Omega}_B,\rho_{AB},\sigma_{AB}),{\Lambda_A}\equiv (\tilde{\Lambda}_A,\Tr_B[\rho_{AB}],\Tr_B[\sigma_{AB})],{\Omega}_B\equiv (\tilde{\Omega}_B,\Tr_A[\rho_{AB}],\Tr_A[\sigma_{AB})],~f(\Lambda_{AB})=f(\Lambda_A)+f(\Omega_B)$.
% \end{ax3}
\color{black}
Additionally, the energy cost that we will define here will satisfy the following properties, that allow to interpret the energy cost in the sense of inevitable expanse. 
\begin{pr1} The trivial task, i.e., $\Lambda = \mathds{1}$, does not possess energy cost: 
    $f(\mathds{1})=0$, (zero element exists). 
\end{pr1}
\begin{pr2}The energy cost is non-negative: 
    $\forall \Lambda,~f(\Lambda) \ge 0$, (non-negativity).
\end{pr2}
Furthermore, the fundamental energy cost derived herein does not take into account, at least explicitly, the energy lost from the target system into the environment. This is because we are only interested in the energy expense that is required from the agent(s) performing a quantum task. However, the approach developed here does not constrain protocols for retrieving energy outgoing from the target system $S$ to the mesoscopic device $A$ with the possibility of reusing it, and thus minimizing the.

In order to realize the operation $\Lambda$, assuming some fixed physical realization, we make the carrier system $S$ of state $\rho_S$ interact with some degrees of freedom of a mesoscopic device $A$ initially in state $\tau_A$.  In a generic protocol that performs the operation $\Lambda$, assuming that the initial state of the mesoscopic device is $\tau_A$, we have the following steps:
\begin{enumerate}
    \item  Prepare the quantum state of the relevant degrees of freedom of the mesoscopic device.
    \item Switch on the interaction between systems S and A.
    \item Let the systems evolve during some predetermined time window.
    \item Switch off the interaction between system S and A.
    \item Read the state of mesoscopic device and prepare its relevant degrees of freedom again.
    \item Repeat previous steps (1-5) a desired number of times.
    \item Reset the state of mesoscopic device by preparing it in state $\tau_A$.
\end{enumerate}
% a) We prepare the quantum state of the relevant degrees of freedom of the mesoscopic device. b) We switch on the interaction between systems S and A. c) We let the systems evolve during some predetermined time window. d) We switch off the interaction between system S and A. e) We read the state of mesoscopic device and prepare its relevant degrees of freedom again. f) We repeat the previous steps a desired number of times. g) We reset the state of mesoscopic device by preparing it in state $\tau_A$. 
The switching on(off) the interaction steps are optional, since for some physical realizations switching off the interaction is impossible (e.g. we can not switch off electromagnetic field). \textcolor{black}{In general such a process is described by a Liouvillian in the form}

% In The Hamiltonian \textcolor{red}{, the Liouvillian and the evolution} of such a process reads:
\color{black}
\begin{align}
    &\dot{\rho}_{SA}= \mathcal{L}_{SA}(t)[\rho_{SA}] = -i \left[H_{SA}(t),\rho_{SA}\right] + \mathcal{D}_{A}(t)\otimes\mathds{1}_S[\rho_{SA}],\\
    &H_{SA}(t)=H_S^{(0)} \otimes \mathds{1}_A + \mathds{1}_S \otimes  H_A(t) + \sum_i H_{SA}^{(i)}(t), 
    % &  \textcolor{blue}{\mathcal{L}_{SA}(t) \equiv \mathcal{L}_{S_1A_1}(t) \otimes \mathcal{L}_{S_2A_2}(t) \otimes \cdots \otimes \mathcal{L}_{S_NA_N}(t)},
\end{align}
\color{black}
where $H_S^{(0)}$ is the free Hamiltonian of system S, $H_A(t)$ is the classically driven Hamiltonian of the auxiliary mesoscopic system\textcolor{black}{, and $\mathcal{D}_{A}(t)$ is a dissipator acting on system $A$ responsible for the heat flows.} 
%This setup is visualized in Fig.~\ref{Currents}. 
The operators take the form
\begin{align}
    &H_A(0)=H_A(t^*)=H_A^{(0)},~~H_A(t)=H_A^{(0)}+\left(\sum_i V_i(t)\times \hat{A}_i + \mathrm{h.c.} \right),~~V_i(t):~\mathbb{R}_+ \to \mathbb{C},
    % ~~\prod_i V_i(t)=0,
    \\
    \color{black}
    & \mathcal{D}_{A}(t) = \sum_i W_i(t) D_{A,_{\beta_i}}^{i},~~\mathcal{D}_{A}(0)=\mathcal{D}_{A}(t^*)=0,~~W_i(t):~\mathbb{R}_+ \to \{0,1\},
\end{align}
% \textcolor{red}{
% Here, we can explain that index $i$ does not iterate parties. 
% We should lift up/or describe the case of many parties.}
\color{black}where $H_A^{(0)}$ is the free Hamiltonian of mesoscopic interface system $A$, time-dependent functions $V_i(t)$ describe the "temporal strength" of the driving fields $\hat{A}_i$, and $\mathrm{h.c}$ stands for Hermitian conjugate. Herein, we assume that driving fields $\hat{A}_i$ are independent in the sense that they act non-trivially on distinct degrees of freedom, which also implies that operators $\hat{A}_i$ commute, i.e. $\forall_{i\neq j}~[\hat{A}_i, \hat{A}_j] =[\hat{A}_i, \hat{A}_j^\dagger] = 0$. This assumption allows us to incorporate multipartite scenario in our regime. Indeed, since each party holds locally its own degrees of freedom, then driving fields act on them independently.
\textcolor{black}{Furthermore, functions $W_i(t)$ describe whether (and how) at time $t$ the auxiliary system $A$ is coupled to a heat bath at inverse temperature $\beta_i$. The former corresponds therefore to the work performed on the system, while the latter describes the heat flows}. Finally, $H_{SA}^{(i)}(t)$ are the interaction Hamiltonian between systems $A$ and $S$ which time-dependence is only in the form of switching on(off) the interaction, namely,
\begin{align}
    % H_{SA}^{\mathrm{int}}(t)\equiv\eta(t-\gamma(t))H_{SA}^{\mathrm{int}}
        H_{SA}^{(i)}(t)\equiv \gamma_i(t)H_{SA}^{(i)}
\end{align}
for some functions $\gamma_i:~\mathbb{R}_+ \to \{0,1\}$ that encode the timing of switching on(off) interactions.
% \color{blue}
% \sout{or in the case if multi-partite device}
% \begin{align}
%     H_{SA}^{\mathrm{int}}(t)\equiv \sum_i\gamma_i(t)H_{S_iA_i}^{\mathrm{int}}
% \end{align}
% for some functions $\gamma_i:~\mathbb{R}_+ \to \{0,1\}$ that \textcolor{black}{encodes the timing of} switching on(off) interactions. 
\color{black}
Here, $H_{SA}^{(i)}$ at r.h.s. is time-independent.
% , and $\eta$ is the Heaviside step function.
We later account work required to switch on(off) the interaction for the work done by the classical drive.

\begin{remark}
    We emphasize that the exact form of the functions $W_{i}(t)$, $V_{i}(t)$ and $\gamma_{i}(t)$ may be determined through a particular run of the protocol.
\end{remark}

The energy cost of a protocol is associated solely to the energy transfer (current) between the classical drive and the mesoscopic device. In the standard approach, one could define the energy current as
\begin{align}
    J_E^{\mathrm{drive}}(t) = \partial_t\left< H_{SA}(t)\right>_{\rho_{SA}(t)} =\underbrace{\left<\partial_t H_{SA}(t)\right>_{\rho_{SA}(t)}}_{\text{work current}}+\underbrace{\left< H_{SA}(t)\right>_{\partial_t\rho_{SA}(t)}}_{\text{heat current}},
\end{align}
which can be decomposed into heat and work currents (heat flux and power) that are independent in their nature. In this approach both energy flowing into and out of the mesoscopic device is taken into account.
% \begin{align}
%     &J_E^{\mathrm{drive}}(t) =J_W^{\mathrm{drive}}(t)  + J_Q^{\mathrm{drive}}(t),\\
%     &J_W^{\mathrm{drive}}(t) =\left<\partial_t H_{SA}(t)\right>_{\rho_{SA}(t)},\\
%     &J_Q^{\mathrm{drive}}(t) = \left< H_{SA}(t)\right>_{\partial_t\rho_{SA}(t)}.
% \end{align}
% where we identify the positive currents with the energy flowing into the mesoscopic device from the environment. 
However, our aim is to quantify only energy that flows into (we associate it with a positive energy current).
This motivates us to precisely define the positive ($\rightarrow$) and negative currents ($\leftarrow$) in the way that exploits assumed structure of Liouvillian and takes into account contribution of independent driving fields separately.

\begin{definition}[Of positive and negative currents]
\label{def:positive-and-negative-currents}
Let $\eta$ denotes a Heaviside step function
\begin{equation}
    \eta(x) = \left\{\begin{array}{c}
        1, x\geq 0\\
        0, x < 0
    \end{array} \right. 
\end{equation}
We define positive ($\rightarrow$) and negative currents ($\leftarrow$) associated with a Hamiltonian model protocol performing a quantum task as follows:
\begin{enumerate}
    \item For each independent driving field $\hat{A}_i$ we define
    \begin{align}
        & J_{A_i}^{\mathrm{drive}}(t) = \tr \left[ \frac{\partial V_i(t)}{\partial t} \hat{A}_i \rho(t)\right] + \mathrm{c.c.} = \frac{\partial V_i(t)}{\partial t} \tr \left[A_i \rho(t) \right]+  \mathrm{c.c.},\\
        & J_{A_i}^{\mathrm{drive}\stackrel{\raisebox{-0.5ex}{$\rightarrow$}}{\raisebox{-0.5ex}{$\leftarrow$}}}(t)  = \eta\left( \pm J_{A_i}^{\mathrm{drive}}(t) \right) \cdot J_{A_i}^{\mathrm{drive}}(t),
        % \eta\left( \pm \frac{\partial V_i(t)}{\partial t} \tr \left[A_i \rho(t) \right]\right) \frac{\partial V_i(t)}{\partial t} \tr \left[A_i \rho(t) \right]
    \end{align}
     where $\mathrm{c.c}$ stands for complex conjugate, and analogously for the interaction Hamiltonians
    \begin{align}
        & J_{H_i}^{\mathrm{drive}}(t)  = \tr \left[ \frac{\partial H_{SA}^{(i)}(t)}{\partial t}\rho(t) \right],\\
        & J_{H_i}^{\mathrm{drive}\stackrel{\raisebox{-0.5ex}{$\rightarrow$}}{\raisebox{-0.5ex}{$\leftarrow$}}}(t)  = \eta \left( \pm J_{H_i}^{\mathrm{drive}}(t) \right) \cdot J_{H_i}^{\mathrm{drive}}(t).
        % \eta\left( \pm \tr \left[ \frac{\partial H_{SA}^{\mathrm{int}}(t)}{\partial t}\rho(t) \right]\right) \tr \left[ \frac{\partial H_{SA}^{\mathrm{int}}(t)}{\partial t}\rho(t) \right]
    \end{align}
    We define {\it work current} as a sum of currents of each independent driving fields and to work current corresponding to the interaction part of Hamiltonian.
    \begin{align}
        & J_W^{\mathrm{drive}}(t)  = \sum_i J_{H_i}^{\mathrm{drive}}(t)  + \sum_i J_{A_i}^{\mathrm{drive}}(t) \\
        & J_{W}^{\mathrm{drive}\stackrel{\raisebox{-0.5ex}{$\rightarrow$}}{\raisebox{-0.5ex}{$\leftarrow$}}}(t)  = 
        \sum_i J_{H_i}^{\mathrm{drive}\stackrel{\raisebox{-0.5ex}{$\rightarrow$}}{\raisebox{-0.5ex}{$\leftarrow$}}}(t)  + \sum_i J_{A_i}^{\mathrm{drive}\stackrel{\raisebox{-0.5ex}{$\rightarrow$}}{\raisebox{-0.5ex}{$\leftarrow$}}}(t).
    \end{align}
    \item For each dissipator $D_{A,\beta_i}^i$ we define its independent heat current as
    \begin{align}
        & J_{D_i}^{\mathrm{drive}}(t)  = \tr \left[ H_{SA}(t)W_i(t)D_{A,\beta_i}^i[\rho(t)] \right],\\
        & J_{D_i}^{\mathrm{drive}\stackrel{\raisebox{-0.5ex}{$\rightarrow$}}{\raisebox{-0.5ex}{$\leftarrow$}}}(t)  = \eta \left( \pm J_{D_i}^{\mathrm{drive}}(t) \right) \cdot J_{D_i}^{\mathrm{drive}}(t).
        % \eta\left( \pm \tr \left[ H_{SA}(t)W_i(t)D_{A,\beta_i}^i[\rho(t)] \right] \right)\tr \left[ H_{SA}(t)W_i(t)D_{A,\beta_i}^i[\rho(t)] \right]
    \end{align}
    We define heat current as a sum of all independent heat currents 
    \begin{align}
        & J_Q^{\mathrm{drive}}(t)  = \sum_i J_{D_i}^{\mathrm{drive}}(t) \\
        & J_Q^{\mathrm{drive}\stackrel{\raisebox{-0.5ex}{$\rightarrow$}}{\raisebox{-0.5ex}{$\leftarrow$}}}(t)  = \sum_i  J_{D_i}^{\mathrm{drive}\stackrel{\raisebox{-0.5ex}{$\rightarrow$}}{\raisebox{-0.5ex}{$\leftarrow$}}}(t).
    \end{align}
    \item Finally, we define energy current as follows
    \begin{align}
        & J_E^{\mathrm{drive}}(t) = J_Q^{\mathrm{drive}}(t)  + J_{W}^{\mathrm{drive}}(t)\\
        & J_E^{\mathrm{drive}\stackrel{\raisebox{-0.5ex}{$\rightarrow$}}{\raisebox{-0.5ex}{$\leftarrow$}}}(t)  = J_Q^{\mathrm{drive}\stackrel{\raisebox{-0.5ex}{$\rightarrow$}}{\raisebox{-0.5ex}{$\leftarrow$}}}(t)  + J_{W}^{\mathrm{drive}\stackrel{\raisebox{-0.5ex}{$\rightarrow$}}{\raisebox{-0.5ex}{$\leftarrow$}}}(t) .
    \end{align}
    \end{enumerate}
    If the choice of Hamiltonian model protocol performing quantum task is not clear from context, we write 
    \begin{equation}
        \left. J_X^{\mathrm{drive}\stackrel{\raisebox{-0.5ex}{$\rightarrow$}}{\raisebox{-0.5ex}{$\leftarrow$}}}(t) \right|_{\mathcal{P}},
    \end{equation}
    to clarify that we consider currents corresponding to the protocol $\mathcal{P}$.
\end{definition}

% \color{blue}
% \begin{definition}[Of local positive and negative currents] We define positive ($\rightarrow$) and negative local currents ($\leftarrow$) associated with a Hamiltonian protocol $\mathcal{P}$ with
%     \begin{align}
% \left. J_X^{\mathrm{drive}\stackrel{\raisebox{-0.5ex}{$\rightarrow$}}{\raisebox{-0.5ex}{$\leftarrow$}}}(t) \right|_{\mathcal{P}}
%  = \sum_i \left. \eta \left( \pm J_X^{\mathrm{drive}}(t) \right)J_X^{\mathrm{drive}}(t) \right|_{\mathcal{P}_{i}},
% \end{align}
% where $X \in \{E,W,Q\}$, and $\mathcal{P}_i$ denotes the reduced protocol of the $i$-th party. {\color{red} I think, this definition should be removed.}
% \end{definition}
% \color{black}

%\begin{figure}[!ht]
%    \centering
%    \includegraphics[width=0.5\textwidth]{image-from-presentation-made-by-pawel.png}
%    \caption{Graphical representation of the model we adopt to define the energy cost of quantum operations. It consists of a quantum system, a mesoscopic interface device and classical controls. To quantify the energy cost of an operation $\Lambda$ performed by the mesoscopic device on the system, we take into account heat and work currents directed from classical control to the mesoscopic device.}\label{Currents}
%    \label{fig:Heat}
%\end{figure}

Ultimately, we want to quantify the fundamental energy cost, not the flows of energy, heat and work. Therefore we assume that we can not recover the energy that flows out of the mesoscopic device. This assumption is motivated by the following arguments
\begin{itemize}
    \item  It is reasonable to assume that the classical drive does not possess a secondary, dual functionality to capture and store the energy flowing out of the mesoscopic interface. Furthermore, suboptimal protocols can waste a lot of energy, especially in realistic implementations. This can be never described in a conservative setup in which no energy is truly lost. Indeed, from the very beginning we assumed that we only have access to a few degrees of freedom of our mesoscopic device, leaving many of them uncontrollable. This lack of control leads to irreversibility in the description (see, e.g. Ref.~\cite{Zukowski2024} for a discussion). \textcolor{black}{Still, we allow for arbitrary energy circulation (complying with the protocol) inside the mesoscopic device, hence this assumption is not too strict.}
    \item \textcolor{black}{Suppose we imagine that we} implement the task $\Lambda$ by considering systems $S$ and $A$ as parts of an autonomous quantum machine~\cite{BohrBrask2015,Woods2018}. Furthermore, assume that system $A$ comprises a quantum \textcolor{black}{battery(-ies) and a finite quasi-}reservoir(-s))\footnote{\textcolor{black}{Here, by finite quasi-reservoir we mean a system with discrete spectrum, however large enough (in dimension), so that behaves as a good approximation of the proper reservoir at the time scale of quantum system $S$ yet, at the larger timescale of the protocol its individual degrees of freedom can be controlled.} } as its subsystems, initially prepared in a desired state. In this setup, the heat and work required for system $A$ to perform the protocol $\mathcal{P}$ can be supplied by these distinct subsystems. The role of the classical drive is merely to reheat (or cool) the reservoir\textcolor{black}{(-s)} and to charge (or discharge) the battery\textcolor{black}{(-ies)} when the protocol is completed. \textcolor{black}{Additionally it needs to reset the control state of the mesoscopic device.} Notice that every protocol can be encapsulated in this way. In this scenario, the mesoscopic system can act as both a source and a sink for heat and work at the end of the protocol. However, an appropriate choice of the initial state of the mesoscopic device would allow the heat and work currents, required to reset the device, to have the same sign ($\leftrightarrow$). The above description of perfect setup suggests that, from the perspective of fundamental energy cost, we should consider only the positive currents of heat and work, and consider them independent. The above idea shows that the optimal protocol (in intuitive sense) doesn't necessarily need (or be able) to recover energy from the negative currents, while neglecting them can be used to eliminate suboptimal protocols. 
    \item In principle, the negative currents can be effectively utilized to some extent, which reflects the unitary nature of quantum mechanical evolution. \textcolor{black}{In particular, the extend to which the negative heat current can be utilized is limited by the laws od thermodynamics. In particular, the second and the third law.} However, any alternative approach that aims to harness these negative currents—beyond being technologically speculative—must also \textcolor{black}{finally} take into account the state of the laboratory, or at least its parameters such as temperature. This adds another "layer" to our setup, and it is difficult to estimate how many such additional layers—akin to a quantum onion structure—would be necessary for a fully satisfying description. Finally, to describe the setup to the full extent, another quantifier "energy produced" should be considered. However, this one is not of our recent interest and will be studied elsewhere \cite{future}. 
\end{itemize}
 In this place, we consequently propose the following quantifiers of the energy cost.
 \begin{definition} The energy cost of execution of a protocol $\mathcal{P}$ that implements a quantum task $\Lambda$ is defined as
\begin{align}
    E(\Lambda_\mathcal{P}) := \int_0^{t^*}dt  \left(J_W^{\mathrm{drive}\rightarrow}(t)+J_Q^{\mathrm{drive}\rightarrow}(t)\right).
\end{align}\label{def:EconsumP}
 \end{definition}
\begin{observation}
The energy cost of the execution of protocol $\mathcal{P}$ realizing $\Lambda$, denoted by $E(\Lambda_\mathcal{P})$, is non-negative and additive. 
\label{obs:nonNeg}
\end{observation}

\begin{proof}
Let us start with non-negativity. By definition, the energy cost $E(\Lambda_\mathcal{P})$ is given by the definite integral of a non-negative function over real-valued time interval. Since the integrand is non-negative at all times, the value of the integral—and thus the total energy cost is also non-negative.
The second property is proved by inspection. 
\begin{align}
      &E(\Lambda_{\mathcal{P}_1}) + E(\Lambda_{\mathcal{P}_2}) =\int_0^{t_1}dt  \left(J_W^{\mathrm{drive}\rightarrow}(t)+J_Q^{\mathrm{drive}\rightarrow}(t)\right)_{\mathcal{P}_1} +\int_0^{t_2}dt  \left(J_W^{\mathrm{drive}\rightarrow}(t)+J_Q^{\mathrm{drive}\rightarrow}(t)\right)_{\mathcal{P}_2}\\
      &=\int_0^{t_1+t_2}dt  \left(J_W^{\mathrm{drive}\rightarrow}(t)+J_Q^{\mathrm{drive}\rightarrow}(t)\right)_{\mathcal{P}_2\circ\mathcal{P}_1} = E(\Lambda_{\mathcal{P}_2 \circ \mathcal{P}_1}). \label{eq:redefinitionHamiltonian},
\end{align}
where the first equality in equation~\eqref{eq:redefinitionHamiltonian} is by introducing a new Hamiltonian model protocol being temporal composition of $\mathcal{P}_1$ and $\mathcal{P}_2$.
\end{proof}
\begin{remark}
    The energy cost of the execution of a protocol $\mathcal{P}$ that implements a quantum task $\Lambda=\mathds{1}$, i.e. $E(\Lambda_{\mathcal{P}})$, is not necessarily~0. This is because for every protocol $\mathcal{P}_1$, with $E(\Lambda_{\mathcal{P}_1}) > 0$,which realizes some task $\Lambda_1 \neq \mathds{1}$ we can always find a protocol $\mathcal{P}_2$ which realizes $\Lambda_2$ such that $\Lambda_2 \circ \Lambda_1 = \mathds{1}$. Yet, the energy cost $E(\Lambda_{\mathcal{P}_2} \circ \Lambda_{\mathcal{P}_2})$ stays strictly positive. The composition $\mathcal{P}\equiv \mathcal{P}_2 \circ \mathcal{P}_1$ is therefore an example of such protocol for which $E(\mathds{1}_{\mathcal{P}})>0$.
\end{remark}

We define the architecture-dependent minimal energy cost of the execution of the task $\Lambda$ on $\rho$ as the infimum over energy costs $E(\Lambda_\mathcal{P})$ of protocols from a given class $\mathcal{A}$.
  \begin{definition} The architecture-dependent minimal energy cost of implementing quantum task $\Lambda$ via the set of possible protocols $\mathcal{A}=\{\mathcal{P}\}$, containing trivial task (i.e., $\mathds{1} \in \mathcal{A}$), is defined as
\begin{align}
E_\mathcal{A}(\Lambda):=\inf_{\mathcal{P} \in \mathcal{A}}E(\Lambda_\mathcal{P}).
\end{align}\label{def:EconsumArchitecture}
 \end{definition}
 We define the fundamental energy of executing the task $\Lambda$ on $\rho$ as the infimum over energy costs of all Hamiltonian model protocols realizing $\Lambda(\rho)$.
\begin{definition} The fundamental energy cost of quantum task $\Lambda$ is defined as
\begin{align}  
E_{\mathcal{F}}(\Lambda) \equiv E(\Lambda,\rho):=\inf_{\mathcal{P}}E(\Lambda_\mathcal{P}),
\end{align}\label{def:EconsumFundamental}
where the infimum is taken over all protocols $\mathcal{P}$ that conform to a Hamiltonian model.
 \end{definition}
 The above definition is one of central notions employed in our results in Section~\ref{section:UB}. This is because of the following observation.
 \begin{observation}
     The fundamental energy cost $ E(\Lambda,\rho)$ constitutes a lower bound on the energy cost of any protocol realizing task $\Lambda$.
 \end{observation}
 \begin{proof}
     The proof follows directly from the definition of infimum.
 \end{proof}
% \begin{remark} We leave open the question whether the fundamental energy consumption depends on the platform of physical realization (PR). The answer to this question is out of scope of this contribution.
% \end{remark}

\color{black}
    The quantities in Definitions \ref{def:EconsumArchitecture} and \ref{def:EconsumFundamental} can be interpreted as the minimal energy that has to be invested (supplied to the system) in order to perform some task, while seemingly neglecting the energy outgoing from the mesoscopic interface as a byproduct. Still, in principle energy leaving the mesoscopic interface can be at least partially reused. In principle, this can be done by constructing another mesoscopic interface that is supplied (not exclusively) with the negative energy flow of the original device, classical control modifies the state of the secondary device, and finally it powers back the original device with some part of its negative currents. Similarly, we can introduce tertiary, quaternary, quinary, and higher-order interfaces. The amount of useful energy that can ultimately be retrieved is therefore limited only by the laws of thermodynamics — specifically, the second and third laws. However, the quantities we define are formulated as infima over such interfaces as parts of protocols, which means that the possibility of internal energy circulation (i.e., retrieval and reuse) is inherently accounted for in their definitions. This line of reasoning also shows that negative currents are unavoidable when only classical control is employed, regardless of how complex the mesoscopic device may be.
\color{black}
 
\begin{remark}
    In a similar manner one can define the following (non-negative) quantities:
    \begin{itemize}
        \item Net-energy cost: $E_\mathrm{net}(\Lambda_\mathcal{P}) \equiv E^{(+)}_\mathrm{net}(\Lambda_\mathcal{P}) := \int_0^{t^*}dt  J_E^{\mathrm{drive}\rightarrow}(t)$.
        \item Work cost: $W(\Lambda_\mathcal{P}) \equiv W^{(+)}(\Lambda_\mathcal{P})  := \int_0^{t^*}dt  J_W^{\mathrm{drive}\rightarrow}$(t).
        \item Heat cost: $Q(\Lambda_\mathcal{P}) \equiv Q^{(+)}(\Lambda_\mathcal{P})  := \int_0^{t^*}dt  J_Q^{\mathrm{drive}\rightarrow}$(t),
    \end{itemize}
    Furthermore, in a direct analogy to Definitions~\ref{def:EconsumArchitecture} and ~\ref{def:EconsumFundamental} one can define their architecture-dependent and fundamental versions
    
\color{black}
    Additionally, we can define the versions of the above quantities related to the negative currents.
    \begin{itemize}
        \item Net-energy gain: ${E}^{(-)}_\mathrm{net}(\Lambda_\mathcal{P}) := \int_0^{t^*}dt  J_E^{\mathrm{drive}\leftarrow}(t)$.
        \item Work gain: $W^{(-)}(\Lambda_\mathcal{P}) := \int_0^{t^*}dt  J_W^{\mathrm{drive}\leftarrow}$(t).
        \item Heat gain: $Q^{(-)}(\Lambda_\mathcal{P}) := \int_0^{t^*}dt  J_Q^{\mathrm{drive}\leftarrow}$(t),
    \end{itemize}
    which are always non-positive. Furthermore, one can define their architecture-dependent and fundamental versions, as those quantities evaluated on protocols that saturate the infima in their "cost" counterparts. 
\color{black}
\end{remark}

Having the notions central to this section defined, we have the following proposition.
\begin{proposition} The quantities in Definition~\ref{def:EconsumArchitecture}, and~\ref{def:EconsumFundamental} satisfy Properties \ref{property:1}, \ref{property:2}, and Axioms 1, 2. Hence, they are good quantifiers of energy cost in quantum tasks. \\
% The energy consumption measure in Definition~\eqref{def:EconsumP} satisfies Axiom~\ref{axiom:3} with equality.  
\label{prop:SatisfyAxioms}
\end{proposition}
\begin{proof} It is enough to prove that the quantity in Definition~\ref{def:EconsumArchitecture} satisfies the Axioms and the Properties for all possible quantum tasks (CPTP maps). To obtain the proof for the fundamental energy cost in Definition~\ref{def:EconsumFundamental} it is enough to extend the set over which we optimize to all protocols exhibiting a Hamiltonian model.

The compliance with Property~\ref{property:2} is trivial. The infimum is taken over a set of non-negative elements (see Observation~\ref{obs:nonNeg}). 

In order to prove compliance with Property~\ref{property:1}, we observe that we can realize $\mathds{1}$ by keeping the systems $S$ separated. Therefore, there is no energy cost, i.e., $E_\mathcal{A}(\mathds{1})=0$, due to infimum in the Definition~\ref{def:EconsumArchitecture}.

For the compliance with Axiom~1, upon fixed architecture $\mathcal{A}$, let us assume two quantum tasks $\Lambda_1$ and $\Lambda_2$ with output of $\Lambda_1$ compatible with input of $\Lambda_2$. Furthermore, let $\mathcal{P}_1^*$ and $\mathcal{P}_2^*$ be the optimal protocols for realizing $\Lambda_1$ and $\Lambda_2$. Then:
\begin{align}
      &E_\mathcal{A}(\Lambda_1)+E_\mathcal{A}(\Lambda_2)  = E_\mathcal{A}(\Lambda_{\mathcal{P}_1^*}) + E_\mathcal{A}(\Lambda_{\mathcal{P}_2^*})\\
      &=\int_0^{t^*_1}dt  \left(J_W^{\mathrm{drive}\rightarrow}(t)+J_Q^{\mathrm{drive}\rightarrow}(t)\right)_{\mathcal{P}_1^*} +\int_0^{t^*_2}dt  \left(J_W^{\mathrm{drive}\rightarrow}(t)+J_Q^{\mathrm{drive}\rightarrow}(t)\right)_{\mathcal{P}_2^*}\\
      &=\int_0^{t^*_1+t_2^*}dt  \left(J_W^{\mathrm{drive}\rightarrow}(t)+J_Q^{\mathrm{drive}\rightarrow}(t)\right)_{\mathcal{P}_2^*\circ\mathcal{P}_1^*} = E_\mathcal{A}(\Lambda_{\mathcal{P}_2^*\circ\mathcal{P}_1^*} ) \ge E_\mathcal{A}(\Lambda_2 \circ \Lambda_1),\label{eq:redefinitionHamiltonian2}
\end{align}
where the first equality in equation~\eqref{eq:redefinitionHamiltonian2} is by introducing a new Hamiltonian model protocol being temporal composition of $\mathcal{P}_1^*$ and $\mathcal{P}_2^*$. The last inequality is because we take infimum over a lager set in which the optimal protocol $\mathcal{P}^*$ for realizing $\Lambda_2 \circ \Lambda_1$ is non necessarily decomposable. Because we didn't specify $\Lambda$ the above holds for all $\Lambda \in \mathcal{A}$, and in particular for all $\Lambda$ exhibiting a Hamiltonian model.

Now it remains to check the compliance with Axiom ~2.
    Let $\Lambda_{AB} \equiv (\tilde{\Lambda} \otimes \mathds{1}_B ,\rho_{AB},\sigma_{AB})$ be a quantum task, of two spatially separated parties $A$ and $B$, that involves a separable CPTP map $\tilde{\Lambda} \otimes \mathds{1}_B $. Let us consider a reduced task $\Lambda_{A} \equiv (\tilde{\Lambda}, \Tr_B[\rho_{AB}],\Tr_B[\sigma_{AB}]$), of party $A$.

    By inspection we have
    \begin{align}
        &E_\mathcal{A}(\Lambda_{AB}) = \inf_{\mathcal{P} \in \mathcal{A}} E(\left.\Lambda_{AB}\right|_{\mathcal{P}}) = \inf_{\mathcal{P} \in \mathcal{A}} E\left(\left.\Lambda_{AB}\right|_{\mathcal{P} \equiv (\mathcal{L}_{AA^\prime BB^\prime}(t),t^*,A^\prime B^\prime,\tau_{A^\prime B^\prime},\rho_{AB},\sigma_{AB})} \right)\\
        &\stackrel{(I)}{=} \inf_{\mathcal{P} \in \mathcal{A}} E\left(\left.\Lambda_{AB}\right|_{\mathcal{P} \equiv (\mathcal{L}_{AA^\prime}(t)\otimes \mathds{1}_{BB^\prime},t^*,A^\prime B^\prime,\tau_{A^\prime B^\prime},\rho_{AB},\sigma_{AB})} \right)\\
        &= \inf_{\mathcal{P} \in \mathcal{A}}  \left. \int_0^{t^*}dt~\left( J_W^{\mathrm{drive\rightarrow}}(t)+J_Q^{\mathrm{drive\rightarrow}}(t) \right)\right|_{\mathcal{P} \equiv (\mathcal{L}_{AA^\prime}(t)\otimes \mathds{1}_{BB^\prime},t^*,A^\prime B^\prime,\tau_{A^\prime B^\prime},\rho_{AB},\sigma_{AB})}\\
        &\stackrel{(II)}{=} \inf_{\mathcal{P} \in \mathcal{A}} \left. \int_0^{t^*}dt~\left( J_W^{\mathrm{drive\rightarrow}}(t)+J_Q^{\mathrm{drive\rightarrow}}(t) \right)\right|_{\mathcal{P} \equiv (\mathcal{L}_{AA^\prime}(t),t^*,A^\prime,\Tr_{B^\prime}\tau_{A^\prime B^\prime},\Tr_B[\rho_{AB}],\Tr_B[\sigma_{AB}])}\\
        &= \inf_{\mathcal{P} \in \mathcal{A}} E(\left.\Lambda_{A}\right|_{\mathcal{P}}) = E_\mathcal{A}(\Lambda_{A}),
    \end{align}
    where in $(I)$ we notice that optimal protocols are the ones that have trivial dynamics at the side of $B$ party. In $(II)$ we use the fact that the heat and work currents at the side $B$ vanish due to trivial dynamics of party $B$.
\end{proof}

\begin{observation}
    Let $\Lambda$ and $\Omega$ be quantum tasks. Then $E_\mathcal{A}$ and $E_\mathcal{F}$ satisfy the following equality
    \begin{equation}
        E_\mathcal{A,F}(\Lambda \otimes \Omega) = E_\mathcal{A,F}(\Lambda) + E_\mathcal{A,F}(\Omega).
    \end{equation}
\end{observation}
\begin{proof}
    Since both $E_\mathcal{A}$ and $E_\mathcal{F}$ (which we denote as $E_\mathcal{A,F}$) satisfy axioms 1 and 2, then by Observation \ref{obs:axiom-sub-additivity-in-space} we have that $E_\mathcal{A}(\Lambda \otimes \Omega) \leq E_\mathcal{A}(\Lambda) + E_\mathcal{A}(\Omega)$, which is weaker than the claimed equality. Nevertheless, we can prove it by direct inspection
    \begin{align}
          &E_\mathcal{A}(\Lambda \otimes \Omega) = \inf_{\mathcal{P}_1 \otimes \mathcal{P}_2 \in \mathcal{A}} E\left( \Lambda_{\mathcal{P}_1} \otimes \Omega_{\mathcal{P}_2} \right) = \int_0^{\max\{t^*_1,t^*_2\}}dt  \left(J_W^{\mathrm{drive}\rightarrow}(t)+J_Q^{\mathrm{drive}\rightarrow}(t)\right)_{\mathcal{P}_1^* \otimes \mathcal{P}_1^*} \\
          &= \int_0^{\max\{t^*_1,t^*_2\}}dt  \left(J_W^{\mathrm{drive}\rightarrow}(t)+J_Q^{\mathrm{drive}\rightarrow}(t)\right)_{\mathcal{L}^{(1)}_{S_1A_1}(t) \otimes \mathds{1}_{S_2A_2} + \mathds{1}_{S_1A_1} \otimes \mathcal{L}^{(2)}_{S_2A_2}(t)} \\
          &= \int_0^{t^*_1}dt  \left(J_W^{\mathrm{drive}\rightarrow}(t)+J_Q^{\mathrm{drive}\rightarrow}(t)\right)_{\mathcal{L}^{(1)}_{S_1A_1}(t)} +  \int_0^{t^*_2}dt  \left(J_W^{\mathrm{drive}\rightarrow}(t)+J_Q^{\mathrm{drive}\rightarrow}(t)\right)_{\mathcal{L}^{(2)}_{S_2A_2}(t)}\\
          &= E_\mathcal{A}(\Lambda) + E_\mathcal{A}(\Omega),
    \end{align}
    where in the fourth equality we used the fact that currents vanish for subsystems governed by trivial dynamics,~i.e.,~$\mathds{1}$.\\
\end{proof}
% The above proof does not hold for the quantity in Definition~\ref{def:EconsumP}. More precisely, $E(\Lambda_\mathcal{P})$ can be greater than zero for $\Lambda = \mathds{1}$. This is because we can always find a protocol that first realizes operation $\Lambda_1 \neq \mathds{1}$ with $E(\Lambda_{\mathcal{P}_1})>0$, and next it recovers the initial state of systems $S$ and $A$. In this example, the joint protocol realizes $\mathds{1}$ while $E(\Lambda_\mathcal{P})$ stays strictly positive.

We have an immediate corollary.
\begin{corollary} The quantities in Definition~\ref{def:EconsumArchitecture}, and~\ref{def:EconsumFundamental} satisfy the triangle-like inequality.  
\label{corollary:triangle}
\end{corollary}
\begin{proof} For the proof is enough to notice that the compliance with Axiom~1 in Proposition~\ref{prop:SatisfyAxioms} holds for any decomposition of any $\Lambda$ into compatible $\Lambda_1$ and $\Lambda_2$. Therefore
\begin{align}
    \forall_\rho\forall_{\Lambda}\forall_{\Lambda_1,\Lambda_2~:~\Lambda_2\circ\Lambda_1=\Lambda}~~E(\Lambda,\rho) \le E(\Lambda_1,\rho)+E(\Lambda_2,\Lambda_1(\rho) ),
\end{align}
    mutatis mutandis for the quantity of architecture-dependent minimal energy cost.
\end{proof}

For the sake of comparison, we define the energy balance associated with performing task~$\Lambda$. This simplified measure ignores the assumption that energy recovery \textcolor{black}{(by classical drive)} is impossible (hence it lower bounds the actual energetic cost).
\begin{definition}
     The energy balance of quantum task $\Lambda$ is defined as
\begin{align}   
     E_\mathrm{balance}(\Lambda) \equiv E_\mathrm{balance}(\Lambda_\mathcal{P}) := \int_0^{t^*}dt  J_E^{\mathrm{drive}}(t) = E(\sigma) - E(\rho),
\end{align}
where $\rho$ is the initial and $\sigma$ is the final states of a protocol (see Definition~\ref{def:HamiltonianProtocol}). 
\label{def:EconsumSimple}
\end{definition}
\textcolor{black}{The quantity in the above definition (that can be negative) corresponds to the case in which we can recover all the energy leaving the mesoscopic interface.}

\begin{remark} The energy balance $E_\mathrm{balance}(\Lambda)$ introduced in Definition~\ref{def:EconsumSimple} can take negative values and therefore does not satisfy Property~\ref{property:1}. Moreover, it is invariant under protocol pathway, in the sense that it depends solely on the initial and final states of the system of interest, irrespective of the intermediate dynamics.
\end{remark}
We immediately make the following observation.
\begin{observation}
           The energy balance constitutes a (trivial) lower bound on the fundamental energy cost, i.e., 
    \begin{align}
        E_\mathrm{balance}(\Lambda)\le E(\Lambda,\rho) \le E_\mathcal{A}(\Lambda).
    \end{align}
\end{observation}
\begin{proof}
    The proof follows from the properties of Heaviside step function. Let $\mathcal{P}^*$ be the protocol that saturates the infimum in the fundamental energy cost $E(\Lambda,\rho)$ for the realization of $\Lambda$. Then
    \begin{align}   
     &E_\mathrm{balance}(\Lambda)=E_\mathrm{balance}(\Lambda_{\mathcal{P}^*}) = \int_0^{t^*}dt  J_E^{\mathrm{drive}}(t)\left.\right|_{\mathcal{P}^*} = \int_0^{t^*}dt  \left( J_Q^{\mathrm{drive}}(t) +J_W^{\mathrm{drive}}(t)\right)_{\mathcal{P}^*}\\
     &=\int_0^{t^*}dt  \left(\eta(J_Q^{\mathrm{drive}}(t))+\eta(-J_Q^{\mathrm{drive}}(t))\right)J_Q^{\mathrm{drive}}(t)\left.\right|_{\mathcal{P}^*}+ \int_0^{t^*}dt  \left(\eta(J_W^{\mathrm{drive}}(t))+\eta(-J_W^{\mathrm{drive}}(t))\right)J_W^{\mathrm{drive}}(t)\left.\right|_{\mathcal{P}^*} \\
     &=\int_0^{t^*}dt \left(\eta(J_Q^{\mathrm{drive}}(t))J_Q^{\mathrm{drive}}(t))+\eta(J_W^{\mathrm{drive}}(t))J_W^{\mathrm{drive}}(t))\right)_{\mathcal{P}^*}+\left(\eta(-J_Q^{\mathrm{drive}}(t))J_Q^{\mathrm{drive}}(t))+\eta(-J_W^{\mathrm{drive}}(t))J_W^{\mathrm{drive}}(t))\right)_{\mathcal{P}^*} \\
     &=\int_0^{t^*}dt \left(J_Q^{\mathrm{drive}\rightarrow}(t))+J_W^{\mathrm{drive}\rightarrow}(t))\right)_{\mathcal{P}^*}+\int_0^{t^*}dt\left(J_Q^{\mathrm{drive}\leftarrow}(t))+ J_W^{\mathrm{drive}\leftarrow}(t))\right)_{\mathcal{P}^*} \\
     & = Q^{(+)} (\Lambda_{\mathcal{P}^*}) + W^{(+)} (\Lambda_{\mathcal{P}^*}) + Q^{(-)} (\Lambda_{\mathcal{P}^*}) + W^{(-)} (\Lambda_{\mathcal{P}^*}) \\
     &\le Q^{(+)} (\Lambda_{\mathcal{P}^*}) + W^{(+)} (\Lambda_{\mathcal{P}^*}) = E(\Lambda,\rho) \le E_\mathcal{A}(\Lambda),
\end{align}
where the first inequality is upon neglecting non-positive term. The second inequality is due to restricted set (with respect to $E(\Lambda,\rho)$) over which the infimum in the definition of $E_\mathcal{A}(\Lambda)$ is evaluated.
\end{proof}
\begin{observation} The following relation holds
\begin{align}
    E(\Lambda_\mathcal{P}) =E_\mathrm{balance}(\Lambda)- \int_0^{t^*}dt~\left( J_W^{\mathrm{\leftarrow drive}}(t)+J_Q^{\mathrm{\leftarrow drive}}(t) \right) = E_\mathrm{balance}(\Lambda) - {E}^{(-)}_\mathrm{net}(\Lambda_\mathcal{P}).
\end{align}
\end{observation}
\begin{proof}
At first, we need to prove that $J_E^{\mathrm{drive}}(t) = J_E^{\mathrm{drive\rightarrow}} (t) + J_E^{\mathrm{drive\leftarrow}} (t)$. To achieve this, observe that each function $f:\mathbb{R}\to\mathbb{R}$ can be expressed as $f(x) = \eta(f(x))\cdot f(x) + \eta(-f(x))\cdot f(x)$, where $\eta(\cdot)$ is a Heaviside step function. Now, following Definition \ref{def:positive-and-negative-currents} we have
\begin{align}
    &J_E^{\mathrm{drive}} (t) = J_W^{\mathrm{drive}} (t) + J_Q^{\mathrm{drive}} (t) = \sum_i J_{H_i}^{\mathrm{drive}}(t)  + \sum_i J_{A_i}^{\mathrm{drive}}(t) + \sum_i J_{D_i}^{\mathrm{drive}}(t) = \\
    & \sum_i \left(\eta(J_{H_i}^{\mathrm{drive}}(t))J_{H_i}^{\mathrm{drive}}(t) + \eta(-J_{H_i}^{\mathrm{drive}}(t))J_{H_i}^{\mathrm{drive}}(t)\right)  + \sum_i \left(\eta(J_{A_i}^{\mathrm{drive}}(t))J_{A_i}^{\mathrm{drive}}(t) + \eta(-J_{A_i}^{\mathrm{drive}}(t))J_{A_i}^{\mathrm{drive}}(t)\right) + \\
    & \qquad \qquad \sum_i\left(\eta(J_{D_i}^{\mathrm{drive}}(t))J_{D_i}^{\mathrm{drive}}(t) + \eta(-J_{D_i}^{\mathrm{drive}}(t))J_{D_i}^{\mathrm{drive}}(t)\right) = \\
    & J_W^{\mathrm{drive}\rightarrow} (t) + J_W^{\mathrm{drive}\leftarrow} (t) + J_Q^{\mathrm{drive}\rightarrow} (t) + J_Q^{\mathrm{drive}\leftarrow} (t) = J_E^{\mathrm{drive}\rightarrow} (t) + J_E^{\mathrm{drive}\leftarrow} (t).
\end{align}
Finally, we can use obtained equality to finish the proof
\begin{align}
    &E(\Lambda_\mathcal{P}) =\int_0^{t^*}dtJ_E^{\mathrm{drive\rightarrow}} (t) = \int_0^{t^*}dt \left( J_E^{\mathrm{drive}}(t) - J_E^{\mathrm{drive\leftarrow}} (t) \right)\nonumber \\ 
    &= \left( E(\Lambda(\rho)) - E(\rho) \right) - \int_0^{t^*}dt J_E^{\mathrm{drive\leftarrow}} (t) = E_\mathrm{balance}(\Lambda) - {E}^{(-)}_\mathrm{net}(\Lambda_\mathcal{P}).
\end{align}
\end{proof}

\begin{lemma} Let $\mathcal{N}$ denote the set of tasks such that $E_\mathrm{balance}(\Lambda)=E(\sigma) - E(\rho)\ge0$ then quantities in Definitions~\ref{def:EconsumP},~\ref{def:EconsumArchitecture},~\ref{def:EconsumFundamental} are non-degenerate.
\begin{align}
    \Lambda \in \mathcal{N}~ 	\wedge  ~E(\Lambda,\rho)=0 \Longleftrightarrow \Lambda =\mathds{1}. 
\end{align}
\label{lem:NonDegenerate}
\end{lemma}
\begin{proof} We prove $\Longrightarrow$ first. The conditions $E(\sigma) - E(\rho)\ge0$ and $E(\Lambda,\rho)=0$ imply that there is no energy transfer between the system $A$ and environment, i.e., the joint system $SA$ is isolated from environment. The condition to reset the state of system $A$, in that case, implies trivial dynamics of system $S$, i.e., $\Lambda=\mathds{1}$. The converse implication, i.e,  $\Longleftarrow$ is due to Proposition~\ref{prop:SatisfyAxioms}. We also note that $\mathds{1} \in \mathcal{N}$ due to the definition of $E_\mathrm{balance}(\Lambda)$.
\end{proof}

\begin{corollary} Fundamental energy cost $E(\Lambda,\rho)$ measures ''distance'' from any quantum task $\Lambda \in \mathcal{N}$ to the identity task~$\mathds{1}$.
\end{corollary}
\begin{proof}
    The proof follows from Proposition \ref{prop:SatisfyAxioms}, Corollary \ref{corollary:triangle} and Lemma \ref{lem:NonDegenerate}, as they show that in the set $\mathcal{N}$ the fundamental energy cost $E(\Lambda,\rho)$ is positive, non-degenerate, and satisfy the triangle inequality.
\end{proof}

Examples of setups consisting of well-distinguished quantum system $S$ and mesoscopic interface $A$:
\begin{itemize}
    \item Circuit QED. In this case system $S$ is a transmon qubit, the mesoscopic system is a microwave transmission line, the relevant degrees of freedom (quantum system $A$) of mesoscopic device are modes of the electromagnetic field in the waveguide. The classical driving is the preparation of the pulses that drive the qubit. The interaction between systems $S$ and $A$, in the simplest case, is present all the time.
    % \item Same with laser and atom. Here the waveguide is just free space (air/vacuum). The atom and the field interact all the time, but coherent driving is attained with preparation of laser pulses. In the scenario in which atom is inside a cavity we can perhaps perform switching on and off the interaction.
    \item Quantum photonics. The system $S$ is defined by selected, e.g. polarization, degrees of freedom of electromagnetic field\footnote{In this case, system $S$ is possibly defined on Fock space.}. The mesoscopic system are any optical elements having spatial overlap with system $S$, e.g., beam splitters, photon detectors, that implement quantum operations, e.g., CNOT gates. The mesoscopic system can consist also of auxiliary modes of electromagnetic field. Quantum system $A$ are the degrees of freedom of system $A$ interacting (directly or indirectly) with $S$. If the optical elements are passive then classical driving reduces to preparation of auxilliary states and redout from detectors.  Tasks which naturally employs this use case are entangled states distillation tasks (e.g. considered in the main text protocols \cite{bennett1996purification,deutsch1996quantum,miguel2018efficient} ).
    % \item NV centers (but depends what we want to do with them): the mesoscopic device can be optical fiber, relevant degrees of freedom are electromagnetic modes inside, and free space modes for radio and microwave frequency signals. Preparation of different laser, RF, MF pulses, for protocol and readout is the energetic cost of driving.
    \item NV centers in diamond. The system $S$ is a spin degree of freedom of the electronic state of the NV center. The mesoscopic device includes optical fibers and/or microwave antennas used to deliver control fields. The quantum system $A$ consists of selected modes of the electromagnetic field in free space or confined in the fiber, which mediate interaction with $S$. Classical driving consists of the preparation of optical, radio-frequency (RF), or microwave-frequency (MF) pulses used to control, manipulate, and read out the state of $S$. The interaction between $S$ and $A$ can in principle be switched on and off by shaping and timing these control pulses. The precise structure of the mesoscopic system depends on the protocol under consideration (e.g., quantum memory, sensing, or communication).
    % \item {NMR in liquid. Similarly to NV centers (and other solid state NMR architectures), free space electromagnetic pulses are applied in order to implement desired gates on spins, taking into account the internal interactions between the spins. Inducing these pulses and detection of response signals constitutes for the energetic cost of protocol implementation. }
    \item NMR in liquid. The system $S$ consists of nuclear spin degrees of freedom of molecules dissolved in a liquid. The mesoscopic device includes coils and electronics responsible for generating and detecting electromagnetic pulses. The quantum system $A$ corresponds to selected modes of the electromagnetic field in free space that interact with the spins. Classical driving involves the preparation and application of radio-frequency pulses used to implement desired quantum gates, while also accounting for internal spin-spin interactions. The energetic cost of the protocol arises from both pulse generation and signal detection.

\end{itemize}

\color{black}

% \section{Observations}

% Pawel: \cite{Linpeng_2022} - energy cost in measuremengt transmon qubits - for us, we encode qubits in POlarization, so measureement doees not changee eenergy of thee state
% \cite{Stolz_2022} - new non-linear approach
% \cite{Ralph_2002} - scheme with probability 1/9 
% \cite{Kieling_2010} - 1/9 is upper bound for CNOT in lineear opics
% centrone - sprawdzic

% \begin{rem}
%     Our previous claim saying that it is enough just to compare single shot quantities in order to compare asymptotic ones is not true. This is evident when one looks at 
% \end{rem}

% Let us compare single-shot quantities for fixed $\rho$ and $\epsilon$. Let $\Lambda^*$, $\mathcal{L}^*$ and $\gamma_{d_k,d_s}^*$ be the optimal choices such that
% \begin{align}
%     \exists_{\gamma_{d_k^\prime}} ~~&\frac{1}{2} \norm{\Lambda^*(\rho) - \gamma_{d_k^\prime}}_1 \le \epsilon,\\
%     &\frac{1}{2} \norm{\rho - \mathcal{L}^*(\gamma^*_{d_k,d_s)}}_1 \le \epsilon,
% \end{align}
% hence
% \begin{align}
%     &\kappa_D^{(1,\epsilon)}(\rho) =  d_k^\prime,\\
%     &\kappa_C^{(1,\epsilon)}(\rho) =  d_k, 
% \end{align}
\subsection{Upper  bound on energy consumption rate for entanglement encoded in energetic degrees of freedom}
\label{sec:upper_bounds}
\begin{lemma}[Asymptotic continuity of single particle average energy]  Let $  {\mathrm H}  \ge 0\in  \mathcal{L}(\mathcal{H}) $ be a single particle self-adjoint Hamiltonian, and let $\mathcal{E} \subseteq \mathcal{D}(\mathcal{H})$ be a set of quantum states entirely supported in the subspace of $\mathcal{H}$ corresponding to energies not exceeding $0< E_{\max} < + \infty$, i.e., $\rho \in \mathcal{E} \implies \forall E > E_{\max}, ~\Pi_{E}\rho\Pi_{E} = 0$, where $\Pi_E = \ketbra{\psi_E}{\psi_E}:~~{\mathrm H} \ket{\psi_E}=E \ket{\psi_E}$. Let $\rho, \rho^\prime \in \mathcal{E}$ be such that $\norm{\rho-\rho^\prime}_1 < \epsilon$ for some $\epsilon>0$. Then, energies $E$ and $E^\prime$ associated with states $\rho$, $\rho^\prime$ satisfy
\begin{align}
    \abs{E-E^\prime} \le \epsilon {E_{\max}},
\end{align}
where $E=\Tr[{\mathrm H} \rho ]$ and $E^{\prime}=\Tr[{\mathrm H} \rho^{\prime}]$.
\label{lem:1particleEnegyContinuity}
\end{lemma}
\begin{proof}
    The proof is readily obtained with a series of equalities and inequalities.
    \begin{align}
        &\abs{E-E^\prime} 
        \stackrel{(I)}{=} \abs{\Tr[{\mathrm H}\rho]-\Tr[{\mathrm H}\rho^\prime]} 
        \stackrel{(II)}{=}  \abs{\Tr[{\mathrm H}\Pi_\mathcal{E}\rho \Pi_\mathcal{E}]-\Tr[{\mathrm H}\Pi_\mathcal{E}\rho^\prime\Pi_\mathcal{E}]} 
        \stackrel{(III)}{=}  \abs{\Tr[\Pi_\mathcal{E} {\mathrm H}\Pi_\mathcal{E}\rho ]-\Tr[\Pi_\mathcal{E} {\mathrm H}\Pi_\mathcal{E}\rho^\prime ]} \nonumber \\
        &\stackrel{(IV)}{=}  \abs{\Tr[{\mathrm H}_\mathcal{E}\rho ]-\Tr[{\mathrm H}_\mathcal{E}\rho^\prime ]}
        \stackrel{(V)}{=} \abs{\Tr[{\mathrm H}_\mathcal{E}(\rho-\rho^\prime)]}  
        \stackrel{(VI)}{=} {E_{\max}} \abs{ \Tr[\frac{{\mathrm H}_\mathcal{E}}{E_{\max}}(\rho-\rho^\prime)]} \nonumber \\
        &\stackrel{(VII)}{\le} {E_{\max}} \max_{0\le \Omega \le \mathbf{1}}  \Tr[\Omega(\rho-\rho^\prime)] \stackrel{(VIII)}{=} \frac 12 {E_{\max}} \norm{\rho - \rho^{\prime} }_1  \le \frac 12 \epsilon E_{\max},
    \end{align}
    where $\Pi_\mathcal{E} = \sum_{E\le E_{\max}} \Pi_E$ and the truncated Hamiltonian ${\mathrm H}_\mathcal{E} \equiv \Pi_\mathcal{E} {\mathrm H}\Pi_\mathcal{E}$ satisfies $0 \le {\mathrm H}_\mathcal{E} \le E_{\max} \mathbf{1}$ by construction. In $(I)$ we employed the definition of expected value of Hamiltonian (aka average energy). In $(II)$ we used a fact that the states $\rho$ and $\rho^\prime$ are supported on a specific subspace. In $(III)$ we employed the cyclic property of the trace, to define and denote the truncated Hamiltonian $H_{\mathcal{E}}$ in $(IV)$. In $(V)$ linearity of trace was used. In $(VI)$ we normalized the truncated Hamiltonian, by dividing it by its maximal eigenvalue, i.e., $E_{\max}$. Consequently, the inequality in $(VII)$ is due to the fact that the normalized truncated Hamiltonian does not (in generic case) saturate the introduced supremum. Finally in $(VIII)$, we employed Lemma~9.1.1. of Ref.~\cite{MarkWildeBook}
    \begin{align}
        \frac 12 \norm{\rho-\rho^\prime}_1 =  \max_{0\le \Omega \le \mathbf{1}}  \Tr[\Omega(\rho-\rho^\prime)].
    \end{align}
\end{proof}
% \rj{the above lemma is quite clear but I would just comment on the notation : $\Lambda$ was already used before so it might be a bit confusing.}
% {\mw You ask why I require $H \ge 0$? This is purely technical for the sake of using equation 11 directly. Anyway, I can also shift my Hamiltoian and the shift does not affect the physics. $Lmbda$ here is POVM, that is a standard notation in this context (We can obviously change it). Equation 11 is not a definition (however, it would be quite a good definition), it is just a fact from the book of Mark Wilde.  }
\begin{lemma}[Multi-particle proximity vB] Let $\mathcal{F}$ be an either symmetric or anti-symmetric Fock space describing a collection of identical particles and $\mathcal{E}_\mathcal{F}^{(m,n)}$ be a subset of $\mathcal{D}(\mathcal{F})$ entirely supported in the subspace of $m$- to $n$-particle states with single particle of energy not exceeding $E_{max}$.  Furthermore, let ${\mathrm H}_F \in \mathcal{L}(\mathcal{F})$ be second-quantized (free) Hamiltonian of non-interacting particles constructed using single-particle (self-adjoint) Hamiltonian $0 \le {\mathrm H} \in \mathcal{L}(\mathcal{H})$. For $\rho \in \mathcal{E}_\mathcal{F}^{(m,n)} \subseteq \mathcal{D}(\mathcal{F})$, we have that
\begin{align}
    &\rho \in \mathcal{E}_\mathcal{F}^{(m,n)} \implies \forall l:~0 \le l <m ~\vee~n<l, \forall\ket{\Psi_l} \in \mathcal{F}:~{\mathrm N} \ket{\Psi_l}=l \ket{\Psi_l}~  \Pi_l \rho \Pi_l =0,~\text{where}~\Pi_l = \ketbra{\Psi_l}{\Psi_l},
\end{align}
where $N$ is the total number of particles operator, and
\begin{align}
    \rho \in \mathcal{E}_\mathcal{F}^{(m,n)} \implies \forall E > E_{\max},~ \forall 1 \le i \le n,~  \Pi_E\Tr_{\neq i}[\rho]\Pi_E =0,~\text{where}~\Pi_E = \ketbra{\psi_E}{\psi_E}:~{\mathrm H} \ket{\psi_E}=E \ket{\psi_E}.
\end{align}
Furthermore, let $\rho,~\rho^\prime$, be such that $\norm{\rho-\rho^\prime}_1 < \epsilon$ for some $\epsilon >0$. Then energies $E$ and $E^\prime$ associated with states $\rho$, $\rho^\prime$ satisfy
\begin{align}
    \abs{E-E^\prime} \le \epsilon  n {E_{\max}},
\end{align}
where $E=\Tr[{\mathrm H}_F \rho ]$, and $E^{\prime}=\Tr[{\mathrm H}_F \rho^{\prime}]$.
\label{lem:1orMoreparticleEnegyContinuityB}
\end{lemma}
\begin{proof}
    The proof follows similar lines as the proof of Lemma~\ref{lem:1particleEnegyContinuity}, up to some modifications. First, let us define the projection on the subspace corresponding to the set $\mathcal{E}_\mathcal{F}^{(m,n)}$.
    \begin{align}
        \Pi_{\max}^{(m,n)} =  \left(\bigoplus_{l=m}^{n} \left(\int_0^{E_{\max}}dE\Pi_E \right)^{\otimes l}\right) \left(\bigoplus_{i=0}^{+\infty} \Pi^{(m,n)}_i\right),
    \end{align}
    where $\Pi^{(m,n)}_i \in \mathcal{H}^{\otimes i} $ such that $\Pi^{(m,n)}_i = \mathbf{1}$ when $m \le i \le n$ and $\Pi^{(m,n)}_i = 0$ otherwise. Then
    \begin{align}
        &\abs{E-E^\prime} 
        \stackrel{(I)}{=} \abs{\Tr[{\mathrm H}_F\rho]-\Tr[{\mathrm H}_F\rho^\prime]} 
        \stackrel{(II)}{=}  \abs{\Tr[{\mathrm H}\Pi_{\max}^{(m,n)}\rho \Pi_{\max}^{(m,n)}]-\Tr[{\mathrm H}\Pi_{\max}^{(m,n)}\rho^\prime\Pi_{\max}^{(m,n)}]} \nonumber\\
        &\stackrel{(III)}{=}  \abs{\Tr[\Pi_{\max}^{(m,n)} {\mathrm H}\Pi_{\max}^{(m,n)}\rho ]-\Tr[\Pi_{\max}^{(m,n)}{\mathrm H}\Pi_{\max}^{(m,n)}\rho^\prime ]} \nonumber \\
        &\stackrel{(IV)}{=}  \abs{\Tr[{\mathrm H}_\mathcal{E}\rho ]-\Tr[{\mathrm H}_\mathcal{E}\rho^\prime ]}
        \stackrel{(V)}{=} \abs{\Tr[{\mathrm H}_\mathcal{E}(\rho-\rho^\prime)]}  
        \stackrel{(VI)}{=} \left(\max \sigma\left( \mathrm{H}_{\mathcal{E}}\right) \right) \abs{ \Tr[\frac{{\mathrm H}_\mathcal{E}}{(\max \sigma\left( \mathrm{H}_{\mathcal{E}}\right)}(\rho-\rho^\prime)]} \nonumber \\
        &\stackrel{(VII)}{\le} \left(\max \sigma\left( \mathrm{H}_{\mathcal{E}}\right)\right) \max_{0\le \Omega \le \mathbf{1}}  \Tr[\Omega(\rho-\rho^\prime)] \stackrel{(VIII)}{=} \left(\max \sigma\left( \mathrm{H}_{\mathcal{E}}\right) \right) \frac 12 \norm{\rho - \rho^{\prime} }_1  \le \frac 12 \epsilon \left(\max \sigma\left( \mathrm{H}_{\mathcal{E}}\right) \right),
    \end{align}
    where the truncated Hamiltonian ${\mathrm H}_\mathcal{E} \equiv \Pi_{\max}^{(m,n)} {\mathrm H}\Pi_{\max}^{(m,n)}$ satisfies $0 \le {\mathrm H}_\mathcal{E} \le E_{\max} \mathbf{1}$ by construction. In $(I)$ we employed the definition of expected value of Hamiltonian (aka average energy). In $(II)$ we used a fact that the states $\rho$ and $\rho^\prime$ are supported on a specific subspace. In $(III)$ we employed the cyclic property of the trace, to define and denote the truncated Hamiltonian $\mathrm{H}_\mathcal{E}$ in $(IV)$. In $(V)$ the linearity of trace was used. In $(VI)$ we normalized the truncated Hamiltonian, by dividing it by its maximal eigenvalue. Consequently, the inequality in $(VII)$ is due to the fact that the normalized truncated Hamiltonian does not (in generic case) saturate the introduced supremum. Finally in $(VIII)$, we employed Lemma~9.1.1. of Ref.~\cite{MarkWildeBook}
    \begin{align}
        \frac 12 \norm{\rho-\rho^\prime}_1 =  \sup_{0\le \Omega \le \mathbf{1}}  \Tr[\Omega(\rho-\rho^\prime)].
    \end{align}
        The maximal energy of $m$ to $n$ non-interacting particles with maximal energy of single particle not exceeding $E_{\max}$ is $n \times E_{\max}$, therefore
    \begin{align}
        \max \sigma\left( \mathrm{H}_{\mathcal{E}}\right)  = n E_{\max}.
    \end{align}
        Finally, we obtain:
    \begin{align}
        \abs{E-E^\prime} \le \epsilon n E_{\max}.
    \end{align}
\end{proof}

\begin{corollary}
    Let the Hilbert space of a single particle be isomorphic to the following product of Hilbert spaces
    \begin{align}
        \mathcal{H} \cong \mathcal{H}^\prime \otimes \mathbf{C}^d,~~0<d \le + \infty,
    \end{align}
    (typically $\mathcal{H}^\prime \cong \bf{L}^2(\mathbf{R}^3)$), where $\mathbf{C}^d$ corresponds to some discrete degree of freedom. Then $n$ of Lemma~\ref{lem:1orMoreparticleEnegyContinuityB} above satisfies $n=\log_d \mathrm{dim}_L (\rho)$, where $\mathrm{dim}_L (\rho)$ is the maximal dimension (with respect to number of particles on which state $\rho$ is supported) of multi-particle Hilbert space associated with the mentioned distinguished, discrete degree of freedom, i.e., the logical subspace. In the above case the final inequality of Lemma~\ref{lem:1orMoreparticleEnegyContinuityB} takes the form
    \begin{align}
    \abs{E-E^\prime} \le \frac{\epsilon}{2} \log_d \mathrm{dim}_L (\rho) {E_{\max}},
\end{align}
    and in this shapes constitutes the asymptotic continuity property for the expectation value of energy.
\end{corollary}
We finish the above considerations on the "energetic proximity" with the following Lemma
\begin{rem}
    In the realistic situation $E_{\max}$ is always finite. This is because typically the full dynamics of the experiment/routine prohibits such states, e.g., optical fiber filters out photons with energies above some threshold limit, or trapped-ion with too much energy would escape the potential well, so that they would not be trapped any longer. The alternative and complementary approach is the one in which the effective Hamiltonian of the system, that incorporates only the selected degrees of freedom, is finite-dimensional and therefore bounded from above.
\end{rem}
The "energetic proximity" of states will be an important tool in our forthcoming considerations.

%Assumption 1: For reasonable Hamiltonians and encoding of entanglement in photons, energy is continuous so $E(\rho^{out}_{AB})/\lfloor \log d_{out} \rfloor\geq E(\psi^+_{d_{out}})/\lfloor\log d_{out}\rfloor-f(\epsilon)$
%with $f(\epsilon)\rightarrow 0$ for $\epsilon\rightarrow 0$. [Marek says:] In some cases, e.g. when entanglement is encoded in polarization, $f(\epsilon)=0$ [Marek pls expand] \textcolor{red}{Marek: sure, wkiipedia $\to$ quantization of EM field $\to$ free Hamiltonian $\to$ Energy does  not depend on polarization degree of freeedom.}
\begin{theorem}
The fundamental energy consumption during entanglement generation via quantum channel $\Phi$ admits a bound
\begin{equation}
C^\epsilon(\Phi|\mathrm{Ent})\leq 
\inf_{\mathrm{PR}}\,\inf_{\substack{d_{out}({\mathrm{PR}})\in \mathrm{N}_+ \,\\\Lambda({\mathrm{PR}}) \in LOCC}} \,
    \left[\frac{ E(\psi^+_{d_{in}})+ E(\Lambda,\id\otimes \Phi(\psi_{d_{in}}^+)) -E(\psi^+_{d_{out}})}{\log_2 d_{out}} +\epsilon E_{max},:
    \Lambda(\Phi (\psi^+_{d_{in}}))\approx_\epsilon \psi^+_{d_{out}}\right],
    \label{eq:en_dissip_photon}
\end{equation}
where $|\psi^+_{d_{out}}\>=\frac{1}{\sqrt{d_{out}}} \sum_{i=0}^{d_{out}-1}|ii\>$ is the singlet state, and $E_{max}$ is a maximal energy of a single particle in our setup. 
\label{thm:main}
\end{theorem}
\begin{proof}
Let us first recall that by Definition \ref{def:ent-main},
\begin{align}
C^{\epsilon}(\Phi|\mathrm{Ent}):=\inf_{\mathrm{PR}}\,\inf_{\substack{d_{out}({\mathrm{PR}})\in \mathrm{N}_+ \,\\\Lambda({\mathrm{PR}}) \in LOCC, \,\\\rho_{A''B''}^{(in)}({\mathrm{PR}})}} \,
    &\left[\frac{ E(\rho_{A''B''}^{(in)})+ E(\Lambda,\rho^{(int)}_{A'B'}) -E(\rho_{AB}^{(out)})}{\log_2 d_{out}}\,:\rho^{out}_{AB}:=
    \Lambda(\Phi (\rho_{A''B''}^{(in)}))\approx_\epsilon \psi^+_{d_{out}}\right],
    \label{eq:intheproof}
    \end{align}
    where the intermediate state $\rho^{int}_{A'B'}$ is $\rho^{int}_{A'B'}:=\Phi(\rho^{(in)}_{A''B''})$, PR means Physical Realizations, LOCC stems for Local Operations and Classical Communication,  and $\Lambda : A'B'\rightarrow AB$ are LOCC applied after the channel $\Phi$.

Now, we first fixed the input state $\rho_{A''B''}^{(in)}$ to be singlet $|\psi^+_{d_{in}}\>$, which increases the RHS of the
above Eq. \eqref{eq:intheproof}, since it narrows the infimum over input states.
We further use the continuity argument of Lemma \ref{lem:1orMoreparticleEnegyContinuityB} to modify the final state. Namely we use the fact, that
\begin{equation}
|E(\rho_{AB}^{(out)})- E(\psi^+_{d_{out}})| \leq \epsilon \log_2 d_{out} E_{max}
\end{equation}
hence, lower bounding $E(\rho_{AB}^{(out)})$ by $E(\psi_{d_{out}}^+) - \epsilon \log_2 d_{out} E_{max}$ we arrived at
  
    \begin{align}
    &C^{\epsilon}(\Phi|\mathrm{Ent})\leq\\&\inf_{\mathrm{PR}}\,\inf_{\substack{d_{out}({\mathrm{PR}})\in \mathrm{N}_+ \,\\\Lambda({\mathrm{PR}}) \in LOCC}} \,
        &\left[\frac{ E(\psi_{d_{in}}^{+})+ E(\Lambda,\rho^{(int)}_{A'B'}) -(E(\psi^+_{d_{out}})-\epsilon\log_2 d_{out}E_{max})}{\log_2 d_{out}}\,:\rho^{out}_{AB}:=
        \Lambda(\Phi (\psi_{d_{in}}^{+}))\approx_\epsilon \psi^+_{d_{out}}\right],
        \label{eq:usinglemma1}
    \end{align}
    
    which is equal to the Eq (\ref{eq:en_dissip_photon}) that we claimed in the thesis.
\end{proof}

%\rj{This theorem proves that there is an upper bound to the fundamental energy consumption of generating entanglement between two parties using a quantum channel. This implies that Alice shares a quantum state with Bob through $\Phi$ and then they perform distillation protocol $\Lambda$ (thus sharing classical information?). Isn't $E_{max}$ making the bound go wild? Are there direct implication to the bound?} {\color{purple} LS: Answer to the comment: I don't think there are direct implications to the bound. $E_{max}$ is a finite constant for any choice of Physical Realization $\mathrm{PR}$, so for each $\mathrm{PR}$ the term $\epsilon E_{max}$ does not go wild.}
\subsection{Technical facts}
\begin{observation}
For a bounded sequence of non-negative numbers $\{a_n\}_n$,  we have
\begin{equation}
\liminf_{n\rightarrow \infty} \frac{1}{a_n} =
\frac{ 1}
{\limsup_{n\rightarrow \infty} a_n}.
\end{equation}
\label{obs:liminf_limsup}
\end{observation}
\begin{proof}
We first note that for any fixed $n$, we have
\begin{align}
\inf\{\frac{1}{a_k} ; k\geq n\} =\frac{1}{\sup\{a_k ; k\geq n\}}
\label{eq:inf_sup}
\end{align}
Taking limit on both sides we get
\begin{align}
\liminf_{n\rightarrow \infty} \frac{1}{a_n} =
\lim_{n\rightarrow \infty}
\inf\{\frac{1}{a_k} ; k\geq n\} =
\label{eq:defliminf}\\
\lim_{n\rightarrow \infty}\frac{1}{\sup\{a_k ; k\geq n\}}= \label{eq:limProp}\\
\frac{1}{\lim\limits_{n\rightarrow \infty}{\sup\{a_k ; k\geq n\}}} =
\frac{1}{\limsup\limits_{n\rightarrow \infty} a_n}
\end{align}
Indeed, the equality (\ref{eq:inf_sup}) follows from the definitions of infimum and supremum.
Next the equality (\ref{eq:defliminf}) is by the definition of liminf, and equality (\ref{eq:limProp}) by property of limits.
In the last equality we use definition of limsup.
%\begin{align}
%\liminf_{n\rightarrow \infty} \frac{a_n}{b_n} = \\
%\sup_{n\geq 0}\inf_{k\geq n} \frac{a_k}{b_k} \geq \\
%\sup_{n\geq 0} \inf_{k\geq n}\frac{\inf_{k\geq n} a_k}{b_k} \geq  \\
%\sup_{n\geq 0} \frac{\inf_{k\geq n} a_k}{\sup_{k\geq n} b_k} 
%\geq \\\sup_{n\geq 0} \frac{\inf_{k\geq n} a_k}{\inf_{n\geq 0}\sup_{k\geq n} b_k} =\\
%\frac{\liminf_{n\rightarrow \infty} a_n}{
%\limsup_{n\rightarrow \infty} b_n
%}
%\end{align}
\end{proof}

%LSBIB\putbib
%LSBIB\end{bibunit}
\bibliography{References}
\end{document}